\documentclass[twocolumn]{AASTeX61}

\usepackage{mathrsfs}
\usepackage[T1]{fontenc}
\usepackage{fourier}
\usepackage{hyperref}
\usepackage[normalem]{ulem}
\newcommand{\cdm}{ } %{\color{blue}}
\newcommand{\cdmnew}{ } %{\color{magenta} }
\newcommand{\cdmnow}{}%{\color{magenta}}
\newcommand{\nil}{}%{\color{purple}}
\newcommand{\rev}{}
\newcommand{\revsec}{}
\newcommand{\revthird}{}%{\textbf}
\newcommand{\lphi}{\ell_{\varphi}}
\newcommand{\rhophi}{\rho_{\varphi}}
\newcommand{\vphi}{v_{\varphi}}
\newcommand{\tsh}{t_s}    % time of shock crossing at a given location... 
\newcommand{\tshock}{t_{\rm sh}}   % shock transition time, not the same
\newcommand{\vsh}{v_s}
\newcommand{\tff}{t_{\rm f-f}}
\newcommand{\vshvec}{{\mathbf v}_s}
\newcommand{\calR}{\mathscr{R}}
\newcommand{\calD}{\mathcal{D}}
\newcommand{\Sigmadiff}{\Sigma_{\rm diff}}
\newcommand{\tdiff}{t_{\rm diff}}
\newcommand{\rdiff}{r_{\rm diff}}
\newcommand{\tdyn}{t_{\rm dyn}}
\newcommand{\nBB}{n_{\rm BB}}
\newcommand{\TBB}{T_{\rm BB}}
\newcommand{\ndotem}{\dot n_{\rm em}}
\newcommand{\Mstar}{M_{\rm ej}} 
\newcommand{\vvec}{{\mathbf v}}
\newcommand{\Cff}{C_{\rm ff}} % free-free emission constant 

\usepackage[english]{babel}															% English language/hyphenation
\usepackage[protrusion=true,expansion=true]{microtype}	
\usepackage{amsmath,amsfonts,amsthm} % Math packages

\usepackage{subfigure}
\usepackage{natbib}
\shorttitle{Aspherical Supernovae: Early Light }
\shortauthors{AFSARIARDCHI \& MATZNER}

\begin{document}


\title{Aspherical Supernovae: Effects on Early Light Curves}
%\title{Hydro Dynamic Simulations of Axisymmetric Supernovae Explosion}


\author{Niloufar Afsariardchi*}

\affil{Department {of Astronomy and Astrophysics, 50 St. George St., Toronto, ON M5S 3H4, Canada}, University of Toronto}
%\affil{Astronomy Department, University of Toronto, Toronto}
\email{*afsariardchi@astro.utoronto.ca}

%:
\author{Christopher D.\ Matzner}
\affil{Department {of Astronomy and Astrophysics, 50 St. George St., Toronto, ON M5S 3H4, Canada}, University of Toronto}
%\affil{Astronomy Department, University of Toronto, Toronto}


\begin{abstract}
Early light from core-collapse supernovae, now detectable in high-cadence surveys, holds clues to a star and its environment just before it explodes.  However, effects that alter the early light have not been fully explored.  We highlight the possibility of non-radial flows at the time of shock breakout. These develop in sufficiently non-spherical explosions if the progenitor is not too diffuse.  When they do develop, non-radial flows limit ejecta speeds and cause ejecta-ejecta collisions.  We explore these phenomena and their observational implications, using global, axisymmetric, non-relativistic FLASH simulations of simplified polytropic progenitors, which we scale to representative stars.   We develop a method to track photon production within the ejecta, enabling us to estimate band-dependent light curves from adiabatic simulations. Immediate breakout emission becomes hidden as an oblique flow develops. Non-spherical effects lead the shock-heated ejecta to release a more constant luminosity at a \revthird{higher}, evolving color temperature at early times, effectively mixing breakout light with the early light curve.   Collisions between non-radial ejecta thermalize a small fraction of the explosion energy; we address emission from these collisions in a subsequent paper.


Online-only material: color figures
\end{abstract}

\section{Introduction}
\label{sec:intro}

The discovery volume for time-domain astronomy is expanding rapidly as new surveys come on line.   For core collapse events, the early supernova (SN) light is a key target, as this carries information about the final moments of a star's evolution -- like its radius, wind state, and terminal activity -- that is difficult or impossible to glean from later epochs.   Furthermore, the theory of early SN emission is fairly mature \citep{Nakar2010,Rabinak2011}.   Insofar as an explosion can be considered spherical,  it evolves as follows.  The shock front that defines the SN explosion speeds up as it crosses the thinning density zones of the star's outer layers.   Photons dominate the post-shock pressure; therefore a shock of speed $v_s$ has a characteristic  width, corresponding to the optical depth $\tau_s \simeq c/v_s$, set by photon diffusion.  At some point where $\tau_s$ falls below the optical depth to free space, photons leak away in a `breakout' flash \citep{Klein1978}.  While detailed predictions require simulations, basic properties of this flash, as well as the ejecta's density and temperature profiles in velocity, are all related to the original stellar structure and its explosion energy in a rather deterministic way \citep{Matzner1999,Tan2001}.    The  light curve begins immediately after breakout; as soon as the ejecta have traveled a few times the original stellar radius, self-similar diffusion leads to a power-law decline of the total velocity \citep{Chevalier1992}.   Larger progenitors tend make longer, redder, more energetic flashes and redder early light curves (with lower maximum speeds).  But, explosions within compact progenitors can become relativistic, leading in some cases to low-luminosity gamma-ray bursts \citep{Tan2001,Nakar2012}.  Behind fast shocks ($v_s/c\gtrsim 0.1$), emission and absorption do not have time to equilibrate \citep{Katz2010,Katz2012, Sapir2011, Sapir2013}.   Finally, massive stars produce winds, and the ejecta-wind collision leads to synchrotron and free-free emission \citep[e.g.,][]{Fransson1996}.  However an optically thick wind, like those surrounding Wolf-Rayet stars, will alter and enhance the breakout flash \citep{Chevalier2011}.  Spherical theory has been applied widely, from the ionization of rings around SN 1987A, to the radio shell around SN 1993J, the inference of relativistic ejecta from SN 1998bw, the x-ray flashes from SN 2008D and XRF 060218, the early light curve of SN 2011dh, and early SN observations from CFHT-SNLS, {\em GALEX}, PTF, and {\em Kepler} surveys, to name a few. 

However spherical theory may be misleading if the underlying event is aspherical in an important way.  This would not be surprising, considering that the central engine is thought to  involve large-scale instability \citep{Blondin2003} or magnetorotational energy extraction \citep{Akiyama2003}.   Linear polarization of the line and continuum emission has long provided evidence of non-spherical ejecta, especially at low velocities \citep{Leonard2001,Leonard2006} and for stripped-envelope progenitors \citep{2005ASPC..342..330L,2008ARA&A..46..433W}.  Likewise, young SN remnants like Cassiopeia A show a complicated distribution of ejecta \citep[e.g.,][]{Lee2017}.  The stellar envelope may also be distorted by rapid rotation or tides from a companion.   Moreover there is a growing list of discrepancies between observations and spherical-theory expectations regarding early SN light (see \S \ref{sec:summary}) and it is important to consider the alternatives. 
{\cdm This is especially clear when unexpected features of the early light curve are attributed to plumes of $^{56}$Ni, as any process that moves matter from the central engine to high ejecta velocities must be very strongly aspherical. }

How might an aspherical explosion alter the early SN light?  Previous studies offer somewhat disparate answers.   \citet{Calzavara2004} considered only variations in the timing and intensity of shock breakout, effects calculated by \citet{Suzuki2010} within a simple model for the evolution of the breakout.   \citet{Couch2009} and \citet{Couch2011} perform adiabatic, axisymmetric simulations of jet-driven explosions and analyze the results to infer properties of the early light.   Examining the breakout dynamics, \citet{Matzner2013} point out that aspherical explosions can undergo a transition to distinctly different behavior, predictions \citet{Salbi2014} verify on the basis of simulations that address a small zone near the stellar surface.    Most recently, \citet{Suzuki2016} simulate a mildly aspherical explosion using a radiation hydrodynamics calculation. 

We present a new set of simulations and model light curves with two goals in mind.  First, we wish to realize and test the predictions of \citet{Matzner2013} and \citet{Salbi2014}, or M13 and S14 for short, with global simulations.  Arguing from a limiting case in which the stellar atmosphere is effectively planar, M13 and S14 make the following predictions:
 \begin{itemize}  
\item[({i})] Outward acceleration of the explosion shock ceases at the depth $\ell_\varphi$, where the (radial) shock speed matches the (horizontal) pattern speed $v_\varphi$ of breakout across the stellar surface.  
\item[({ii})] Matter from this region sprays out in a non-radial fashion (with a specific distribution in angle). 
\item[({iii})] The maximum ejecta velocity is limited to $2 v_\varphi$ (when $v_\varphi$ is non-relativistic).  
\item[({iv})] Non-radial flows collide outside the star, providing a new energy source for early SN emission.  
\item[({v})] `Oblique' breakout is hidden from the observer by an optically thick spray of ejecta, except, in some cases, from certain lines of sight.  
\item[{(vi})] Non-radial flows affect the early luminosity and polarization of the ejecta.  (This prediction is also made by \citeauthor{Couch2009} and \citeauthor{Suzuki2016})  
\item[{(vii)}] None of these effects develop in regions so diffusive that shock breakout occurs below $\ell_\varphi$, or in adiabatic simulations that lack resolution of this scale.   Therefore, non-radial effects should be much stronger for compact progenitors than for red supergiant explosions. 
\end{itemize} 

A second goal of this work is to make specific predictions about the early SN emission that reflect emission and absorption as well as the scattering-dominated diffusion of photons.  These effects are treated in the spherical case by \citet{Nakar2010}, and \citet{Couch2009,Couch2011} introduced an approximate treatment for their aspherical explosion simulations.   We will introduce an improved approximation that applies to the expanding ejecta, and also take care to consider the ejecta collision zone separately.   (We consider only the collisions' dynamics and energetics here, saving the emission from this process for a future paper.) 

We are mindful of two potential pitfalls.  One concerns the many ways  an explosion can be non-spherical: even considering only one model, like jet-driven explosions within spherical stars, there are many possible outcomes depending on parameters such as the jet's structure and duration.  Moreover the physics of shock breakout imply that optical depth affects how the outermost ejecta respond to an aspherical explosion (see prediction (vii) above), in addition to the emission from these ejecta; this introduces still more parameters.   

Another possible pitfall involves the complexity of the radiation processes at work.  Robust predictions of  shock breakout emission, for instance, require multigroup radiation hydrodynamics  simulations \citep[e.g.,][]{Tominaga2011} which are beyond our capability for aspherical flows.   Likewise the collision of non-radial ejecta may involve effects like cosmic ray acceleration that are not easily accommodated within our numerical code. 

For these reasons we adopt an approach similar to that of \citeauthor{Couch2011}, but opt for simplicity in our initial model.  Specifically, we perform axisymmetric, \revthird{non-relativistic, adiabatic} explosions within the Flash hydrodynamics code, ignoring details like self-gravity and role of gas pressure in the equation of state.  We adopt a simplified stellar structure (a spherical, $n=3$ polytrope), and we make the explosion aspherical by fiat rather than driving it with a central jet.   

Such simplifications have obvious drawbacks.  Because our simulations do not include radiation transfer, they do not include the physics \revthird{that limit} ejecta speeds in spherical explosions, or prevent the development of non-radial flows in aspherical explosions of red giants.   Because we use a single polytrope we cannot address shock ejection within realistic stellar atmospheres.  Because we make no attempt to emulate a central engine, we cannot model its effect on the shape of the explosion.  

But simplicity has advantages.  A single simulation can be scaled to the radius, energy and optical depth of a variety of supernovae.  Handling radiation transfer in post-processing allows us to specify exactly which aspects of a simulation (like its non-radial flow) are inconsistent with a given context, and in fact our predicted light curves are not badly affected by this type of inconsistency.  We are able to capture overall differences between an aspherical event and its spherical counterpart that might otherwise be buried in the details.

This work is organized as follows. In $\S$\ref{sec:sim}, we describe the numerical approach and details of our  axisymmetric model. In $\S$\ref{sec:res}, we provide a general description of our simulation results, from the shock propagation to the circumstellar collisions. We also discuss the numerical effect of the resolution and  shock speed derivation method. In $\S$\ref{sec:oblique}, we comment on the conditions required for the formation of oblique shock breakout. $\S$\ref{sec:diff} is dedicated to the observational implications of our work including  
radiation diffusion, thermalization, and light curve modeling. We briefly explore the potential effects of ejecta-ejecta collisions in $\S$\ref{sec:col}. A general summary of our results is provided in $\S$\ref{sec:summary}.  We conclude and set goals for future work in $\S$\ref{sec:conc}.

\section{Physical Problem and Numerical Implementation}
\label{sec:sim}
We construct a global numerical FLASH hydrodynamic  \citep{Fryxell2000} simulation in a two-dimensional uniform grid in \revthird{spherical polar coordinates}. We ignore stellar gravity and hydrostatic pressure because the explosion energy is usually much higher than the binding energy of {the stellar envelope}. We also ignore the diffusion of radiation in the simulation; this {leads to errors in the prediction of some of the fastest ejecta, which we account for in our post-processing.}
 We comment on the radiation processes and diffusion in $\S$\ref{sec:oblique}. Furthermore, we consider the case where the post-shock flow is radiation-dominated (i.e., adiabatic index $\gamma= 4/3$) and the initial density profile is of an  $n=3$  polytrope, mainly because this reasonably represents the envelope structure of various progenitors and its density profile can be specified at arbitrary high resolution.  {\cdm (While $n=3/2$ polytropes are often used to represent the convective envelopes of red supergiants, we fix the structure to highlight the influence of radiation processes and aspherical geometry.)} For simplicity, we assume purely non-relativistic motion. 

With all these simplifications our simulations themselves are scale free, apart from the numerical scales introduced by finite resolution and simulation volume; in physical terms they are described by the arbitrary scales of energy $E_*$, ejected mass $\Mstar$, and progenitor radius $R_*$ and derived quantities like the scales of density, $\rho_* = \Mstar/R_*^3$, velocity, $v_* = \sqrt{E_*/\Mstar}$, and time,  $t_* = R_*/v_*$.  {\cdm  (Additional radiation parameters, such as the electron scattering opacity $\kappa$, will enter our post-processing step.) }  We list characteristic scales for red super giants \revthird{(RSGs)}, blue super giants \revthird{(BSGs)}, and compact Type-Ic models in Table \ref{tab:models}. 

%So long as the fluid motion is non-relativistic and the radiation diffusion is ignored, the simulation is scale free. In all simulations the radius of the progenitor is $R_*=0.5$, stellar mass is $\Mstar=0.005$, explosion energy $E_*=0.01$, and box size is $\mathscr{L}=2$ where values are set in code units. Other characteristic parameters are then defined using progenitor parameters progenitor mass, energy, and radius: average density $\rho_*=\Mstar/R^3_*$, average pressure $P_*=E_*/R^3_*$, characteristic velocity $v_*=\sqrt{E_*/\Mstar}$, and characteristic time $t_*=R_*/v_*$. The simulation can then be easily normalized to the natural units of energy, length, and mass for different progenitors: red super giants (RSG), blue super giants (BSG), and compact Type-Ic models (Table \ref{tab:models}).

\begin{figure}[t]
  \centering
    \includegraphics[scale=0.65]{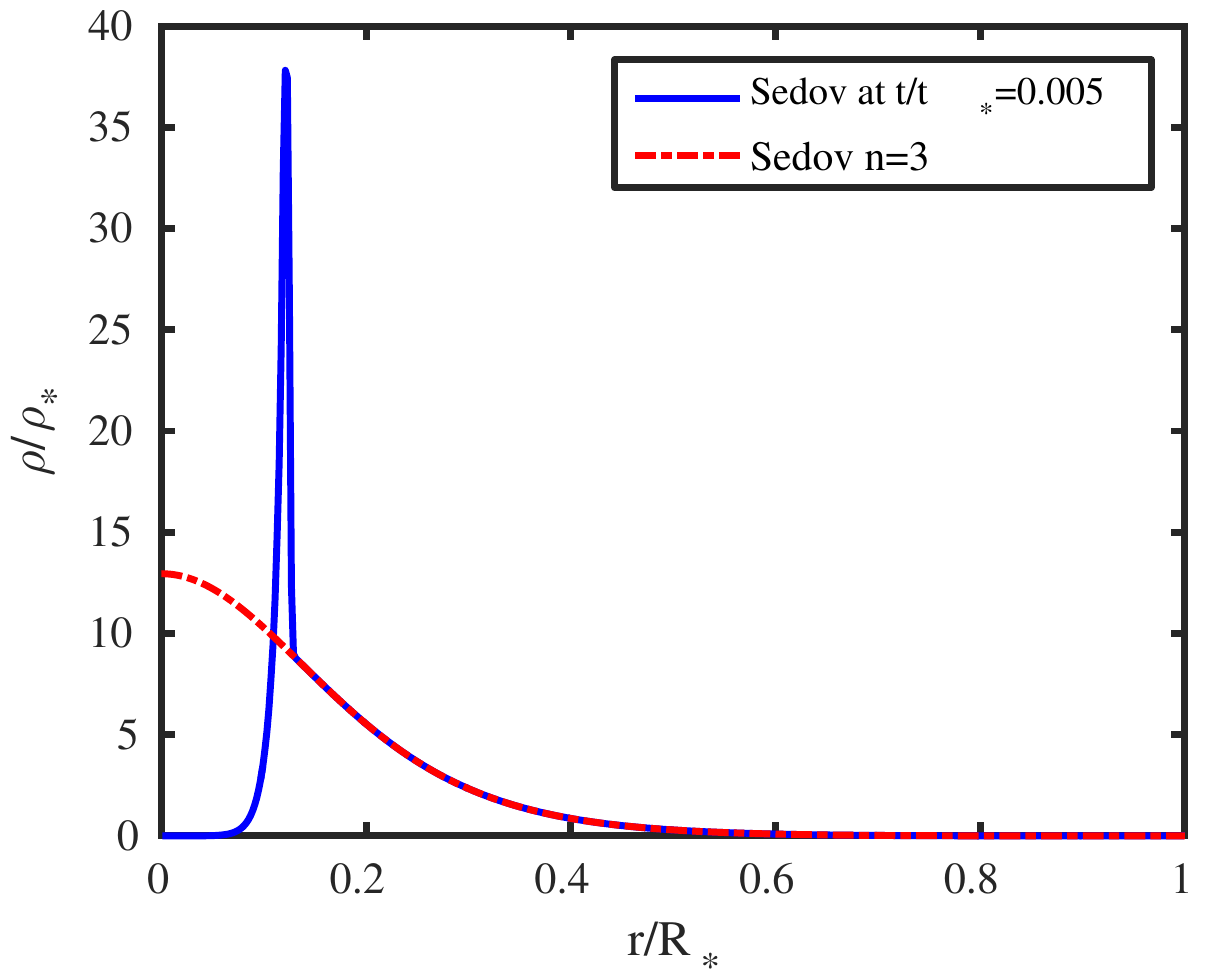}
    \caption{Initial density as a function of radius. Blue curve is the density profile of a $n=3$ polytrope and red curve is the density profile of a Sedov explosion triggered in the center of the polytropic progenitor by depositing a total energy  in cells where radius $r/R_*<0.001$ and letting it evolve until time $t/t_*=0.005$.}
    \label{denprofile}
\end{figure}

A difficulty that arises naturally while simulating asymmetric explosions is having many degrees of freedom in making axisymmetric explosions (e.g., {\cdm finite-duration jet}, off-axis explosion, {\cdm directed} momentum, prolate or oblate progenitor, etc.). The main focus of this work, however, is not exploring the parameter space of aspherical explosion but rather is studying the dynamic of shock emergence and its observational impacts. We thus initiate the asymmetric simulations by adding axisymmetric momentum to the outcome of a purely spherical explosion. 

\begin{deluxetable}{cccccccc}
\tabletypesize{\footnotesize}
\tablecolumns{8} 
\tablewidth{0pt}
 \tablecaption{ Model Core-collapse Supernovae
 \label{tab:models}}
 \tablehead{
 \colhead{Model} & \colhead{$\Mstar$ } & \colhead{$R_*$ } & \colhead{$E_*$ }& \colhead{X} & \colhead{$\kappa$}% &\textbf{n} 
 %& \colhead{}
 \\ 
 \colhead{} & \colhead{($M_\Sun$)} & \colhead{($R_\Sun$)} & \colhead{($10^{51}$ erg)} &  &\colhead{\cdm (cm$^2$ g$^{-1}$)}%& \colhead{} 
 %& \colhead{}
 } 
\startdata 
  RSG & 14 & 400 & 1 & 0.7 & 0.34 & %3 &
  \\ 
BSG & 15 & 49 &  1 & 0.7  &0.34 & % 3 & 
 \\ 
   Ic & 5 & 0.2 & 1 & 1 &0.2  & %& 3
\enddata
 \tablecomments{\cdm Parameters for scaling our model supernovae to represent various explosions.  In each case the dynamical model is a complete $n=3$ polytrope. }
\end{deluxetable}
\begin{figure*}

\gridline{\fig{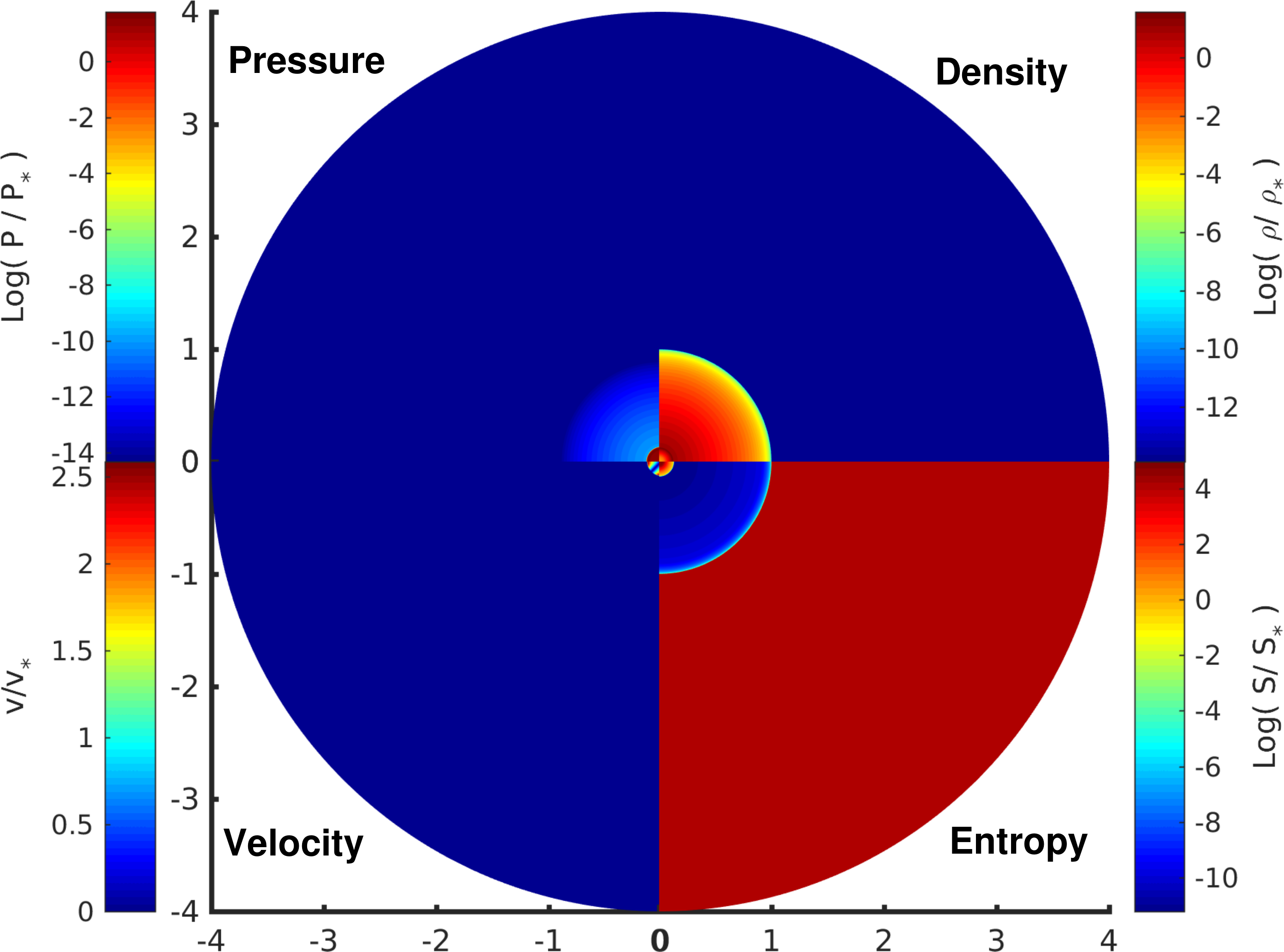}{0.5\textwidth}{(a) Initial conditions at $t/t_*=0$}
		  \fig{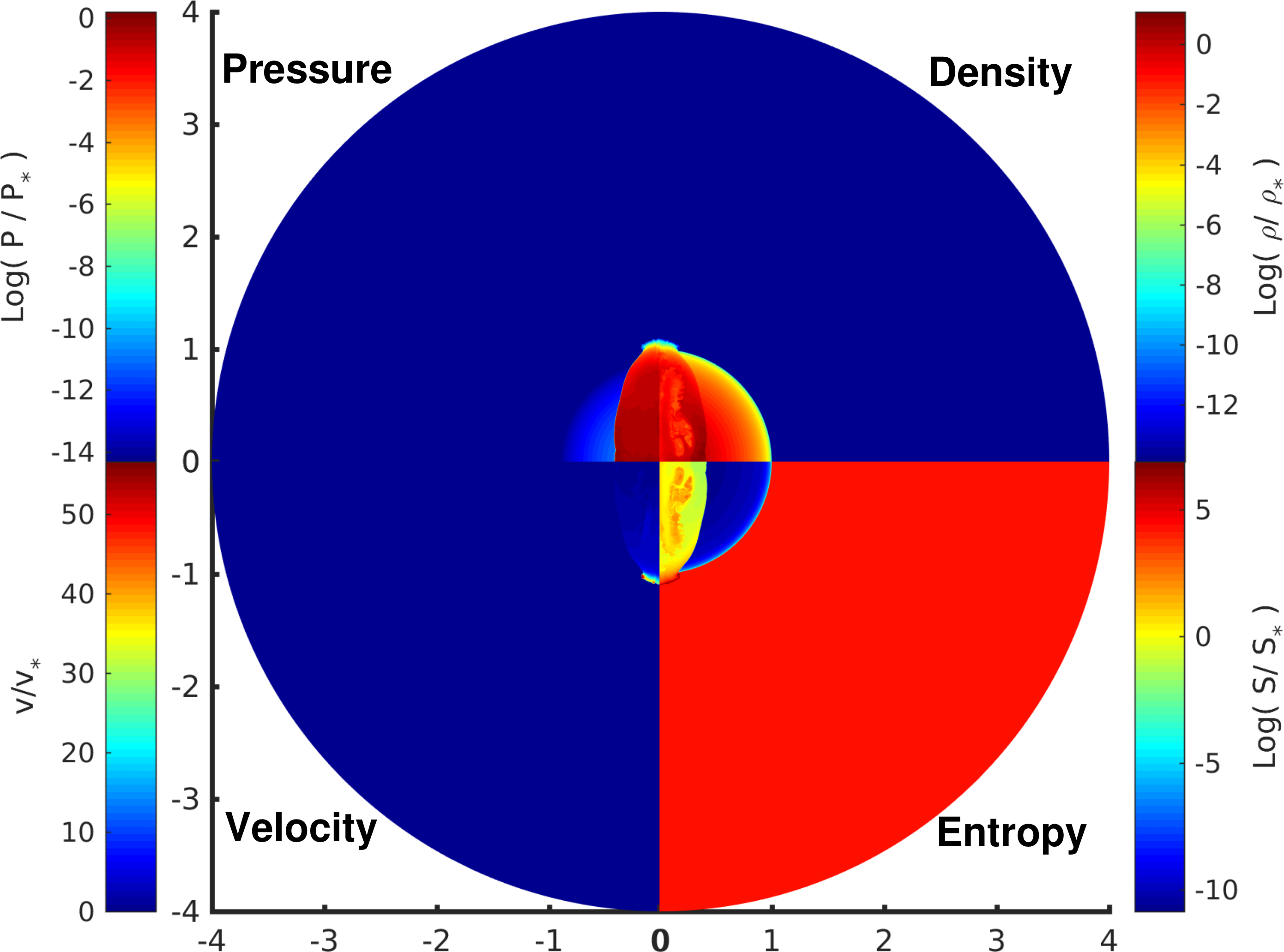}{0.5\textwidth}{(b) Breakout from pole at $t/t_*=0.28$} }

\gridline{\fig{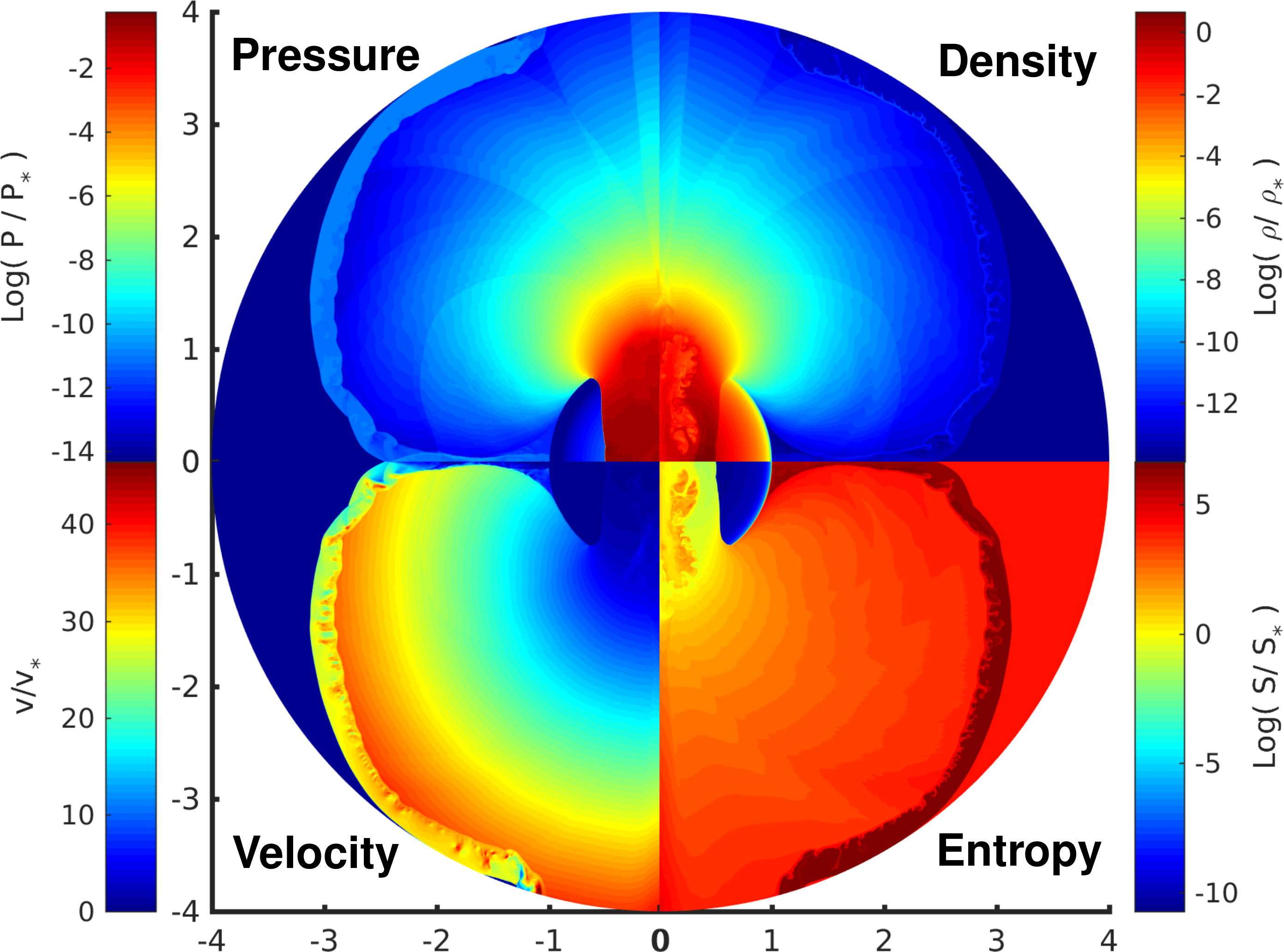}{0.5\textwidth}{(c) Transient evolution at $t/t_*=0.36$}
         \fig{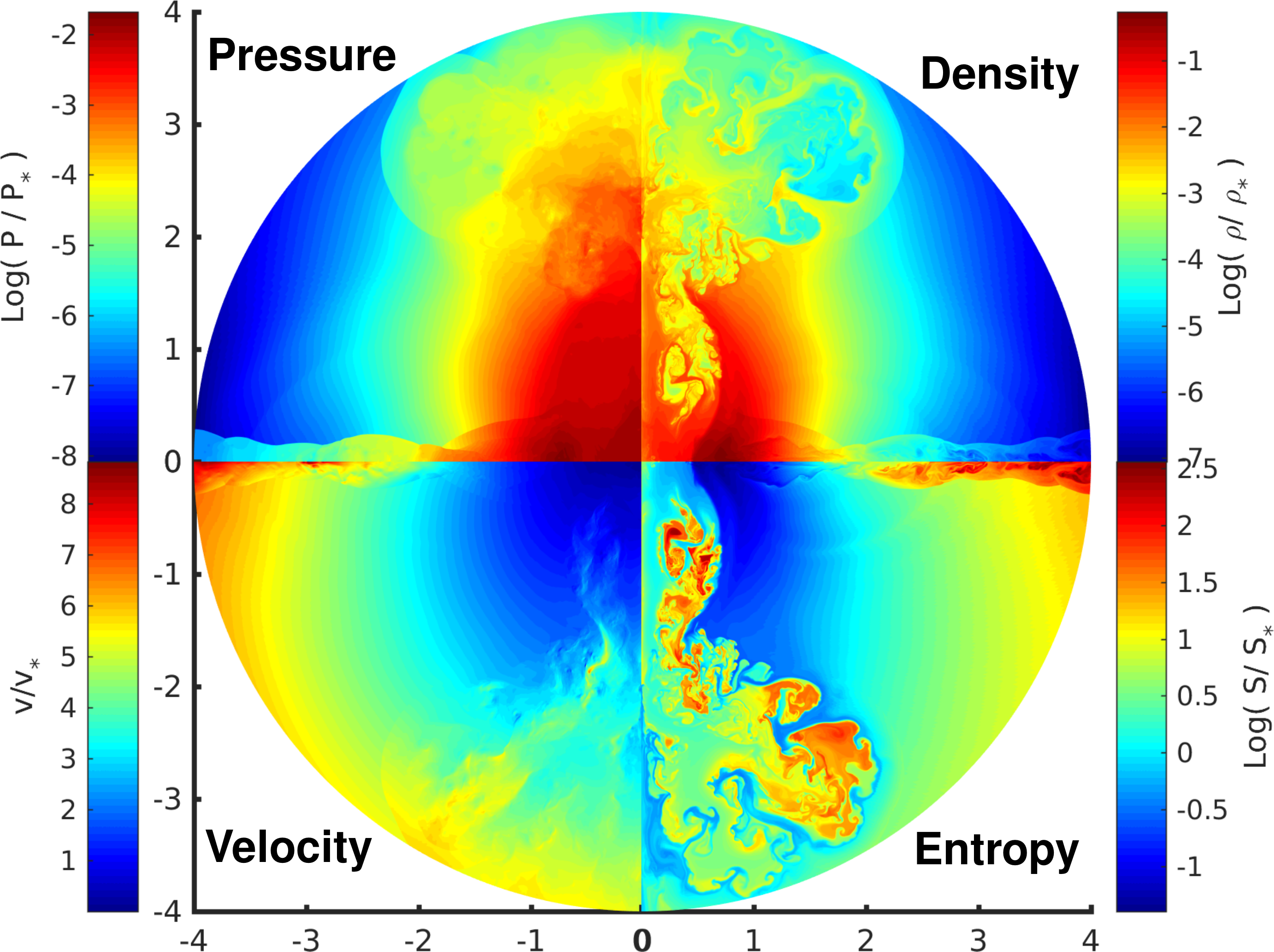}{0.5\textwidth}{(d) Circumstellar collisions at $t/t_*=0.9$}}

\caption{In each panel density distribution, pressure distribution, velocity field, and entropy distribution ($s \equiv P/\rho^{4/3}$) are shown in top right, top left, bottom left, and bottom right respectively for the \revthird{fiducial} run. The top left panel shows the initial density distribution. In top right plot, the bipolar shock is emerged from the pole. The bottom left panel depicts the spray of the ejecta outside the stellar surface and exhibits the kink in the shock front as was previously shown \revthird{by \citet{Salbi2014}}. The bottom right panel marks the time when the shock has completely traversed the progenitor and the circumstellar collisions is happening along the equator.  
%{\cdmcomment At some point it would be good to have big text labels on the four quadrants: $p,\rho,S,v$, so you don't have to read the colorbar to figure out what you're looking at ... but not important now. }
}
\label{fig:evolution}
\end{figure*}

{\cdm We initialize the explosion in a two-step process.  First we model a point explosion within the progenitor, resolving its radius with 2000 zones. }
%The explosion is first triggered spherically in the center of a $n=3$ polytrope with 2000 cells in the radial direction. The initial density profile is shown in Figure \ref{denprofile} (blue curve). The energy of $E_*=0.01$ is deposited in cells where radius $r<0.001$, resulting in a Sedov explosion with a sharp peak in the density and pressure profiles. The outcome of the spherical Sedov explosion is saved at time $t=0.005$ when the shock is at about $0.2R_*$. 
\revsec{ In our fiducial run, we modify the velocity profile  at the moment this explosion reaches $r_{\text{v}}=0.12 R_*$, where $r_{\text{v}}$ is the radius of velocity modification that varies the degree of asphericity (see   $\S$\ref{sub:asphericityParam}). Specifically, the velocity is modified by an angle-dependent scaling factor to induce non-spherical behavior:}
\begin{eqnarray}
\label{eq:velocitysubsonic}
v(r,\theta) = \hat{v}(r) |\cos (\theta)|.
\end{eqnarray}
where $\hat{v}(r)$ is the spherical velocity profile. We used this specific formula to 
%launch a bipolar angle-dependent velocity profile because it 
ensure that the velocity 
{\cdm remains subsonic}; i.e, $v(r,\theta) <= \hat{v}(r)$. {\cdm  The total energy of the modified explosion is then recorded for later scaling to the value of $E_*$ desired. }

For all of our simulations the angular extent of our two-dimensional grid is $\pi/2$ {\cdm and the outer grid boundary is set to allow outflow at $4R_*$}.  In this work, we do not attempt to simulate the effect of wind or mass-loss history.  The ambient density is therefore set to a small value, about $10^{-14} \rho_*$ or $10^{-5.6}$ times below the lowest stellar density, to prevent round-off numerical error. 
{\cdm We run the code until most of the mass has left the grid and record 100 checkpoints at regularly spaced intervals along the way.} 
{\cdm The number of cells in the radial and angular directions are both equal.  We refer to this number as } $\calR$; {\cdm keep in mind that there are $\calR/4$ cells per stellar radius.  We vary $\calR^{\cdm 2}$ from }$128^2$ to $2048^2$ and designate the highest of these resolution as our fiducial run.  
%{\cdmcomment [ It might be better to define $\calR = R_*/\Delta r$, because the star is a physical object while the grid isn't, but that is something we can fix later.  For instance Pegah defined $\calR= \lphi/\Delta x_{\rm min}$.]  } 

For this work, FLASH was compiled with Message Passing Interface (MPI) and HDF5 parallel enabled on up to 256 cores of the GPC cluster at the SciNet facility located at the University of Toronto \citep{Loken2010}. We employed a directionally split Piecewise Parabolic Method (PPM) Riemman solver with a Courant number of 0.8. 

\section{Results}
\label{sec:res}

\begin{figure*}

\gridline{\fig{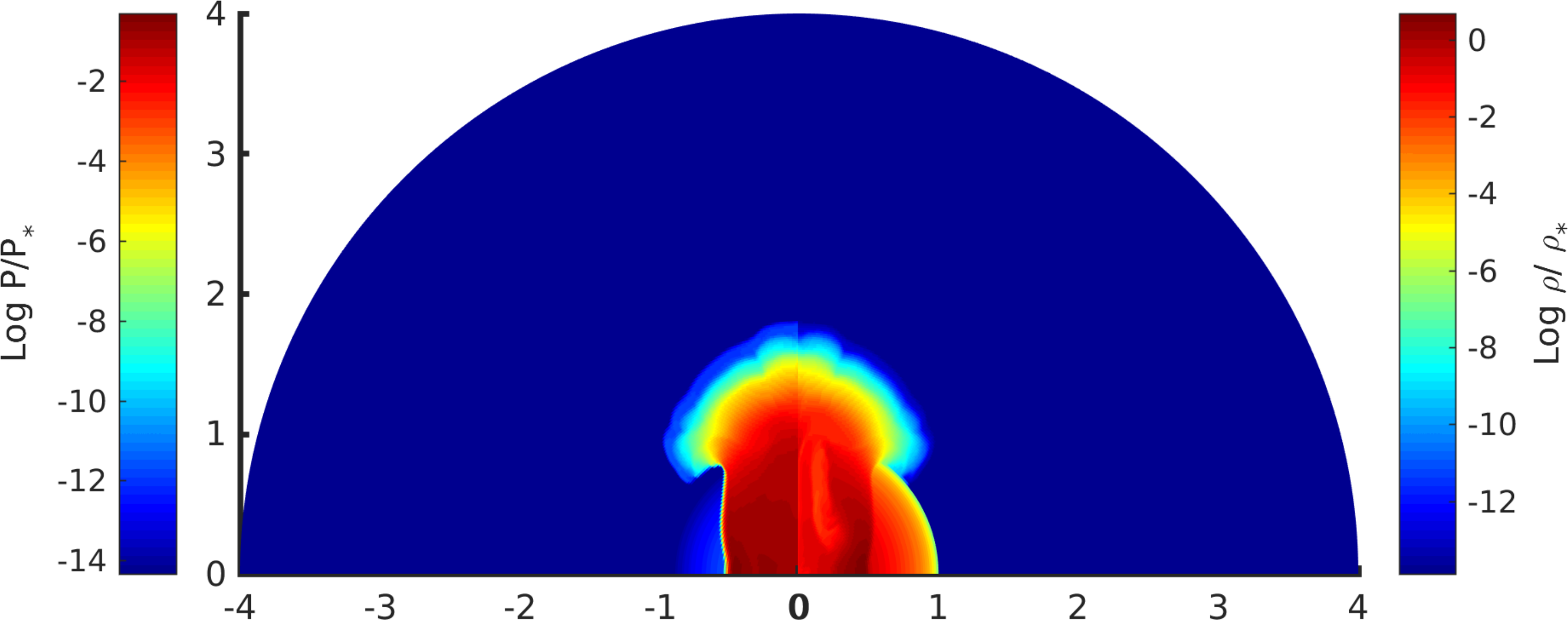}{0.7\textwidth}{(a) Density and Pressure distribution for $\calR=256$}}

\gridline{\fig{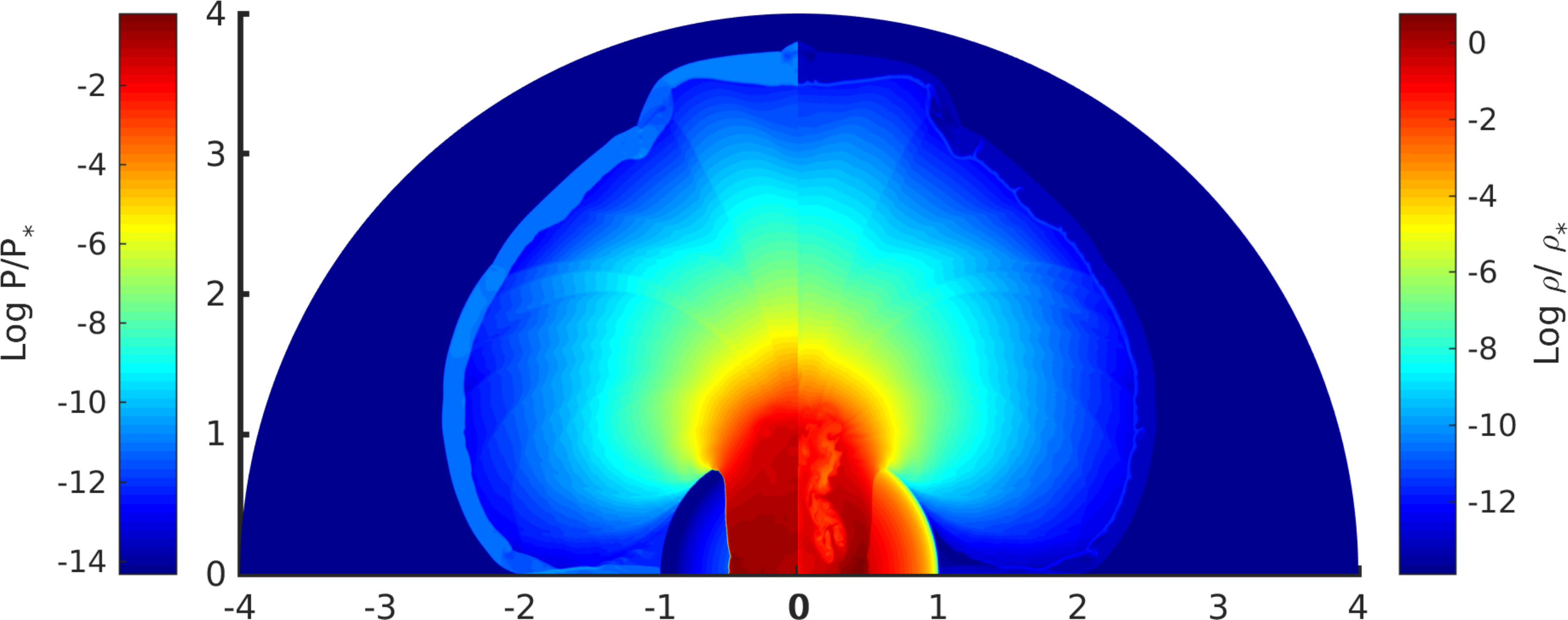}{0.7\textwidth}{(b) Density and Pressure distribution for $\calR=1024$}}
         
\gridline{\fig{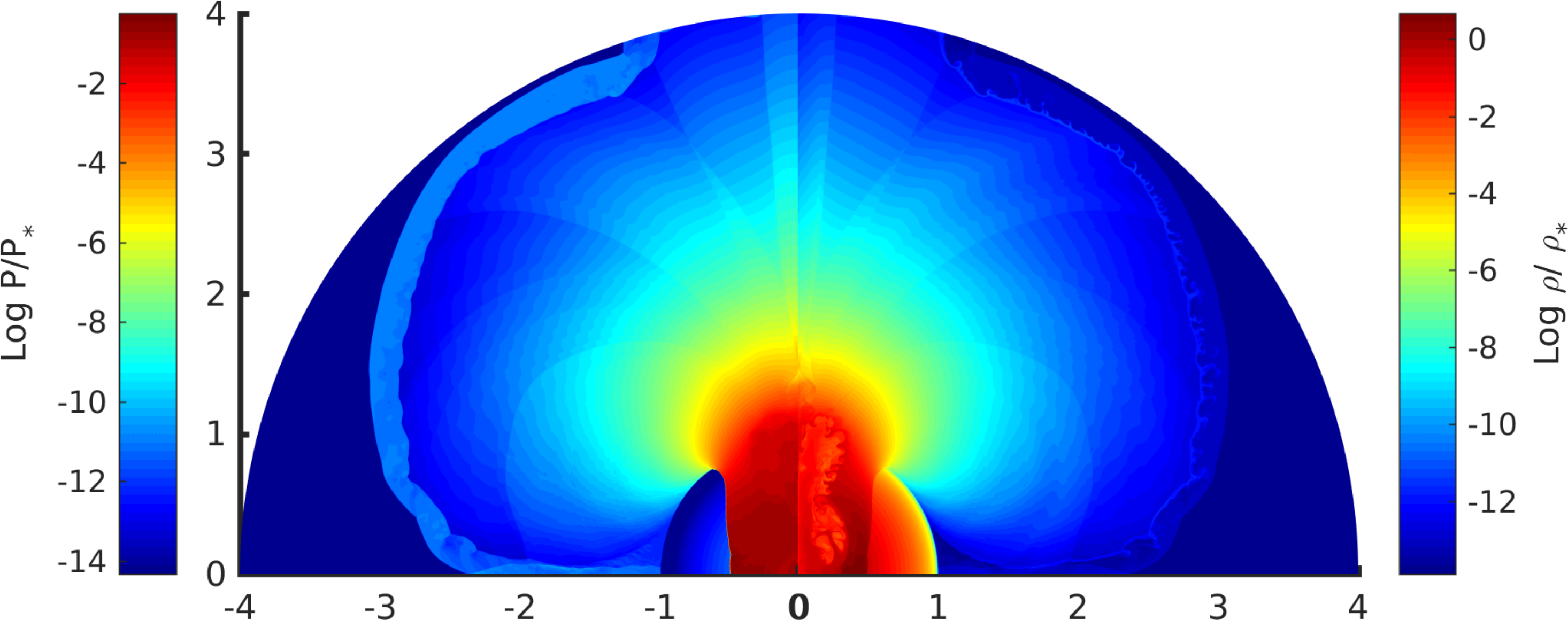}{0.7\textwidth}{(c) Density and Pressure distribution for $\calR=2048$}}

\caption{Snapshots of density and pressure distributions at $t/t_*=0.36$ for different resolutions. The structures develop easily in high resolution simulations due to the growth of KH instabilities and initial supersonic velocities. The pattern speed of the shock and  the speed of ejecta are also resolution dependent. %The resolution of figures is reduced for illustration purpose. 
%{\cdmcomment At some point it would be better to turn $p$ and  $\rho$ profiles into single figures, flipping one and attaching it to the other -- but not important for now. } 
 }
\label{fig:res}
\end{figure*}

\subsection{Fiducial run}
Figure \ref{fig:evolution} shows the evolution of the density distribution of the simulation with $\calR=2048$ 
%{\cdmcomment [It's fair to say the resolution is 2048$^2$, but $\calR$ is defined to be $2048$.]} 
at four different snapshots, representing: initial density distribution, shock breakout, transient evolution, and the end of shock progression inside the progenitor. In Figure \ref{fig:evolution}(a), the initial density condition is depicted. The high density ring inside the progenitor is the Sedov solution taken from the spherical explosion. As expected, the bipolar shock first progresses radially in $\theta=0$ direction. In Figure \ref{fig:evolution}(b), the low density regions in the post-shocked flow are formed by Kelvin$-$Helmholtz (KH) instabilities due to the velocity shear. In this figure, the shock breakout happens when the shock accelerates and reaches the progenitor pole, leading to sprays of ejecta that run into the ambient density.  As shown in Figure \ref{fig:patternvel}, the shock pattern speed  $\vphi$ decreases with distance from the pole, and when the condition $v_s \sim \vphi$ is met, the predictions for oblique flow (M13,S14)  become relevant: the shock meets the stellar surface at a right angle and the post-shock flow spreads out to fill the space above the stellar surface as depicted in Figure \ref{fig:evolution}(c). Finally, the deflected sprays of ejecta collide with each other and form an expanding wedge of re-shocked matter near the equatorial plane (Figure \ref{fig:evolution}(d)).
\begin{figure*}[!ht]
  \centering
    \includegraphics[scale=0.61]{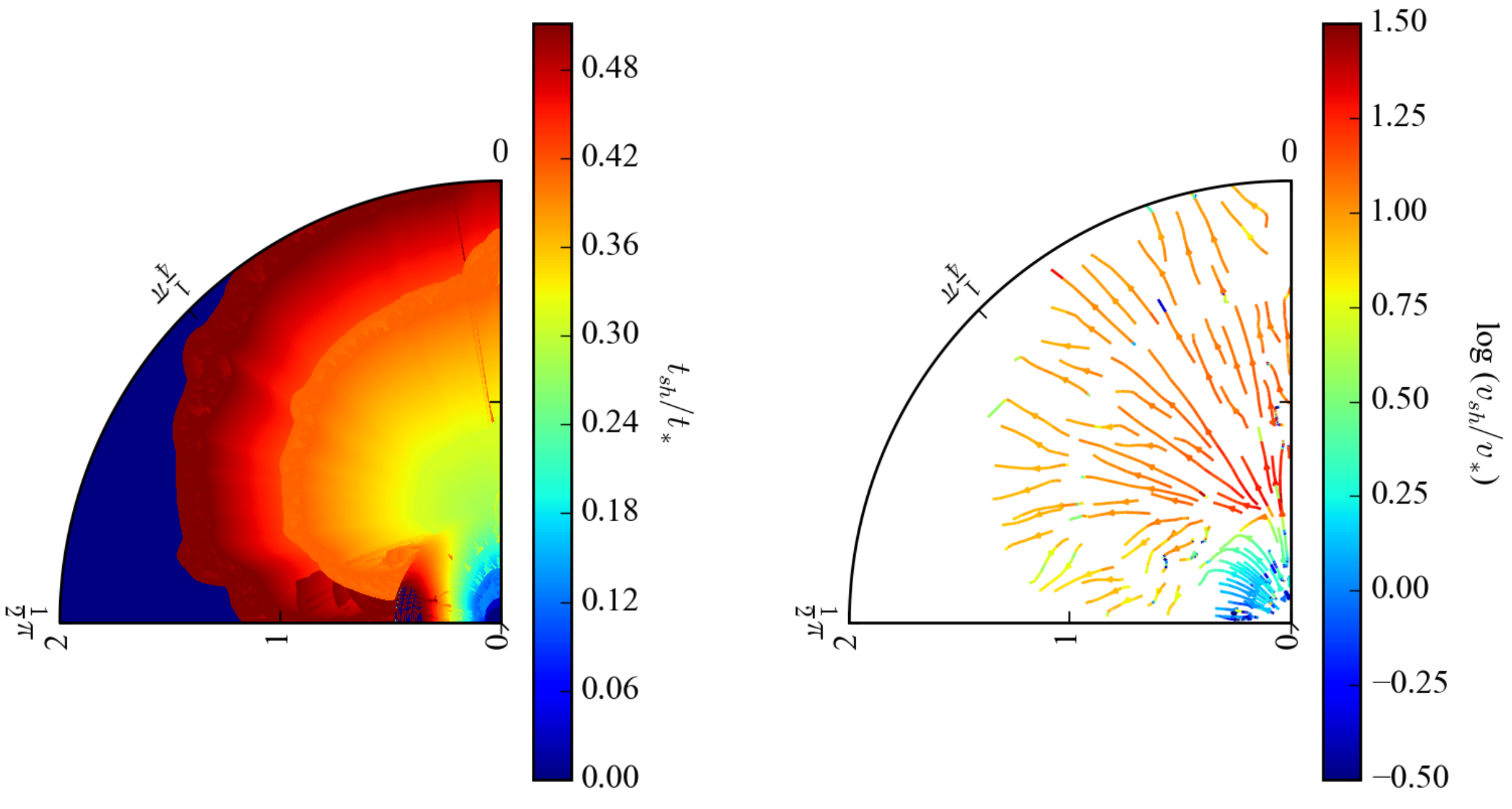} 
\caption{ {\cdm Time of maximum compression $\tsh$ (left) and the corresponding velocity $\vsh$ (right). \revthird{The} star's surface is at $r=0.5$ in these arbitrary units, outside of which $\tsh$ reflects the expanding collision between the ejecta and the ambient density.  Note that the shock normal becomes tangential to the stellar surface where the length scale $\lphi$ is resolved, so that oblique flow can develop.}
%The left panel shows the time of maximum compression rate. This parameter successfully records the time at which shock hits each cell. The mushroom pattern of the shock progression is well captured. In the right panel, the shock velocity field is shown. The stream lines show the direction of velocity vector and the color represents the absolute value of speed. The shock velocity is low inside the progenitor but during the breakout from the pole, the speed considerably increases and is then gradually deflected toward the stellar surface. 
%{\cdmcomment Again, low-priority job: flip and join.  After submission.  Also: why label $\theta$?  Why not quote radii in units of $R_*$, like elsewhere? }  
}
    \label{fig:velocity}
\end{figure*}

\subsection{Resolution Study}
{\cdm We study finite resolution by} changing the number of cells from $\calR=128$ to $\calR=2048$, with results shown in Figure \ref{fig:res}.  {\cdm  Without unphysically large explicit viscosity the outcome of the Kelvin-Helmholz instability is unavoidably resolution dependent, as demonstrated by \citet{Calder2002}; this effect is clearly seen in our results. } 

{\cdm More relevant to our study is that the speed and structure of the fastest ejecta also depend on resolution.  We attribute this to several possible effects.   First, there could be a difference in timing of shock breakout due to slightly different evolutions within the star; however we see little evidence of this.   Second, the highest possible shock speed (realized in a spherical model) depends on the shallowest resolvable depth, thanks to the scaling $v_s\propto (R_*-r)^{-n\beta }$ where $\beta\simeq 0.2$  \citep{Sakurai1960}. 
%{\cdmcomment [No need to define an equation if we only refer to it once.  Likewise I think there's no need to define $\delta$.]}  
%(However this effect will not be visible at sufficiently high $\calR$ because of collisions with the low-density surroundings.)   

Third, the formation of an oblique flow depends on resolving the scale $\lphi$ at which the non-radial motions discussed in \S\,\ref{sec:oblique}) become important.    }
%This is expected outcome, considering that that acceleration of the fluid is limited by the resolution. This argument becomes more clear given
%\begin{eqnarray}
%\label{eq:vs}
%v_s \propto	 \delta^{-n \beta}
%\end{eqnarray}
%from \citep{Sakurai1960} where $\beta=0.19$ to a good accuracy for $n=3$ polytrope and $\delta$ is normalized stellar depth resolved to the finest radial spacing. Equation \ref{eq:vs} is a good approximation for the regions where $\vphi >> v_s$ (spherical limit) and explains why the dependence on the resolution is expected. Since the shock velocity is not well resolved in the low resolution, its $\theta$ component is not resolved either and therefore we expect the shock advancing in $\theta$ direction with at a lower speed in low resolution runs.
%The formation of the oblique flow is also critically dependent on how well the obliquity depth is resolved: i
%If $\lphi/R_*$ is the same or less than the minimum radial spacing $\delta$, then the oblique flow is suppressed due to limited deflection angle of the ejecta \citep{Matzner2013}.
This consequence of this fact can be seen in the results of $\calR=256$ run in Figure \ref{fig:res} where the spray of ejecta is confined to a limited range of deflection angles, compared to higher resolution runs. 

{\cdm In fact this transition should be visible even {\em within} a well-resolved run, as $\lphi$ is zero at the pole of a bipolar breakout and must therefore be of order the grid spacing at some angle.   A radial feature is apparent in the  high-velocity ejecta, and appears at a smaller angle when  $\calR=2048$ run than when $\calR=1024$, so it is plausibly caused by this effect.  (We note that S14 found subtle differences between resolutions even when $\lphi$ is highly resolved.) }

\section{Oblique Flow Formation}
\label{sec:oblique}

Our simulations neglect radiation diffusion, so they display non-radial flows where the resolution is sufficient to capture them.  In real explosions radiation diffusion limits the development of non-radial flows, just as radiation diffusion (or inadequate numerical resolution) limits the maximum ejecta velocity in a spherical explosion.   This means that the fastest and most non-radial portions of a simulation may be unrealistic for a given progenitor model.  However, with careful attention to the scales of oblique flow and to the effects of radiation diffusion, these effects can be accounted for after the fact.  
For this purpose, it is important to compute the scales on which oblique or non-radial flows develop.  These include a relevant length scale $\lphi$, the density  $\rhophi$ at depth $\lphi$, and the pattern velocity $\vphi$ at each location on the stellar surface.  

\subsection{Oblique Parameters}
Oblique flow involves a rotation of the shock normal away from the radial direction, so the key to deriving these quantities is the shock velocity \revthird{$\mathbf{v}_S$}.  One can retrieve snapshots of  \revthird{$\mathbf{v}_S$} from checkpoint data but the limited number of such files demands a method with higher time resolution. 
%  Therefore, it is important to capture the shock velocity by taking into account each timestep. 
For this reason, {\cdm we record at each timestep the time of maximum compression rate, $\tsh$, at each location}. 
%{\cdmcomment [The verbal description is as good as the formula, and simpler.]} 
%\begin{eqnarray}
%\label{eq:comprate}
%\tsh(r,\theta) = \underset{t}{\max} (- \nabla . \vec{v} |_{r,\theta} ): \quad  r \in [0,2] , \quad \theta \in \left[0,\frac{\pi}{2}\right]
%\end{eqnarray}
%was saved at every timestep, where $\tsh$ is the time of maximum compression rate. 
{\cdm This is a viable proxy for the time at which the shock crosses each location. } As shown in Figure \ref{fig:velocity}(a), $\tsh$ captures the mushroom pattern of the shock progression very well and  only fails in regions where the expanding ejecta along stellar equator runs into the post-shocked lateral flow. The shock velocity vector is determined from $\tsh$: 
\begin{eqnarray}
\label{eq:shockvel}
\vshvec =\frac{ \nabla \tsh }{| \nabla \tsh |^2}. 
\end{eqnarray}
The streamline plot of $\vshvec$ is depicted in Figure \ref{fig:velocity}(b). As expected, all streamlines are in outward direction. The shock velocity is low inside the progenitor but considerably increases during breakout from the pole as the density steeply decreases. The shock streamline is then deflected toward the stellar surface and advances in the $\theta$ direction. 

\begin{figure}
  \centering
    \includegraphics[scale=0.7]{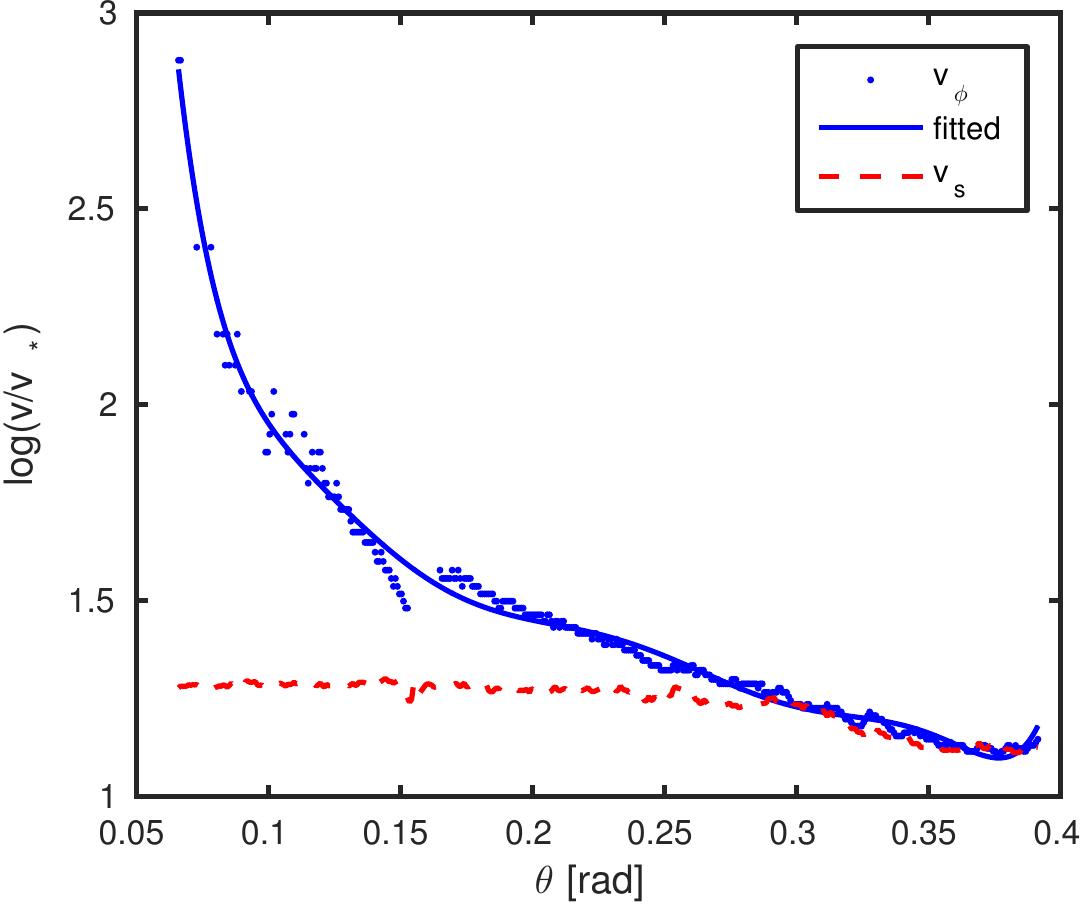}
    \caption{The shock pattern speed $\vphi$ and the {\cdm total shock speed at} the stellar surface $v_s$ are plotted as a function of {\cdm $\theta$}. As expected, $\vphi$ is generally decreasing except for  glitches. Initially, the pattern speed is much higher that the shock speed, reflecting {\cdm the radial nature of breakout at the pole}. The condition $v_s \sim \vphi$ is met for {\cdm $\theta > 0.3$}\,rad, {\cdm where oblique flow is resolved}.  }
    \label{fig:patternvel}
\end{figure}

The shock pattern speed across the stellar surface, $\vphi(\theta)$, can be  derived using the same quantities by
\begin{eqnarray}
\label{eq:shockpat}
\vphi(\theta)= \frac{1}{(\nabla \tsh \cdot \hat{\theta}) _{r=R_*}}.
\end{eqnarray}
{\cdm (The subscript $\varphi$ refers to the lateral shock motion, and should not be confused with the polar angle $\theta$.) }
Figure \ref{fig:patternvel} shows $\vphi$ as a function of {\cdm $\theta$.  The pattern speed diverges at the pole, because shock breakout is simultaneous there (as it is in a spherical explosion). 
However, as  $\tsh$ and its gradient rise away from the pole,} $\vphi$ drops. The shock velocity at the stellar surface ($\vsh$ at $r=R_*$) is also plotted in Figure \ref{fig:patternvel}. 

{\cdm In a fully-developed oblique breakout the shock normal turns parallel to the stellar surface (M13, S14), implying $\vsh=\vphi$.  Within the fiducial simulation this occurs at $\theta\simeq 0.3$ radians, so we interpret our results in terms of a resolved non-radial flow for $\theta>0.3$\,rad.  }
%By comparing $v_s$ and $\vphi$, it becomes apparent that the analysis in M13 and S14 best describes our simulation in the region where $v_s \sim \vphi$ or in other words $\theta > 0.2$ rad. It is worth mentioning that the time of maximum compression rate array is noisy due to many shock-like instabilities and therefore the velocity curves in Figure \ref{fig:velocity} are not smooth.

\begin{figure}[t!]
  \centering
    \includegraphics[scale=0.53]{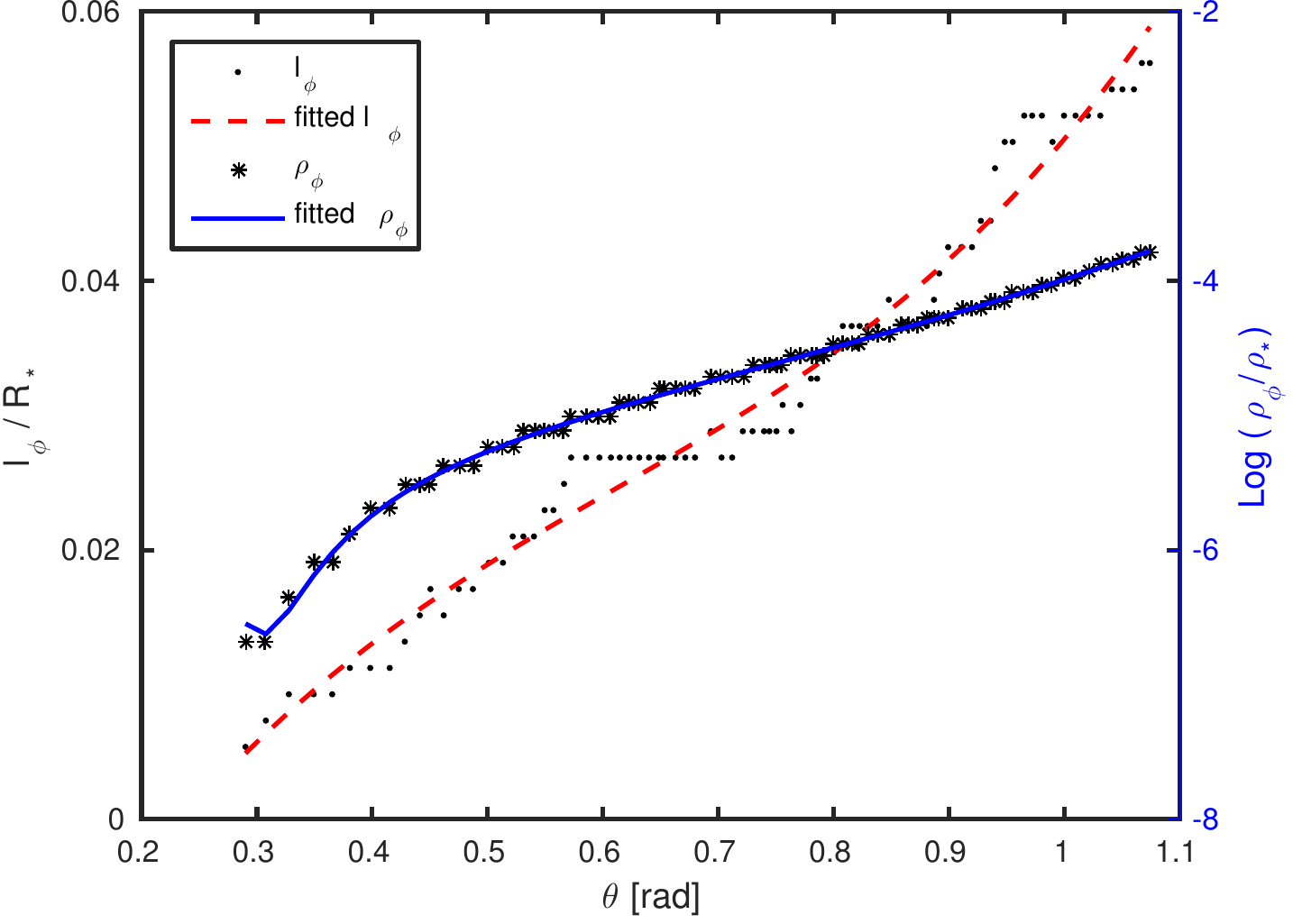}
    \caption{{$\lphi$ denotes the depth from stellar surface that obliquity affects the flow. Here, it is measured as the depth at which the shock turns $45^\circ$ from the radial direction. Where it is resolved, the increasing value of $\lphi$ with $\theta$ shows that the flow originates from deeper layers within the progenitor as $\theta$ increases. We fitted a polynomial (red dashed-curve) to our measurements of $\lphi$ (black points). We also plotted the logarithm of the ratio $\rhophi / \rho_*$ which represents pre-shocked density at depth $\lphi$ scaled by the characteristic density $\rho_*= \sqrt{\Mstar/R_*^3}$. Here, density profile is polytropic with $n=3$. The ratio $\rhophi / \rho_*$ increases with $\theta$ as it is being measured at a higher depth $\lphi$ (starred points). A low-order polynomial fit is also shown (solid blue curve).}}
    
     %{\cdmcomment [Same comment as prev.\ figure.  Actually it would be better for these plots to share an x-axis, i.e.\ be combined into one plot. ]} }
    \label{fig:lphi}
\end{figure}

The oblique length scale $\lphi$, defined in M13, is the depth where the {\cdm outward shock velocity matches the pattern speed, so that a non-radial flow (oblique breakout) develops}. Here it is measured {\cdm as the depth at which}
%from the 
contours of constant shock time $\tsh$
% by identifying the depth where the $\tsh$ contours
deflect $45^\circ$ from the radial direction.  {\cdm This is plotted} 
%The identified $\lphi$ is shown 
in Figure \ref{fig:lphi} {\cdm along with a smoother version (a low-order polynomial fit) 
%for which we fitted a polynomial to be able to interpolate $\lphi$ 
for the region $0.3<\theta<1.1$} where $\lphi$ is resolved.  \revthird{The increase of $\lphi(\theta)/R_*$ toward the equator is expected in bipolar explosions, because $\vphi(\theta)$ declines as the shock progresses away from the poles, and it thus matches the outward shock velocity at greater depths.}
 %is decreasing with $\theta$ as the flow becomes more and more oblique with progression of the shock front and the ejecta originates from the deeper layers of progenitor.
 Once $\lphi$ is inferred from the $\tsh$ array, we retrieve oblique density parameter $\rhophi$ as the progenitor density measured the appropriate location, as shown in Figure \ref{fig:lphi}.

{\cdm We see in Figure \ref{fig:lphi} that the minimum resolvable value of $\lphi$ is about $0.005R_*$, which corresponds to 2.5 zones at $\calR=2048$.  We infer that a few zones are required to resolve the oblique flow.  Applying this rule to the lower-resolution runs, and supposing that changes in resolution do not appreciably change the timing of shock breakout, we expect oblique flow to develop at  $\theta\simeq(0.55, 0.35)$, where $\lphi = (0.02, 0.01)R_*$, respectively, in the runs with $\calR = (512, 1024)$.  }

\subsection{Obliquity and Radiation Diffusion}
{\cdm  \revthird{Sufficiently close to the stellar surface}, oblique breakout has the property that each streamline (in a frame co-moving with the breakout pattern) either traps photons, resulting in adiabatic near-surface flow, or releases them through diffusion (M13).  The streamline in question can be parameterized by the comoving-frame deflection angle $\alpha_f$, or, in the star's frame, by the deflection angle $\delta\theta_f = \theta_f - \theta$ between the fluid's initial and final directions ($\theta$ and $\theta_f$, respectively, for a bipolar eruption).  These angles are conveniently related: $\alpha_f = 2\, \delta\theta_f$ (S14). Using high resolution simulations of the adiabatic, planar, non-relativistic limit of shock breakout, S14 measure the strength of diffusion on each streamline in terms} of the local P\'eclet number

\begin{eqnarray} \label{eq:calD_Peclet}
\calD &=& 3 \kappa \rho L_p \vphi/c, \nonumber \\ 
 &\simeq& D_1\alpha_f^{-\delta_D} \kappa \rhophi \lphi \vphi/c
 \end{eqnarray} 
which is approximately constant on streamlines as seen in the co-moving frame (in the near-surface limit). \revsec{ Here, the radiation pressure scale length is $L_p = P_{\rm rad}/|\nabla P_{\rm rad}|$, $D_1\simeq 126$, and $\delta_D\simeq 6$.  } 
%This formula that is found from the results of a very high resolution local simulation of oblique flow reflects the strength of diffusion in oblique breakout: 
Diffusion is negligible when $\calD>>1$ and strong for $\calD<<1$. The dimensionless parameter $\calD$ scales with the progenitor parameters as

%\begin{figure}[t]
%  \centering
%    \includegraphics[scale=0.6]{rhophi_comp_v3.pdf}
%    \caption{The ratio $\rhophi / \rho_*$ represents pre-shocked density at depth $\lphi$ scaled by the characteristic density $\rho_*= \sqrt{\Mstar/R_*^3}$. Here, density profile is polytropic with $n=3$. The ratio $\rhophi / \rho_*$ increases with $\theta$ as it is being measured at a higher depth $\lphi$. {\cdm A low-order polynomial fit is also shown}. {\cdmcomment [x-label: $\phi\rightarrow \theta$. It looks to me like the y-xaxis should be log.]}    }
%    \label{fig:rhophi}
%\end{figure}
\begin{figure}[ht!]
  \centering
    \includegraphics[scale=0.6]{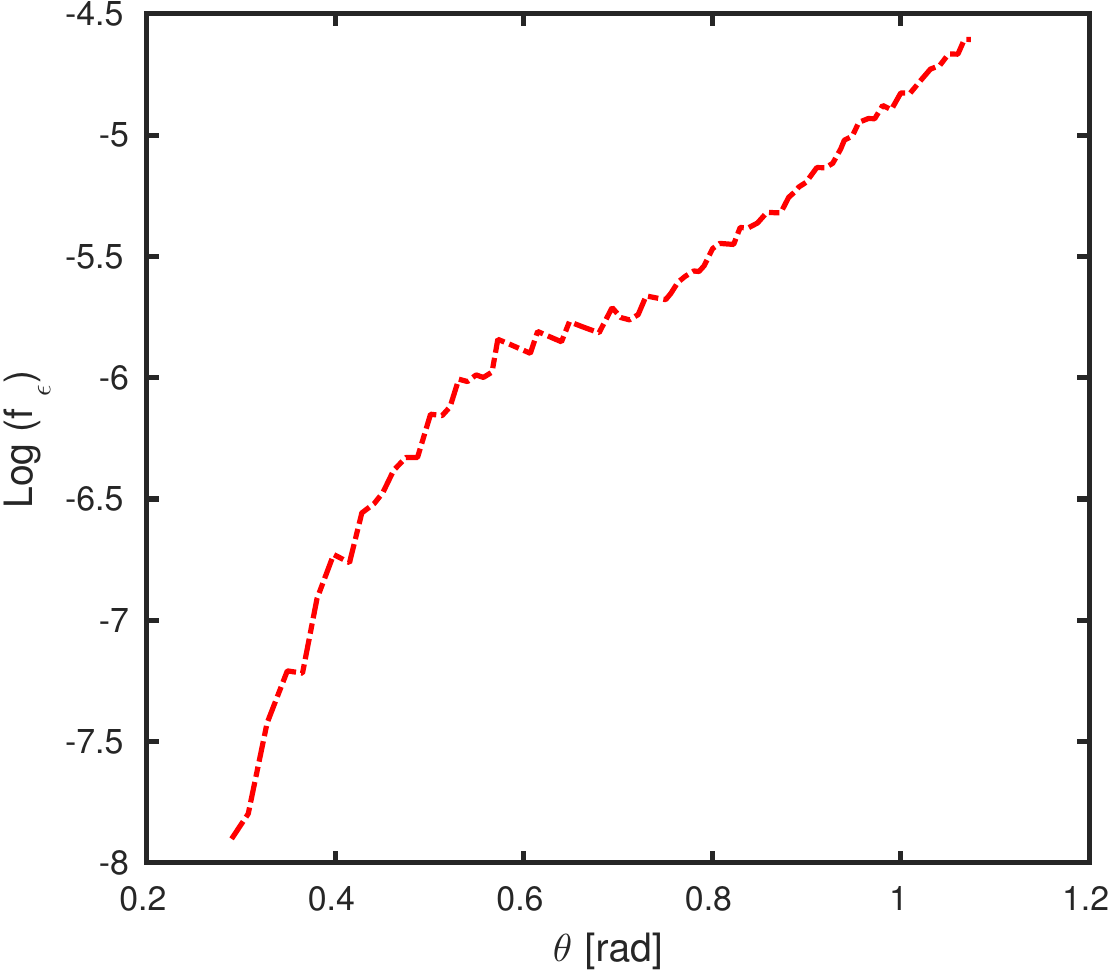}
    \caption{The function $f_\epsilon(\theta)= \frac{\rhophi}{\rho_*} \frac{\lphi}{R_*} \frac{\vphi}{v_*}$, containing main oblique parameters scaled by their characteristic values for  \revthird{our fiducial} initial conditions, which induce an ellipticity $\epsilon=0.26$ at the time of breakout. \revthird{For a chosen progenitor, higher values of $f_\epsilon$ imply a greater trapping of radiation, i.e., weaker diffusion.} }
    \label{fig:fphi}
\end{figure}
\begin{figure}[ht!]
  \centering
    \includegraphics[scale=0.6]{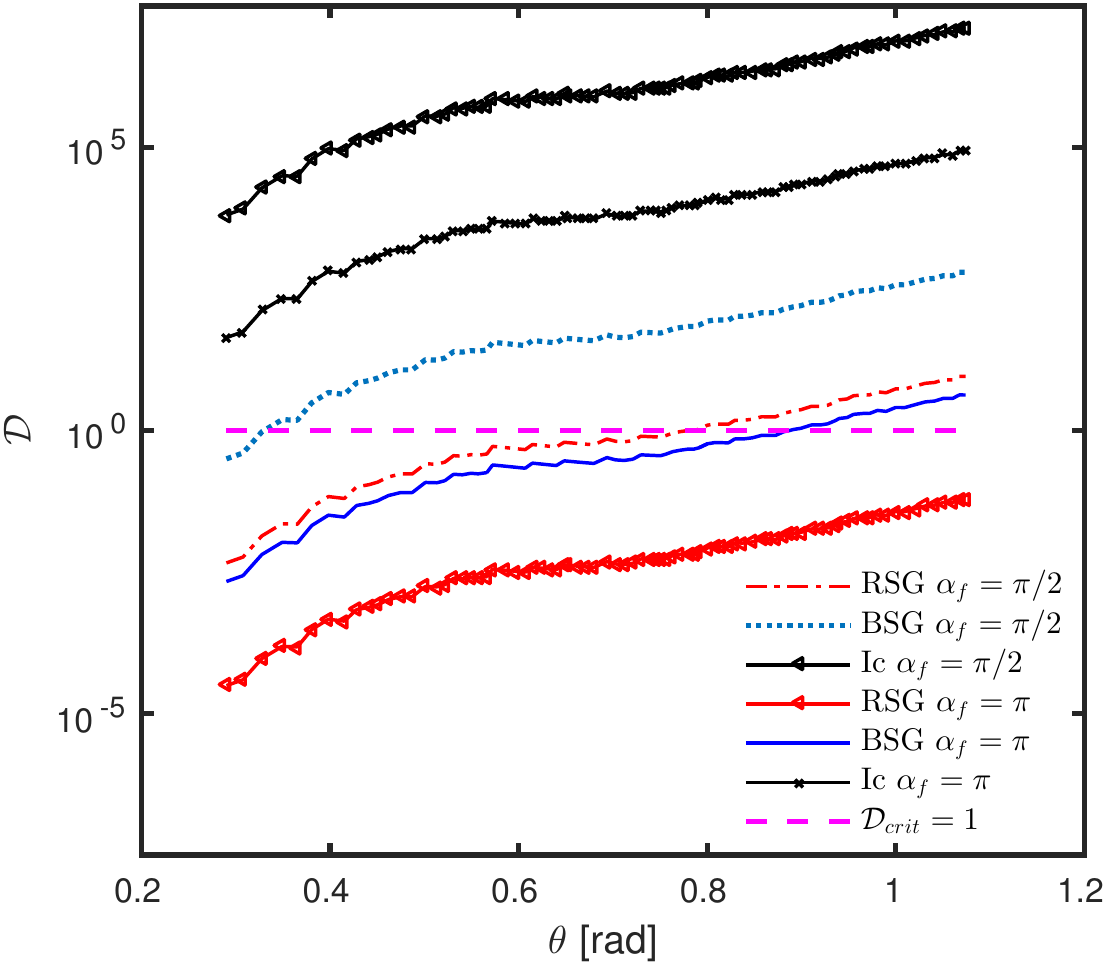}
    \caption{\revthird{Parameter $\calD(\alpha_f,\theta)$ of eq.\ (\ref{eq:dparam}) gauges the importance of radiation diffusion in the development of non-radial flows at the stellar surface.} The curves represent $\calD(\alpha_f=\pi/2,\theta)$ and $\calD(\alpha_f=\pi,\theta)$ for various progenitors models.  \revthird{For each model (denoted by line color), the upper line indicates whether photon trapping affects moderately non-radial flow ($\alpha_f = \pi/2$) whereas the bottom line makes the same comparison for the most strongly deflected ejecta ($\alpha_f = \pi$).  Regions with $\calD<1$ are diffusive. The RSG model 
    %is partly below $\calD_{\text{crit}}=1$ for $\alpha_f=\pi/2$ 
 is strongly diffusive except for modest deflections near the equator, %which means that the diffusion is characteristically strong and 
 so non-radial flow should be strongly suppressed.  
 %we do not expect to see  the oblique shock breakout for $\theta$s with $\calD(\alpha_f=\pi/2,\theta)<1$ if radiation diffusion was assumed. 
 The BSG model develops non-radial flow, but diffusion prevents strong deflections except near the equator. 
% is mostly above $\calD_{\text{crit}}=1$ at $\alpha_f=\pi/2$ but also mostly below $\calD_{\text{crit}}=1$ for $\alpha_f=\pi$, meaning that the oblique shock breakout is expected but radiation diffusion limits the flow to streamlines $\pi/2<\alpha_f<\pi$ where $\calD>1$. 
In the Type Ic model diffusion does not hinder the development of non-radial flow at any of the angles plotted. }
%has $\calD \gg 1$ for all streamlines $\alpha_f$ which proves that our neglect of radiation diffusion in the adiabatic simulation does not cause major changes in the results.
 }
    \label{fig:dall}
\end{figure}

\begin{figure*}[ht!]

\gridline{\fig{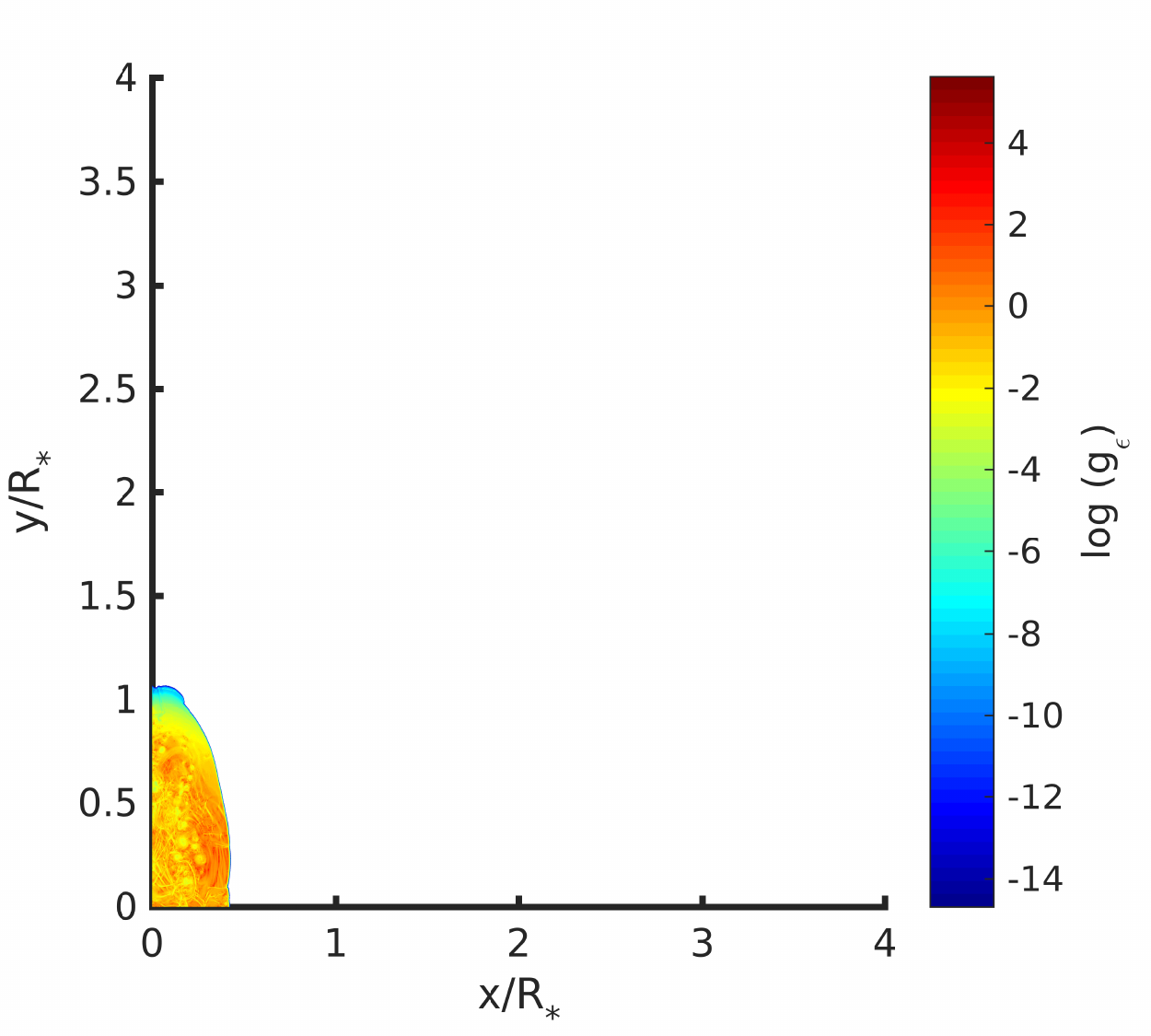}{0.32\textwidth}{(a) $g_\epsilon(r/R_*,\theta,t/t_*=0.28)$}
		  \fig{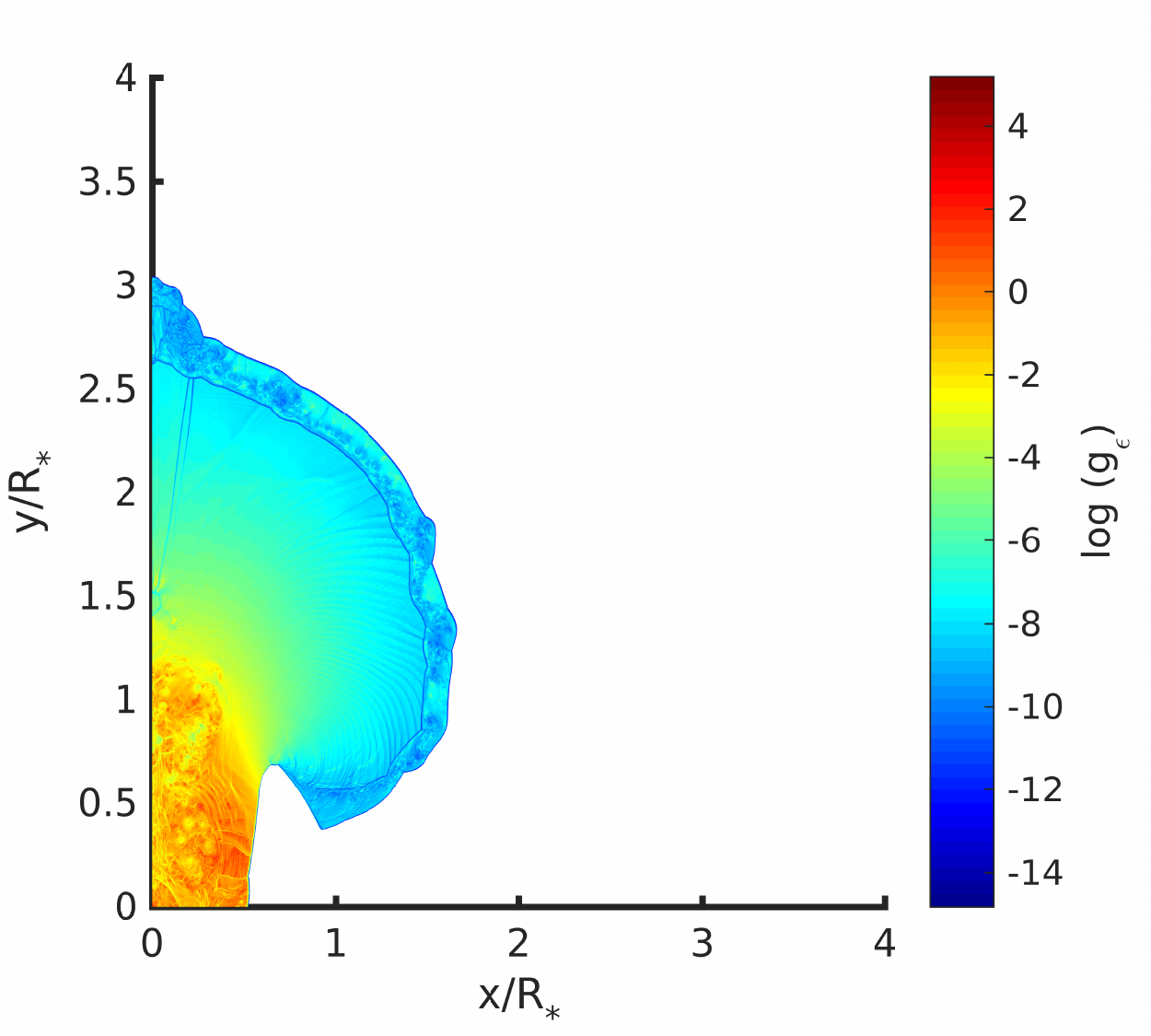}{0.32\textwidth}{(b) $g_\epsilon(r/R_*,\theta,t/t_*=0.38)$} 
		  \fig{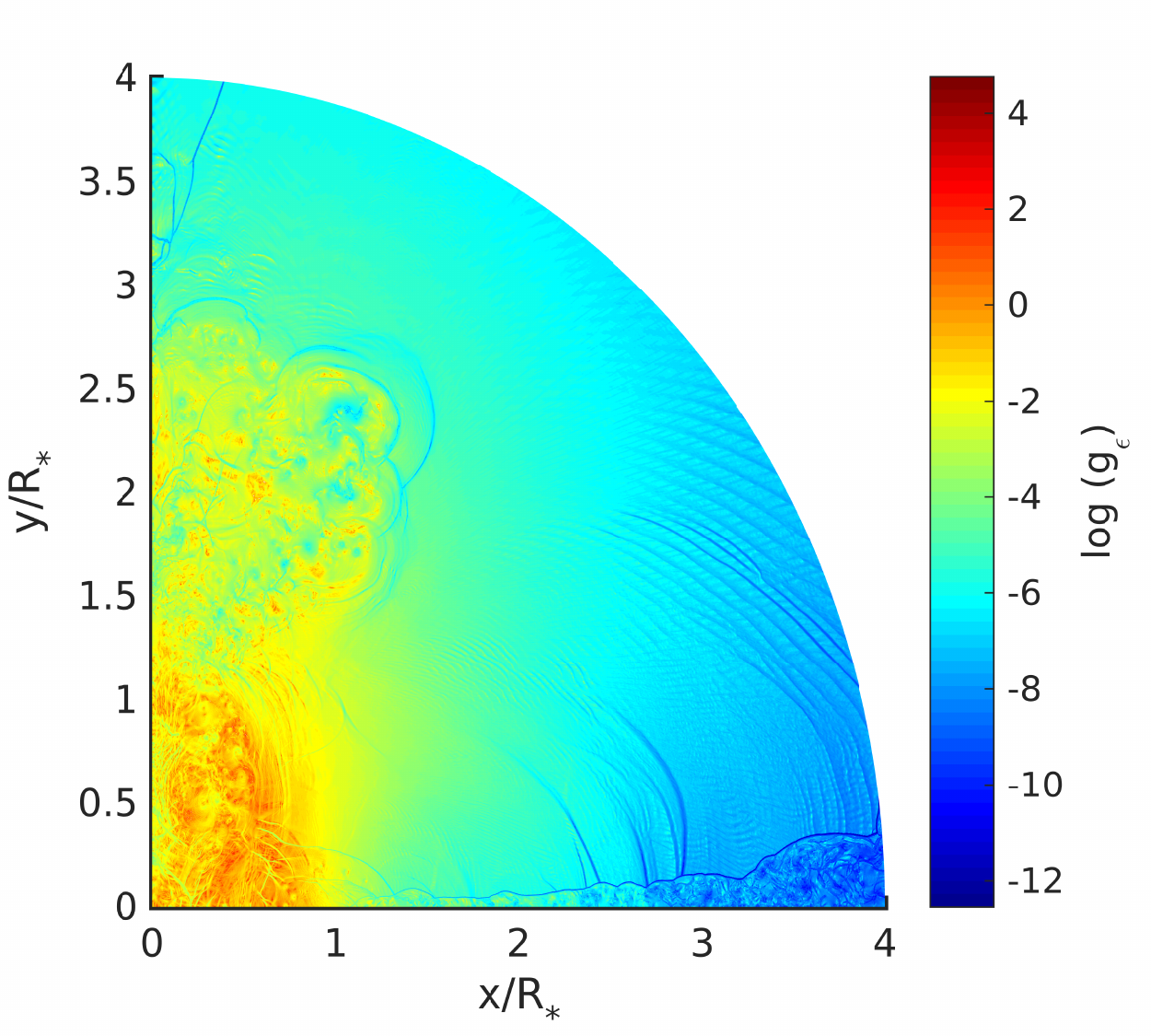}{0.32\textwidth}{(c) $g_\epsilon(r/R_*,\theta,t/t_*=0.71)$} }

	\caption{The evolution of function $g_\epsilon$ for $\epsilon=0.26$ during shock breakout from the pole, half way through, and during circumstellar collisions. This function factors out the simulation output from progenitor parameters in diffusion (equation \ref{eq:diffvel1}). }
	\label{fig:gfunc}
\end{figure*}

\begin{figure*}

\gridline{\fig{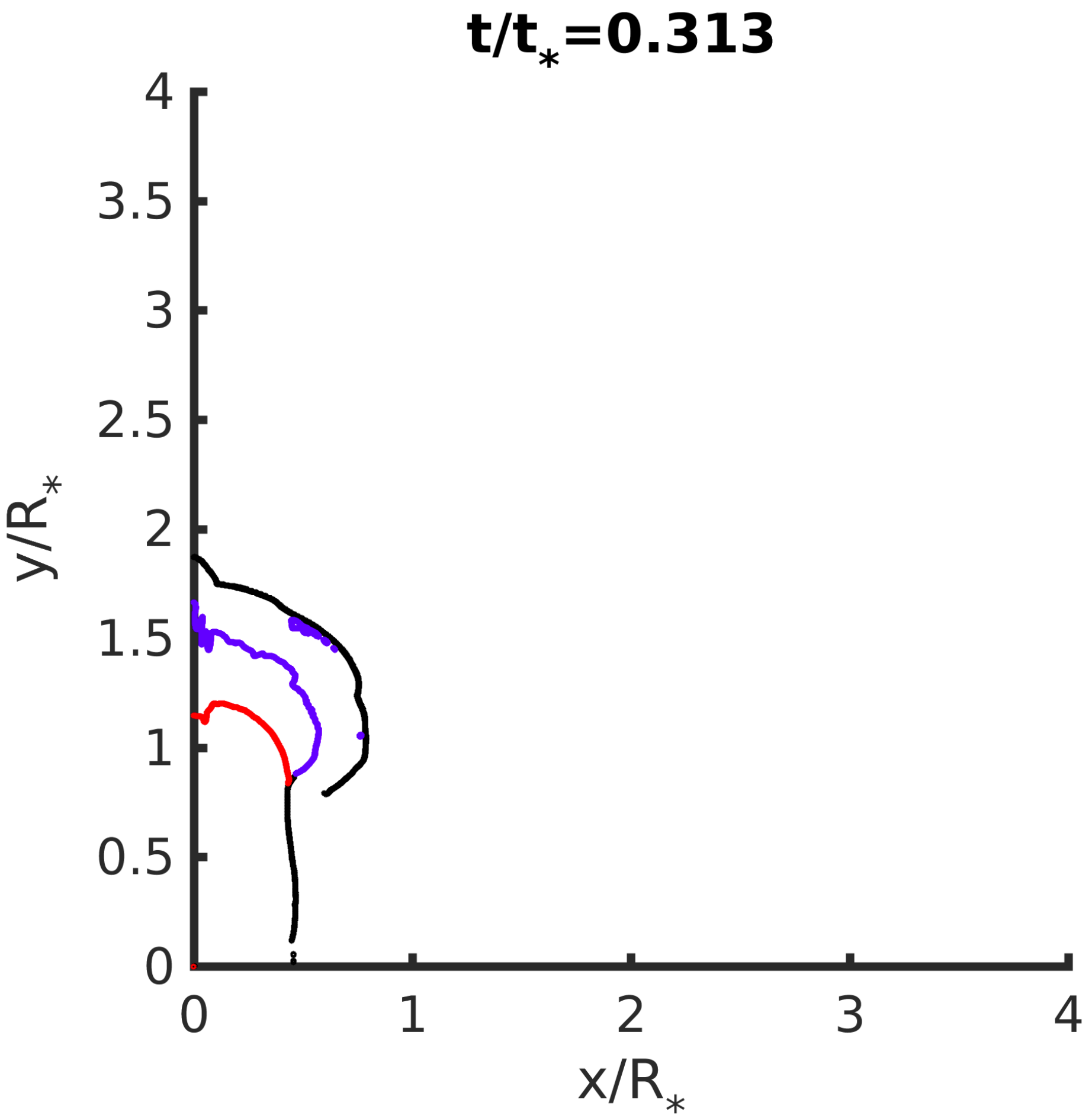}{0.32\textwidth}{(a) Diffusion front at $t/t_*=0.31$}
		  \fig{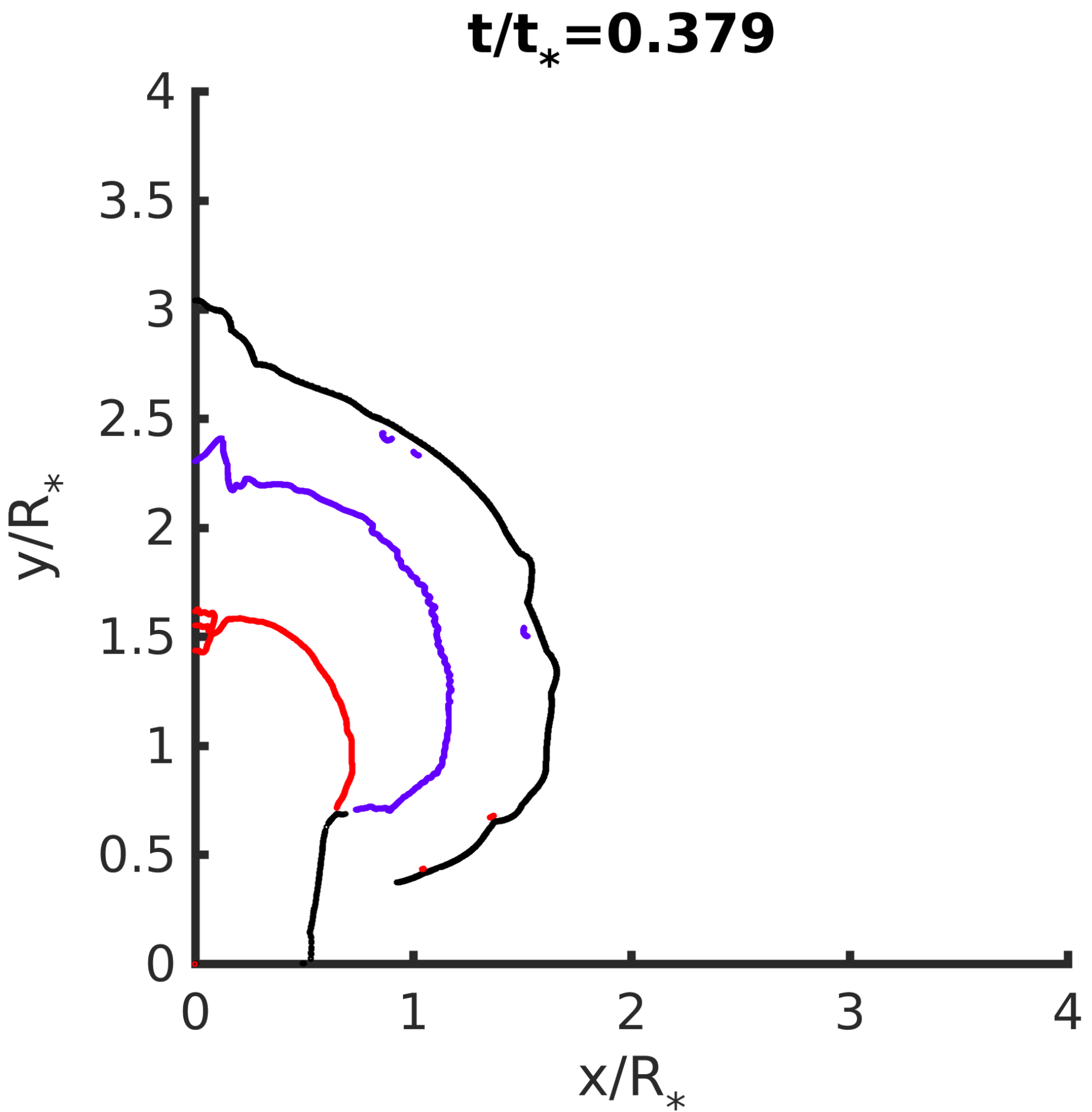}{0.32\textwidth}{(b) Diffusion front at $t/t_*=0.38$} 
		  \fig{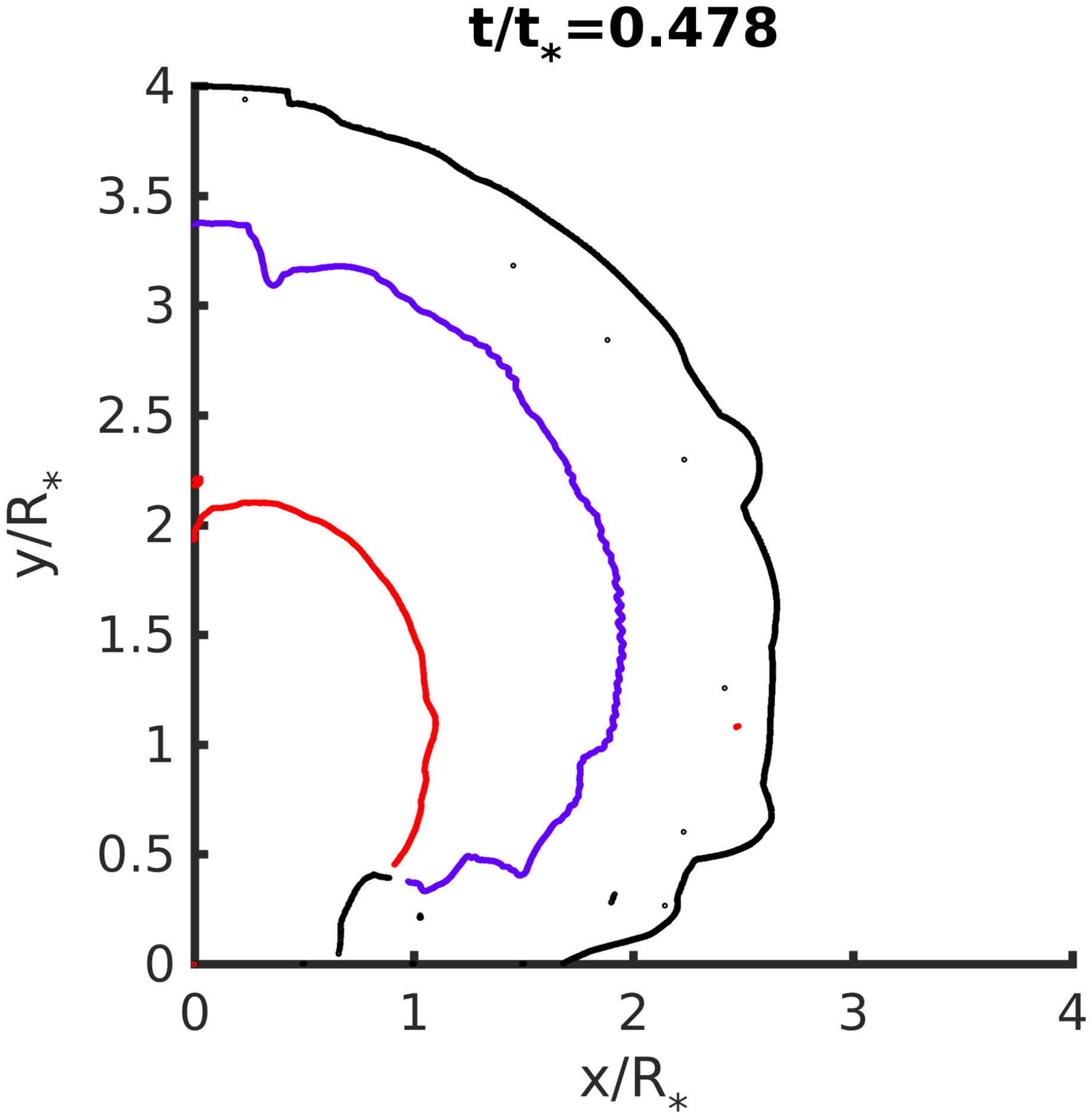}{0.32\textwidth}{(c) Diffusion front at  $t/t_*=0.48$} }

	\caption{The diffusion front is depicted for different progenitor models at three times: red contours represents diffusion front for Type-Ic, blue contours stands for BSG, and green shows it for RSG. The diffusion front moves inward for more extended progenitors. 
	%This shows that our adiabatic simulation are more applicable to compact progenitors as oppose to extended ones. 
	}
	\label{fig:diffall}
\end{figure*}

%% CDM -- don't need intermediate formula, and there should only be one eq.no (and one label) for this.  Also try to avoid more than one level of \frac, 
%%              and don't use more parens than you need.  For parens, use \left( \right) instead of \Big( \Big).  
\begin{eqnarray}
\label{eq:dparam} 
\calD(\alpha_f,\theta) &\simeq & \frac{D_1}{\alpha_f^{\delta_D} c} \kappa \rho_* R_* v_*  \left( \frac{\rhophi}{\rho_*} \frac{\lphi}{R_*} \frac{\vphi}{v_*}\right) \nonumber \\ 
%\label{eq:dparam1}
&= & \frac{D_1}{\alpha_f^{\delta_D}} \left( \frac{\kappa \sqrt{E_* \Mstar}}{R^2_* c} \right) f_\epsilon(\theta) \nonumber\\
\label{eq:dparam2}
&= & \frac{4.33 \times 10^8}{(\alpha_f / \pi)^{\delta_D}}  \frac{\kappa_{0.34} \sqrt{E_{51} \Mstar/M_\odot}}{(R_*/R_\odot)^2} f_\epsilon(\theta)
\end{eqnarray} 
where $f_\epsilon(\theta)= (\rhophi\lphi\vphi)/(\rho_* R_* v_*)$. {\cdm In the second expression,} the parameters in the  parenthesis are the scales of the explosion and the effective opacity $\kappa$ (usually dominated by electron scattering: $0.2(1+X)$\,cm$^2$\,g$^{-1}$ for H mass fraction $X$), which are all progenitor dependent, while the function $f_\epsilon(\theta)$, shown in Figure \ref{fig:fphi}, contains oblique parameters normalized to their characteristic values.  {\cdm These} are directly related to the simulation output as well as the degree of ellipticity 
\begin{eqnarray} 
\label{eq:epsilon}
\epsilon &=& 1-\frac{\tsh(\theta=0, R_*)}{\tsh (\theta=\pi/4, R_*)}
\end{eqnarray} 
defined {\cdm by} M13. In the final line of Equation \ref{eq:dparam2}, progenitor parameters are scaled to their explosion values, where $E_* = 10^{51}E_{51}$\,erg %, $R_{x}=R_*/xR_\Sun$, $M_{x}=\Mstar/xM_\Sun$, 
and $\kappa = 0.34\,\kappa_{0.34}$ cm$^2$ g$^{-1}$. \revsec{ For our fiducial run with $r_{\text{v}}=0.12 R_*$ we find $\epsilon=0.26$; but we also consider a spherical case ($\epsilon=0$) as well as an intermediate case ($r_{\text{v}}=0.06 R_*$ and $\epsilon=0.09$).}  

%Epsilon: $\epsilon$ varepsilon: $\varepsilon$  element: $\in$  Funny E: $\cal{E}$ 

{\cdm \revthird{Diffusion begins to affect the most strongly non-radial  streamlines when $\calD <1$ at $\alpha_f=\pi$;} but to strongly inhibit non-radial motion one must have $\calD <1$ at $\alpha_f=\pi/2$.  These criteria correspond to 
\[R_* \gtrsim (120, 940) \left(\kappa_{0.34} \frac{f_\epsilon}{10^{-5}}\right)^{1/2} \left(E_{51} \frac{\Mstar}{10\,M_\odot}\right)^{1/4} R_\odot,\] 
respectively. }
 %diffusion to strongly affect the flow (i.e., for  $\calD <1$ at $\alpha_f=\pi$, this requires $2.31 \times 10^{-9} > \Big( \frac{\kappa_{0.34} \sqrt{E_{51} M_1}}{R^2_1} \Big) f_\epsilon(\theta)$. 
 We evaluate $\calD(\theta)$ for the progenitor models in Table \ref{tab:models}  in Figure \ref{fig:dall}.  \revthird{Radiation diffusion interferes with non-radial flows in the RSG except near the equator}, {\cdm so we must be very careful when drawing \revthird{RSG flow quantities from our adiabatic simulations}}. Our typical BSG model is moderately affected by radiation, meaning that  diffusion affects the most deflected flow ($\calD<1$ for $\alpha_f=\pi$ but $\calD>1$ for $\alpha_f=\pi/2$) over most of the stellar surface. 
 %We conclude that radiation diffusion will limit the hydrodynamic breakout obliquity to the streamlines $\pi/2<\alpha_f<\pi$ where $\calD \gtrsim 1$ is satisfied \citep{Salbi2014}. 
 {\cdm Finally, type Ic models have $\calD \gg 1$ for all possible $\alpha_f$, implying that radiation diffusion is negligible near the stellar surface.  These conclusions are entirely consistent with the expectations of M13 and S14 (although these authors adopt more sophisticated progenitors). }

%These results show that our simulation results match the outcome of \cite{Salbi2014} oblique breakout in local coordinate. %
\section{Observational Implications}
\label{sec:diff}

{\cdm In the aspherical geometry considered here, early SN emission consists of three components: shock breakout radiation, glow from the expanding ejecta that have been heated by the SN shock, and cooling radiation from the zone of circumstellar ejecta collisions. Whereas a spherical SN produces the first two in sequence and does not have colliding ejecta (in the absence of  circumstellar matter), an aspherical SN can make all three simultaneously.   (We assume there is no relativistic jet that pierces the surface, or the jet cocoon and radioactive matter that would accompany one.)    }

%In M13 and S14, the authors conclude that when the radiation is truly trapped the non-radial flows eliminate the breakout emission. In regions where $\vphi >> v_s$, however, the emission is not effected by non-radial flows. These studies also predict new transients that might be possible by converting the kinetic energy of ejecta to the thermal energy in circumstellar collisions near the equatorial plane. In this section, we quantify the radiation diffusion analytically to be able to make predictions about what would be observed from aspherical explosions. We provide the early light curve and color temperature evolution for oblique shock breakout and the expanding ejecta.

{\cdm 
\subsection{Shock breakout (SBO) emission} \label{sub:LSBO} 

By `SBO emission', we refer to light that is emitted as the shock arrives at the stellar surface, and before as the shock matter has been ejected any significant distance.  In earlier analyses, \citet{Calzavara2004} and \citet{Suzuki2010} recognized that breakout emission would be extended by the time taken for the shock to cross the star's visible surface, and would  change in intensity according to the local shock strength.  However, as discussed in detail by M13 and S14, shock breakout emission becomes obscured by the ejecta spray as a non-radial motions develop.   As a crude approximation we assume the SBO radiation is unaffected by these effects wherever equation \ref{eq:dparam} predicts $\calD(\alpha_f=\pi/2)>1$.   

Where this criterion is not satisfied, SBO emission is inhibited. but there still exists an outer layer of matter, with mass surface density $\Sigmadiff\simeq c/(3 \kappa \vphi)$, for which diffusion is important .  In this layer the diffusion speed, $c/(3\kappa \Sigmadiff)$ equals the pattern speed $\vphi$.  The SBO energy per unit area is approximately the rate which the kinetic energy of this layer ( $\Sigmadiff \vphi^2/2$ per unit area) is consumed in the frame of the advancing shock: $L_{\rm SBO} \simeq \Sigmadiff \vphi^2 \dot A/2$ where $\dot A$ is the rate at which the surface area is shocked.   For a bipolar explosion $\dot A = 4\pi R_* \vphi \sin \theta$, so }{
\begin{eqnarray} \label{eq:L_SBO_estimate}
L_{\rm SBO} &\simeq & {2\pi\over 3} {R_* c \vphi^2 \over \kappa} \sin\theta \\
 %&\simeq & 10^{40.33} \text{erg/s} \Big(\frac{E_*}{E_{51}}\Big) \Big(\frac{R_*}{50 R_\Sun}\Big)\Big(\frac{M_\text{ej}}{15 M_\Sun}\Big)^{-1} \kappa^{-1}_{0.34} \Big( \frac{\vphi}{v_*} \Big)^2 \sin \theta.   \nonumber \\ 
 &\simeq &  10^{40.33} \frac{ E_{51}  (R_*/50\,R_\odot) }{(\Mstar/15\,M_\odot) \kappa_{0.34} } \left(\frac{\vphi}{v_*}\right)^2 \sin\theta\, {\rm erg\,s^{-1} } \nonumber  
\end{eqnarray}
} {\cdm This radiation can escape through the outflow in directions for which $\calD(\alpha_f)\lesssim 1$ (if there are any).   

Because $\calD$ is conserved on each streamline, we note that much of this luminosity may diffuse out from streamlines with $\calD \sim 1$ at radii much larger than $\lphi$; for this reason the SBO emission may blend, to some extent, into the early diffusion luminosity we consider below. 

Note that $L_{\rm SBO}$ scales as $R_* E_*/\kappa \Mstar$ for explosions that share the same eccentricity, i.e., the same pattern of $\vphi/v_*$. 
 }

%Due to the limitations imposed by a local co-moving frame considered in these studies, the energetics of breakout emission and the expanding ejecta are only discussed qualitatively. It is however 
% observational predictions such as light curve and color temperature evolution of the oblique shock breakout was . It is thus important 

%This means that the breakout emission is more localized that one would expect from spherical explosions. In other words, the breakout flash can be shorter than the shock and light crossing times since only a part of the progenitor can participate, this conclusion contradicts \cite{Couch2011}'s claim that the breakout radiation time scale is longer than the spherical time scale $R_*/c$ and set by $R_*/v_{s}$.
\subsection{Diffusion Front}
\label{sub:difffront}
Our simulations {\cdm treat the explosion as} adiabatic; i.e., {\cdm as though the photons that generated post-shock pressure are trapped in the flow} and never get a chance to diffuse upstream. To make observational predictions, {\cdm and to identify physically invalid features of the adiabatic approximation, we must identify} where this condition breaks.  {\cdm For global simulations we cannot rely on a constant flow speed (as S14 did), so we must be careful in parametrizing diffusion in Galilean-invariant way.  For this purpose we construct a new quantity $D=\tdiff/t_\mathrm{dyn}$, where $\tdiff = 3 \kappa \rho L_p^2/c$  is time for diffusion to act across the radiation pressure scale length $L_p$, and $\tdyn = |\nabla \cdot {\mathbf v}|^{-1}$ is an estimate of the local dynamical time. \footnote{In an expanding flow with $\rho(m, t)\propto (t-t_0)^{\pm q}$ this definition corresponds to $t -t_0 = q \tdyn$, so it would be reasonable to adopt $q = 1,2,3$ for planar Hubble, wind or cylindrical Hubble, or  spherical Hubble flow.  We set $q=1$ for simplicity. } With these definitions
} 
% this condition translates to  $\tdiff<t_\mathrm{dyn}$ where $\tdiff$ is  the photon diffusion time and $t_\mathrm{dyn}$ is dynamic time (i.e., time since shock breakout). This means that the diffusion is important when the local parameter $D=\tdiff/t_\mathrm{dyn}$ is less than 1. The diffusion time is approximated by  
%\begin{eqnarray}
%\label{eq:diffvel}
%t_{\text{diff}}= \frac{3 \kappa \rho L^2_p}{c}
%\end{eqnarray}
% where $L_p$ is the local radiation pressure length scale $L_p= \frac{P_{\text{rad}}}{ | \nabla P_{\text{rad}} |}$. Dynamical time is approximated by $\tdyn=|\nabla . v|^{-1}$. Therefore,
\begin{eqnarray}
\label{eq:ddiff}
D =3 \kappa \rho L^2_p |\nabla \cdot {\mathbf v}| / c
\end{eqnarray}
is a diffusion parameter (analogous to S14's $\calD$) that we evaluate in our simulations. In terms of progenitor parameters, 
\begin{eqnarray}
\label{eq:diffvel1}
D &=& \frac{3\kappa (E_* \Mstar)^{1/2}}{R_*^2 c}  ~ g_\epsilon
\end{eqnarray}
where $g_\epsilon = L_p^2 \rho |\nabla\cdot {\mathbf v}|/(R_*^2 \rho_* t_*^{-1})$ can be determined directly from the simulation outputs: $g_\epsilon$ is a function of space and time (i.e. $r/R_*, \theta$, and $t/t_*$) that depends on the details of the simulation -- most importantly, the ellipticity parameter $\epsilon$.   Normalized to  characteristic values, 
\begin{eqnarray}
\label{eq:diffvel2}
D &=&  10^{10.0}\frac{\kappa_{0.34} \sqrt{E_{51} \Mstar/M_\odot}}{R_*^2/R_\odot^2} g_\epsilon.
\end{eqnarray}

 %As shown in Equation \ref{eq:diffvel1}, the diffusion parameter $D$ scales with the progenitor parameters in the same manner as in Equation \ref{eq:dparam2}. The oblique breakout parameter $\calD$ in Equation \ref{eq:dparam}, however, only describes the shock breakout phase and thus is not time-dependent, while diffusion front progresses over time. 
 The function $g_\epsilon$ takes a wide range of values from $10^{-14}$ to $10^{4}$ within our simulation volume (Figure \ref{fig:gfunc}). The condition $D=1$ marks the diffusion surface \revthird{or `luminosity shell'} $\rdiff(\theta, t)$, outside of which photons stream through the ejecta to ultimately be released. In Figure \ref{fig:diffall}, the diffusion front is shown for the progenitor models of Table \ref{tab:models}. As expected from Equation \ref{eq:diffvel2}, the diffusion front moves inward (relative to $R_*$) as the progenitor becomes more extended. 

\begin{figure}[t]
  \centering
    \includegraphics[scale=0.55]{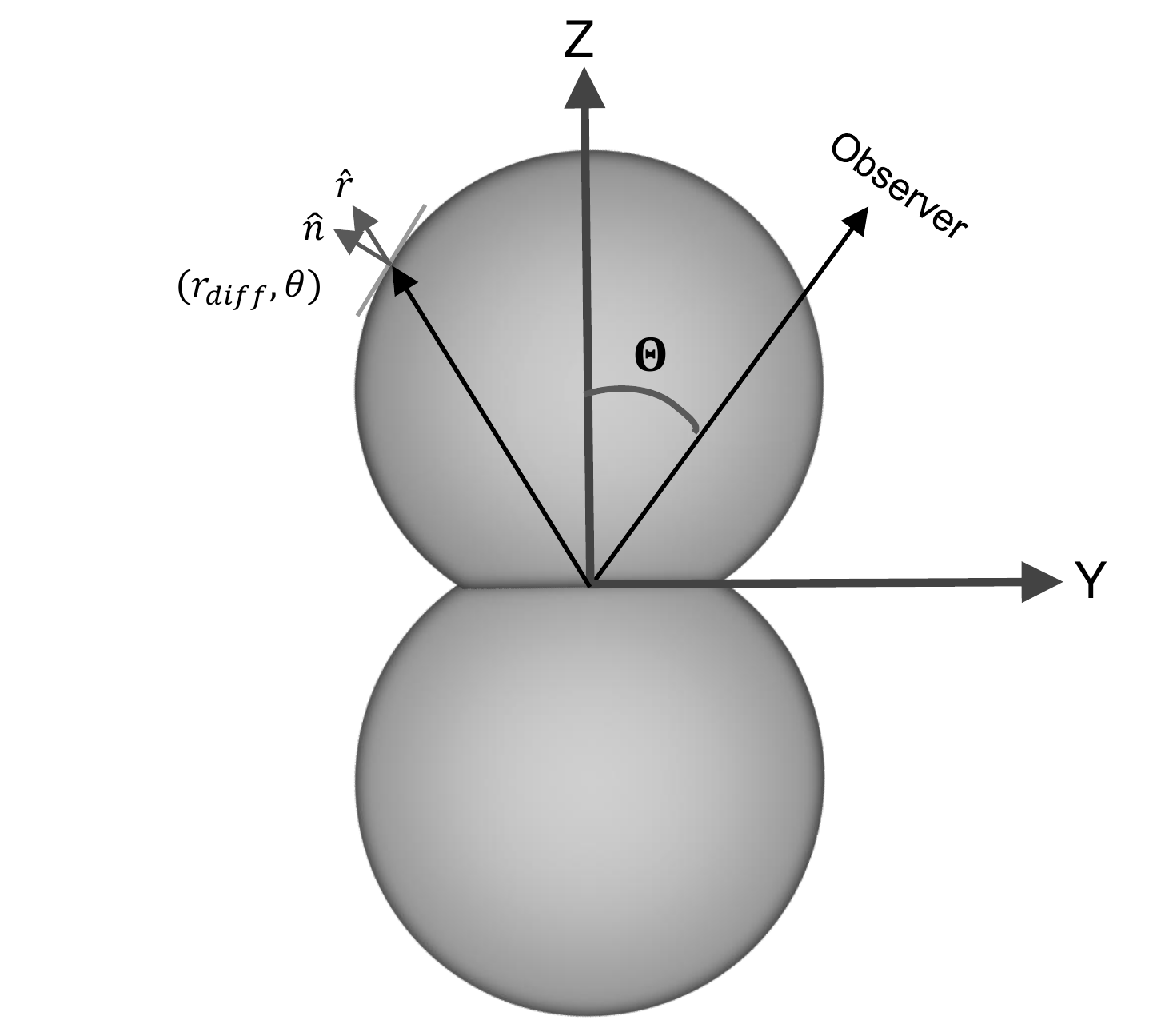}
    \caption{A schematic that demonstrates the geometry of the aspherical SN with regards to an observer. The observed luminosity depends on the angle $\Theta$ between line of sight and the symmetry axis of SN. }
    \label{fig:schematic}
\end{figure}

\begin{figure*}[!ht]
\gridline{\fig{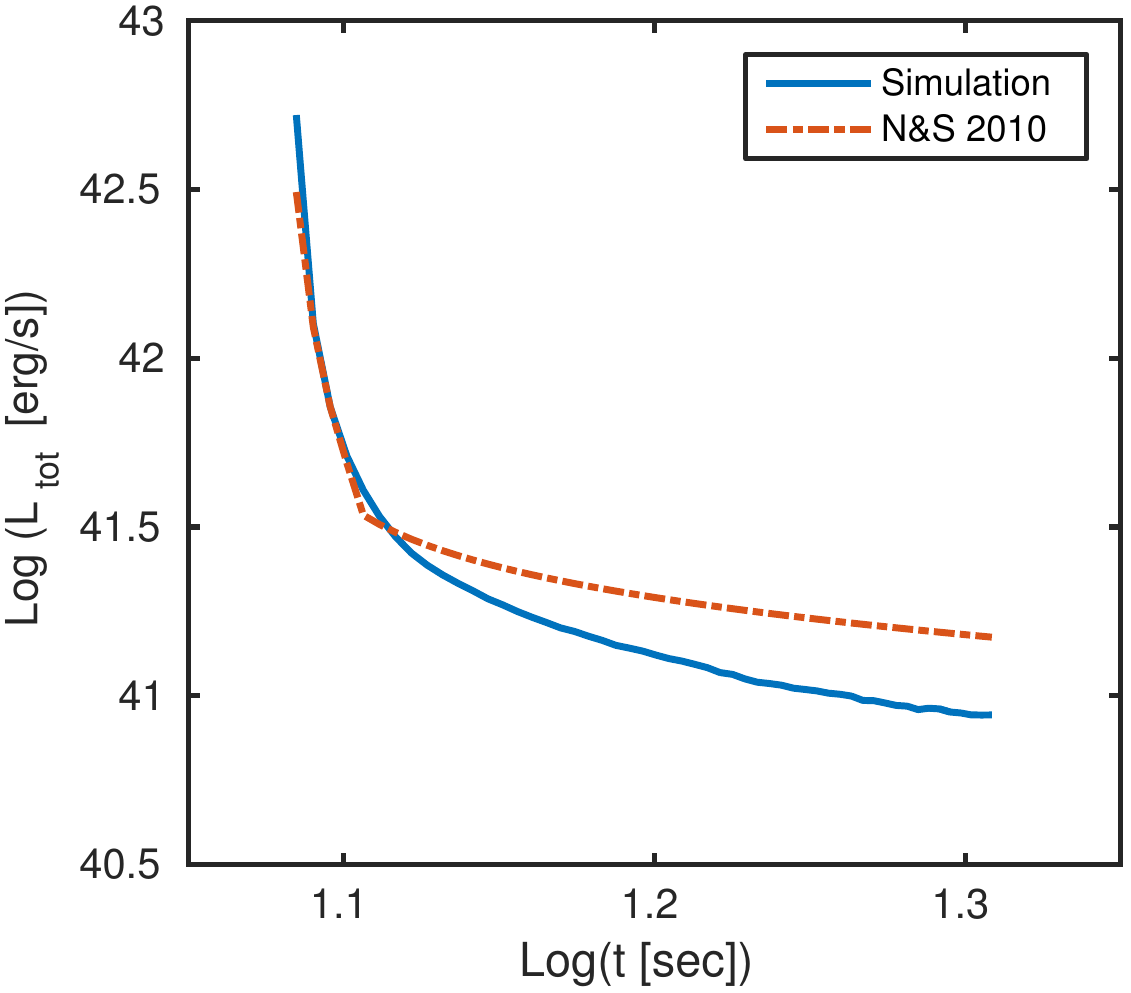}{0.32\textwidth}{(a) Type Ic model}
		  \fig{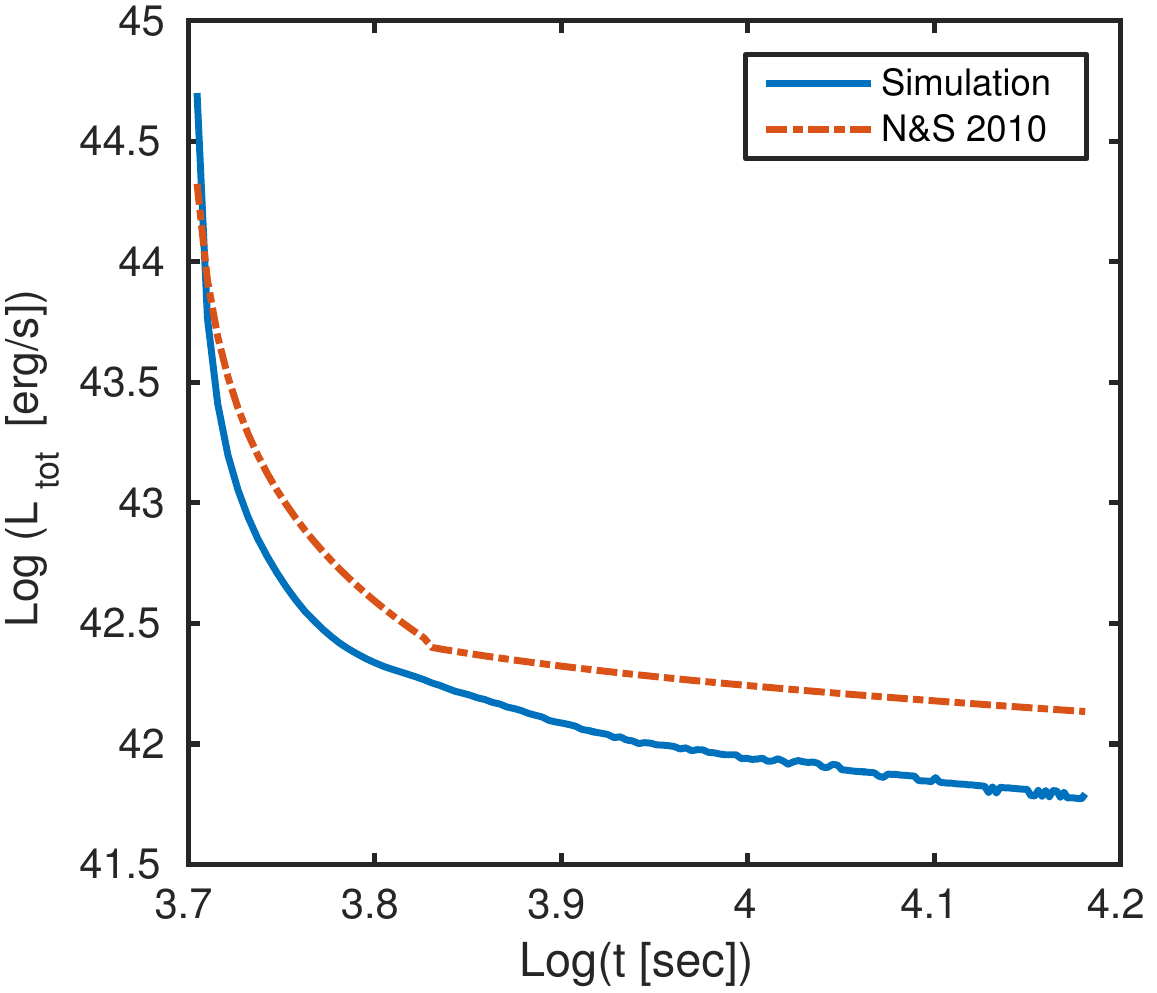}{0.32\textwidth}{(b) BSG model} 
		  \fig{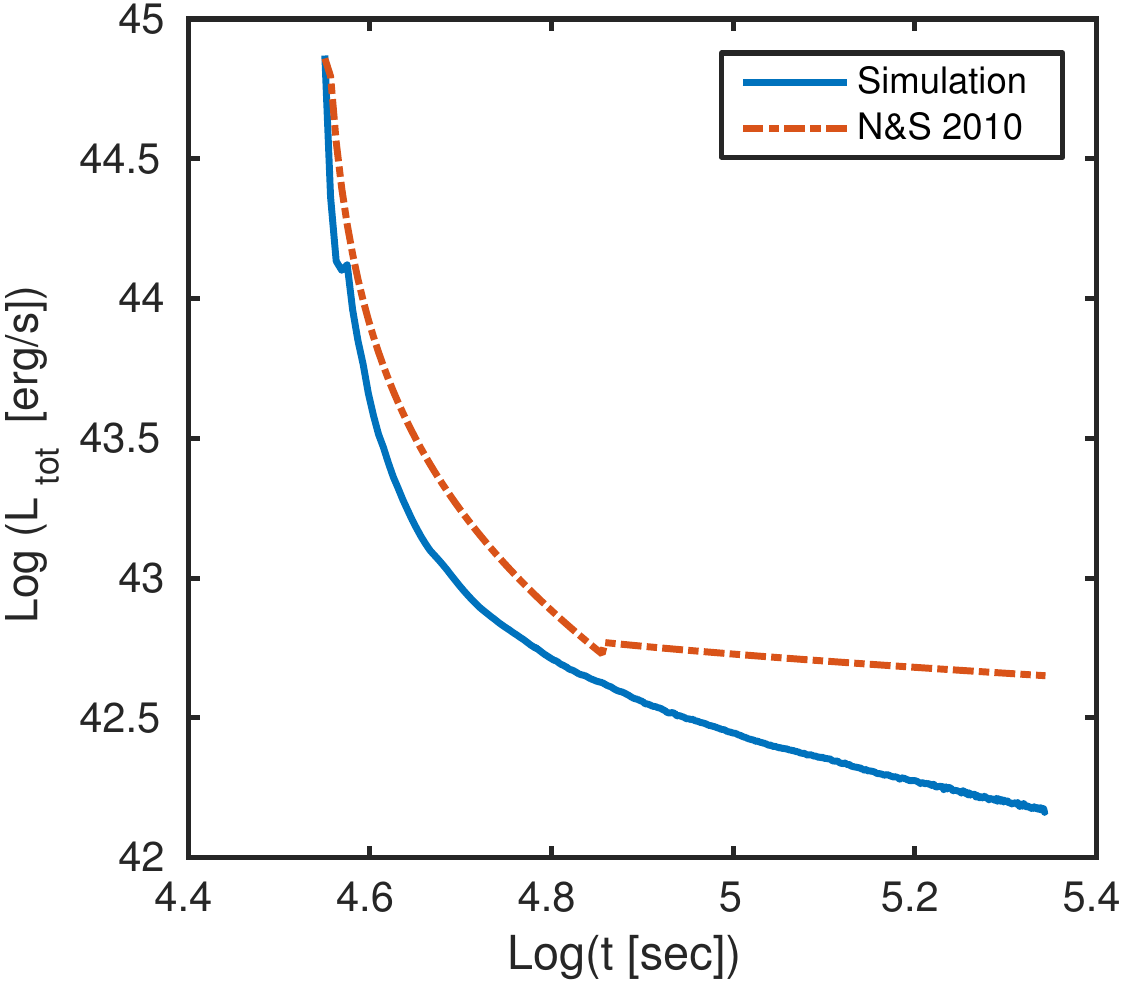}{0.32\textwidth}{(c) RSG model}}

	\caption{The  bolometric light curves of our spherical simulation are plotted against the analytical light curve of \cite{Nakar2010} for three progenitor models.  }
	\label{fig:spherical}
\end{figure*}

\begin{figure*}[!ht]
\gridline{\fig{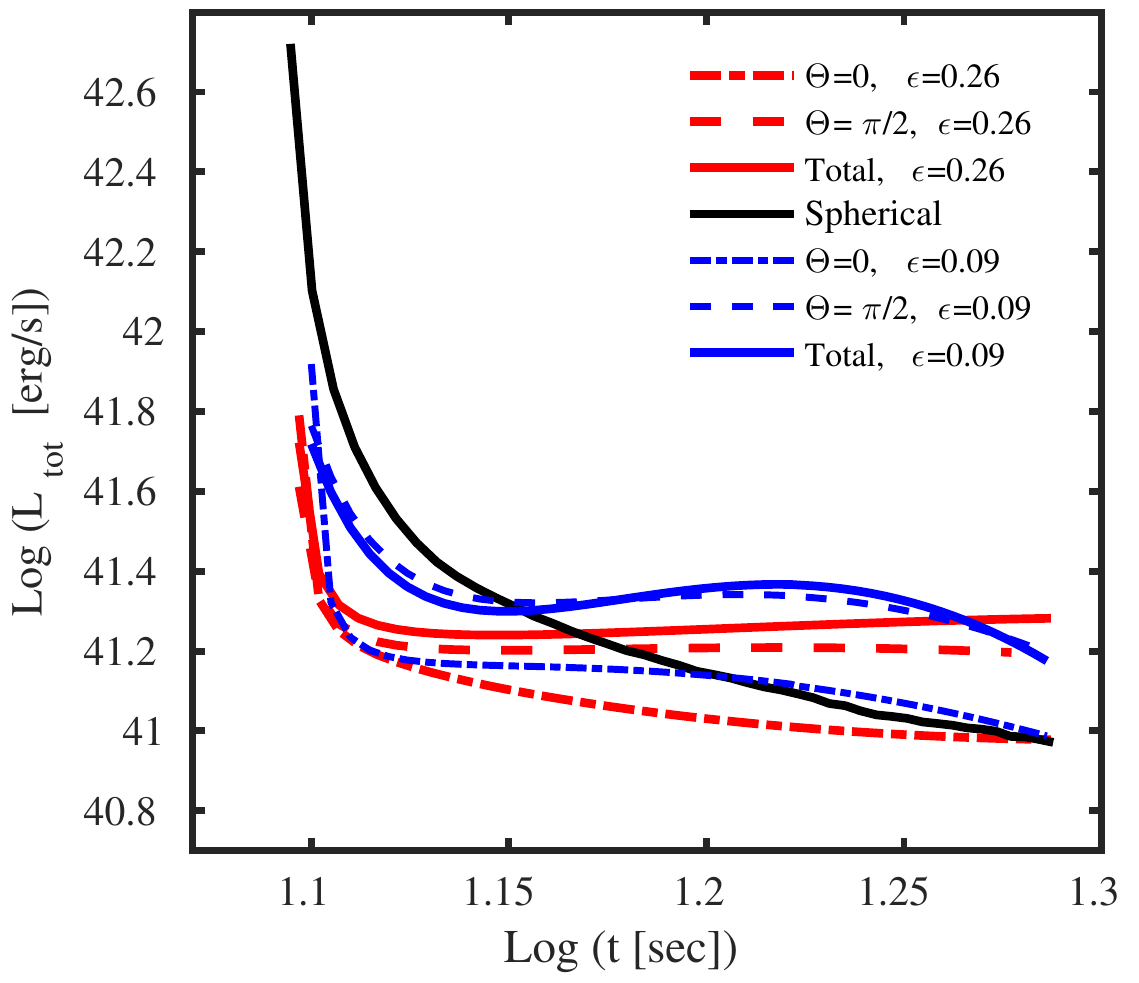}{0.32\textwidth}{(a) Type Ic model}
		  \fig{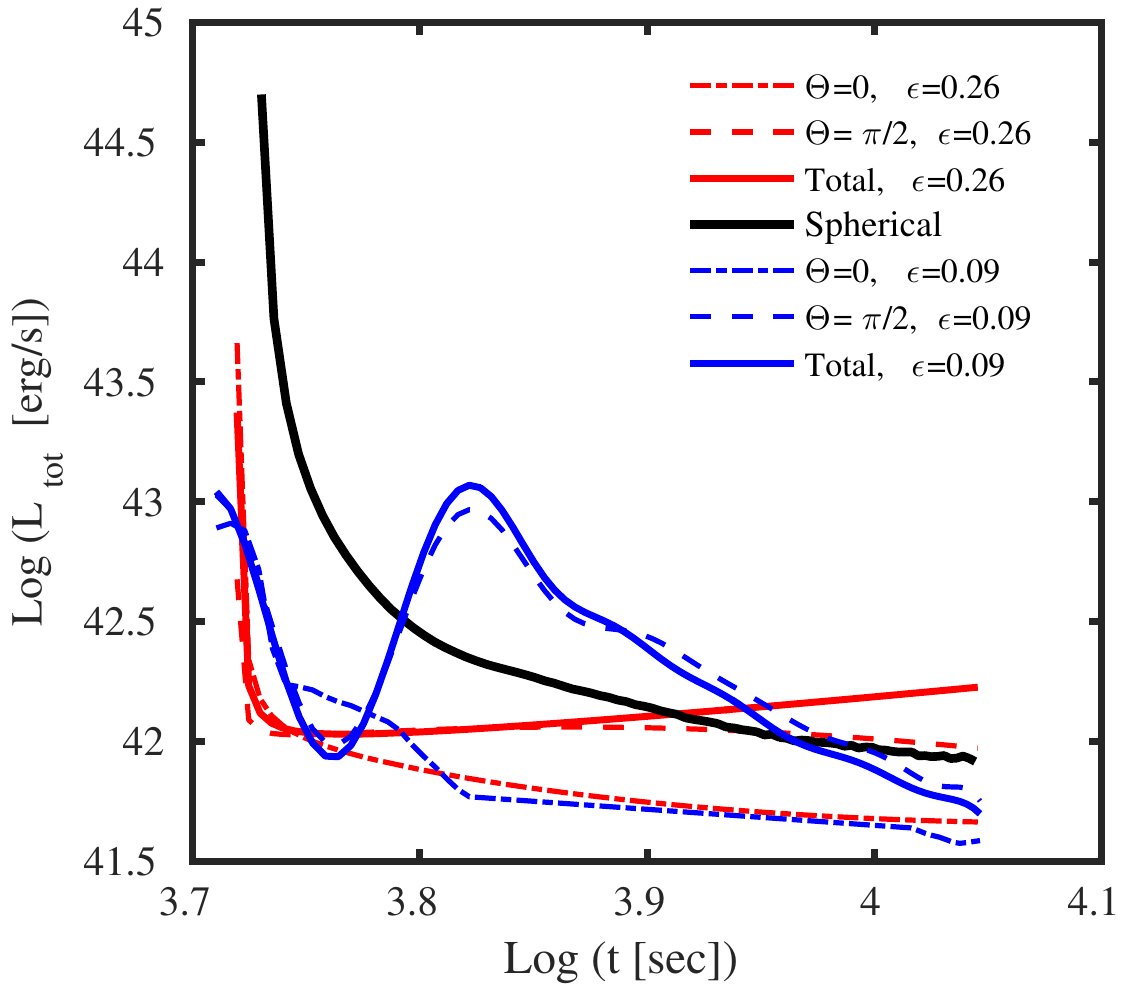}{0.32\textwidth}{(b) BSG model} 
		  \fig{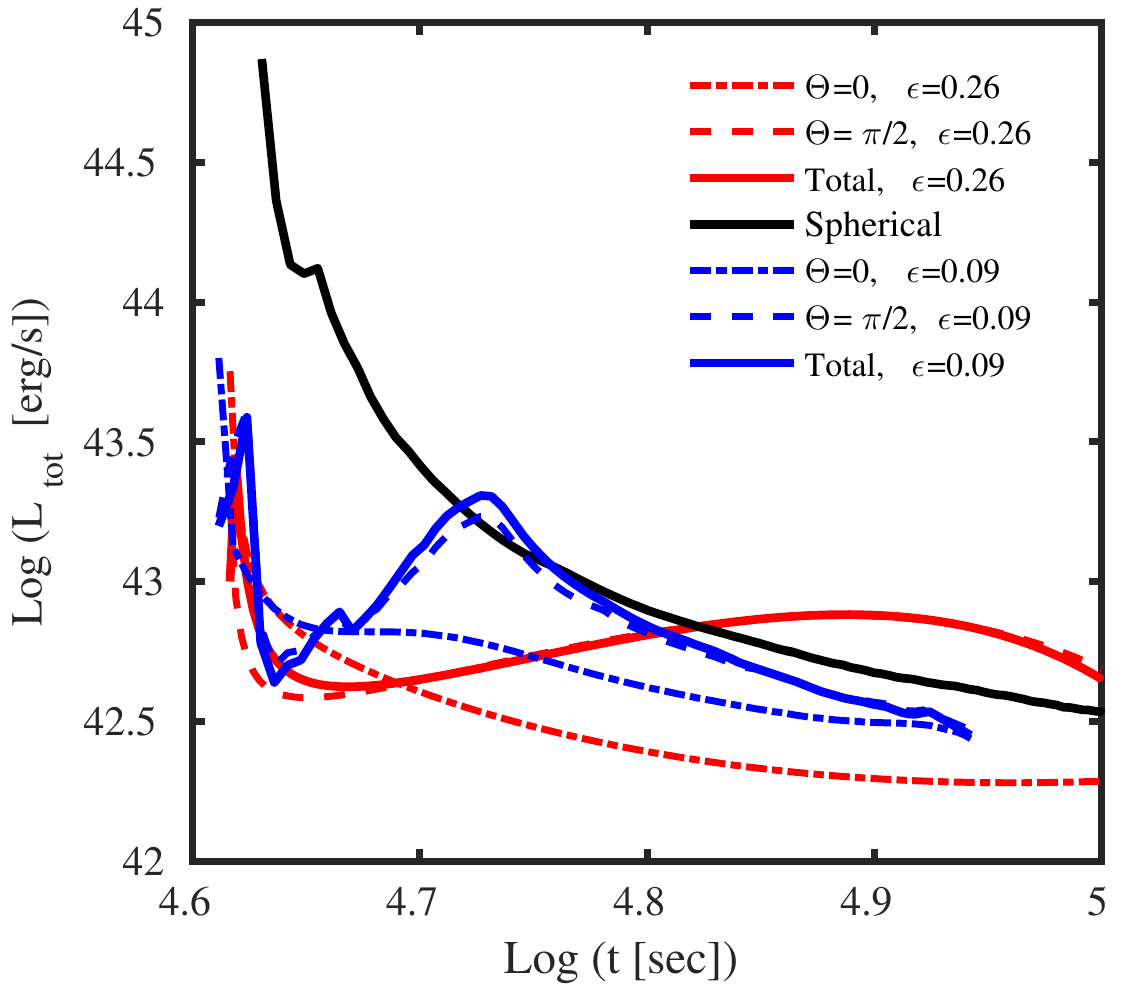}{0.32\textwidth}{(c) RSG model}}

	\caption{\revsec{ The bolometric light curves are presented for three progenitor models; each panel illustrates the light curve for a spherical (thick solid black curve), a mildly aspherical (blue curves, $\epsilon=0.09$), and an oblique (red curves, $\epsilon=0.26$) explosion. Aspherical cases are plotted for an observer along the axis of symmetry ($\Theta=0$, dash-dotted curve, $\epsilon=0.26$), the equator ($\Theta=\pi/2$, dashed curve) as well as the total radiated luminosity (solid curve). {\em Note:} Circumstellar collisions are not included.  }}
	\label{fig:totlum}
\end{figure*}

\subsection{Bolometric Light Curve}
\label{sub:bolLC}
The radiation flux can be expressed from diffusion approximation \citep{Chevalier1992} as
\begin{eqnarray}
 \label{eq:flux}
\mathbf{F}_\text{rad}(\rdiff,\theta,t)&=& -\frac{c}{\kappa \rho(\rdiff,\theta,t)}  \nabla  P_\text{rad}(\rdiff,\theta,t),
\end{eqnarray} 
which takes different values along the diffusion front for aspherical explosions. The total photon luminosity is the rate at which the photon energy is released from the diffusion front \citep{Chevalier1992,Nakar2010} derived in $\S$\ref{sub:difffront}. The total luminosity $L_{\text{tot}}$ at time $t$ is given by
\begin{eqnarray}
 \label{eq:lumtot}
L_{\text{tot}}(t)&=& \int_{S_\text{diff}} \mathbf{F}_\text{rad}(\rdiff,\theta,t) \cdot d\mathbf{A} ,
\end{eqnarray} 
where $S_\text{diff}$ is the diffusion surface and $dA={2\pi} \sin \theta d\theta/({\hat{r} \cdot \hat{n}} )$. As shown in Figure \ref{fig:schematic}, $\hat{r}$ and $\hat{n}$ are unit vectors along radial and normal to $S_\text{diff}$ direction at location $(\rdiff,
 \theta)$ respectively. For spherical SNe, Equation \ref{eq:lumtot} takes the familiar form $L_{\text{tot}}=4 \pi \rdiff^2 F_\text{rad}$.  

The observer dependent luminosity is generated by the geometry depicted in Figure \ref{fig:schematic}. For a distant observer at  $D_\text{obs}$ whose angle from the symmetry axis is $\Theta$, the observed flux is
\begin{eqnarray}
 \label{eq:fluxobs}
F_{\text{obs}}  &=& \frac{1}{D_{\text{obs}}^2} \int_{S_{\text{diff}}} I d S 
\end{eqnarray} 
where $I$ is the intensity, which  we assume is isotropic \revthird{outward at the source:} $I= {F_\text{rad}}/{ \pi}$ for radiative flux $F_\text{rad}$. 
The surface differential projected by the observer is
\begin{eqnarray}
 \label{eq:ds}
dS  = 2 \frac{\rdiff^2}{\hat{r} \cdot \hat{n}} \sin \theta && \left( \phi_\text{max} \cos \Theta \cos \theta_n + \right. \\ \nonumber
&& \left. \sin  \phi_\text{max} \sin \Theta \sin \theta_n \right) d\theta
\end{eqnarray}
where 
\begin{equation}
\label{eq:phimax}
\phi_\text{max} = 
\begin{cases} 
           \pi  & \text{if \ $\cos \theta_n > \sin \theta$  \& $\theta_n>0$} \\ 0 & \text{if \ $\cos \theta_n > \sin \Theta$  \& $\theta_n<0$} \\
           \arccos (-\cot \theta \cot \theta_n) & \text{if \ $\cos \theta_n < \sin \Theta$ } 
\end{cases}.
\end{equation}
and $\theta_n$ is the angle from $\hat{n}$ to the z-axis (i.e., $\cos\theta_n=  \hat{n}\cdot \hat{z}$). According to Equation \ref{eq:ds}, the surface differential takes the simple form $dS= 2 \pi {r^2_{\text{diff}}}/({\hat{r} \cdot \hat{n}} )\sin \theta \cos \theta_n d\theta $ for an observer along the symmetry axis (i.e., $\Theta=0$), while for an equatorial observer ($\Theta={\pi}/{2}$), $dS= 2  {r^2_{\text{diff}}}/(\hat{r} \cdot \hat{n}) \sin \theta \sin \theta_n d\theta$. The observed isotropic-equivalent luminosity at time $t$ is calculated as
\begin{eqnarray}
 \label{eq:lumobs}
 L_{\text{obs}}(\Theta,t) &=& 4 \pi D_\text{obs}^2 F_{\text{obs}}(\Theta,t) \\ 
& =& 4 \int_{S_{\text{diff}}} F_\text{rad}(\rdiff, \theta , t) dS \nonumber .
\end{eqnarray} 
%From Equation \ref{eq:lumtot} and Equation \ref{eq:lumobs}, it is clear that $L_\text{tot}$ and $L_\text{obs}$ are equal for spherical explosions. 

 \rev{ In Figure \ref{fig:spherical} we compare the bolometric luminosity of a spherical explosion obtained from  Equation \ref{eq:lumobs} against the analytical light curves of \cite{Nakar2010}. We note that the peak bolometric luminosities are in good agreement for all three progenitors, but our light curves slowly diverge from the corresponding analytical result in the phase of homologous expansion.  }

\revsec{ Figure \ref{fig:totlum} depicts the bolometric light curves of Type Ic, BSG, and RSG models for spherical (thick black curve), a mildly aspherical explosion with degree of ellipticity $\epsilon=0.09$ according to Equation \ref{eq:epsilon} (blue curves), and oblique shock breakout of $\epsilon=0.26$ (red curves). } In the aspherical cases we consider several views: those of (a) an observer looking along the axis of symmetry ($\Theta=0$) and (b) an observer looking along the equator ($\Theta=\pi/2$), and these are compared with (c) the bolometric luminosity $L_\text{tot}$ defined in Equation \ref{eq:lumtot}. In this section we only discuss the fiducial case, leaving the mildly aspherical results for next subsection. The time $t=0$ is the beginning of the explosion. Note that the time axis is shifted for the light curves of spherical and the mildly aspherical explosion for comparison purposes.

{\nil We first focus on the bolometric light curve, $L_\text{tot}$}. The peak luminosity occurs when the shock breaks from the poles. At this stage, the shock is normal to the surface and oblique breakout is not relevant, so the peak is due to radial diffusion just as in the case of a spherical explosion. Only a small region of the progenitor is hit, however, so light travel time is relatively unimportant in the peak duration. Next, the luminosity rapidly drops as non-radial flows develop and the shock becomes oblique. As discussed in $\S$\ref{sub:LSBO}, when the shock is oblique, a fraction of photons with $\calD(\alpha_f)>1$ cannot immediately escape as they are engulfed in the spray of ejecta. They are released, along with the early diffusion luminosity, at much larger radii when $\calD(\alpha_f)\sim 1$. This can explain the slight increase in $L_\text{tot}$ after the first peak.

%At this phase, there are two sources contributing to the total luminosity: a near-zone glow from the streamlines near the stellar surface (i.e.,  $ \pi/2 \leq \alpha_f \leq \pi$) with $\calD<1$ that let photons diffuse out early, and a far-zone glow from those that trap photons during the breakout and release them later during the adiabatic expansion of the ejecta when the radius is expanded to few $R_*$s.  Our numerical integral in Equation \ref{eq:lumtot} shows that the dominant source in the early light curves of Figure \ref{fig:totlum} is the former. A very crude analytical estimate of the near field-glow can be obtained by taking the mass column density of material on the stellar surface as $\Sigma_\text{diff} = c/(\kappa \vphi)$, and the luminosity as $L_\text{tot} \sim \Sigma_\text{diff} v^2_\theta (dA/dt)$ where $dA /dt \sim 2 R_* \vphi$, we find 
%\begin{eqnarray}
% \label{eq:lumapprox}
% L_{\text{tot}} &\sim & 2 R_* \vphi^2 c \slash \kappa \\ 
%                &\sim & 10^{40.5} \text{erg/s} \ %R_{50}M^{-1}_{10} \kappa^{-1}_{0.34} \frac{\Mstar %\vphi^2}{E_*}.   \nonumber
%\end{eqnarray} 
%This simple estimate gives $L_{\text{tot}} \sim \{ 10^{40.34}, 10^{42.03}, 10^{42.97} \}$ erg/s for our \{Ic,BSG,RSG\} progenitors which nicely confirm our numerical approach in finding the total luminosity in the oblique phase.  

Comparing the total aspherical luminosity with the spherical case shows that the oblique shock breakout luminosity is comparable to cooling envelope emission of the expanding ejecta in the early light curve of spherical SN.

 The light curve depends strongly on the direction to the observer.  Except for the initial breakout peak, which is brightest when observed along the axis, the apparent luminosity  is  higher  when viewed from the equator. This is because (a) the flux along the z-axis is smaller than along y-axis during the plateau phase of the light curve as the regions close to poles have already radiated much of their SBO luminosity, the emission from other regions being greatly suppressed due to viewing angle; and (b) the observer along equator sees the breakout from both progenitor hemispheres, while the one along z-axis can only observe the early light from one. %It worth mentioning that the apparent isotropic-equivalent luminosity can exceed the $L_\text{tot}$, especially at early times for on-axis observers. 
 %observed luminosity along $\theta=0$ is higher than $L_\text{tot}$ in the early evolution as the observed luminosity is isotropic and therefore can be larger when an observer looks along the direction of diffusing photons.

% showing that only a small fraction of photons is released along the equator due to the geometric reduction of effective emission surface near the poles as well as photons being trapped along the streamline with $\calD\gg1$ where radiation diffusion is negligible. It worth mentioning that the observed luminosity along $\theta=0$ is higher than $L_\text{tot}$ in the early evolution as the observed luminosity is isotropic and therefore can be larger when an observer looks along the direction of diffusing photons. Moreover, the bolometric luminosity along the equator ($\theta=\pi/2$) has a double-peaked evolution; the first peak corresponds for shock breakout from the poles and the second one is associated with breakout from the equator.

Finally, note that even though the RSG model does not develop strong oblique flow, we can still obtain its luminosity based on our adiabatic simulations as the hydrodynamic solution must be valid within the diffusion front, where the luminosity is set. 
 
%This mechanism, however, is greatly suppressed during the oblique shock breakout phase.
%It is worth mentioning that the flux is parametrized along the axis of symmetry (un-primed coordinate) and therefore $\theta$ is related to the observer's coordinate as
%%\begin{eqnarray}
%% \label{eq:angle}
%%\cos \theta &=& \cos \theta^\prime \cos \theta + \sin %%\theta^\prime \sin \theta \cos \theta^\prime
%%\end{eqnarray} 

%%$t^\prime = t- \frac{D- \rdiff \cos \theta^\prime}{c}$ takes the effect of light crossing time for an observer at distance $D$ from the center of the explosion into the account. 
\begin{figure}
  \centering
    \includegraphics[scale=0.53]{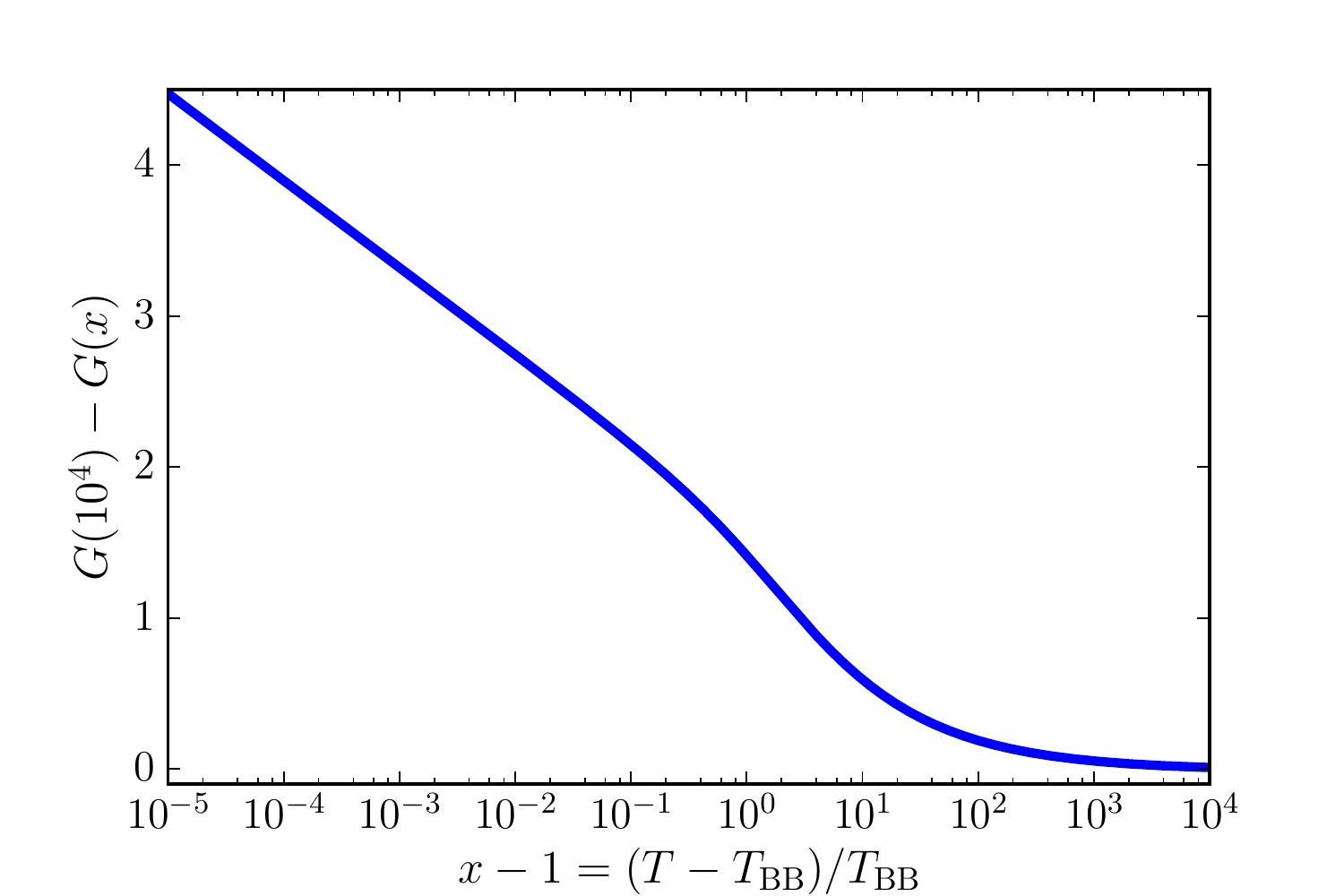}%{Gfun.pdf}
    \caption{The thermalization function $G(x)$ is defined by integrating \revthird{the color temperature evolution, equation (\ref{eq:x})}. Here, $x=T/T_{\rm BB}$ denotes the color temperature excess.}
    \label{fig:Gfun}
\end{figure}

\begin{figure*}[ht!]

\gridline{\fig{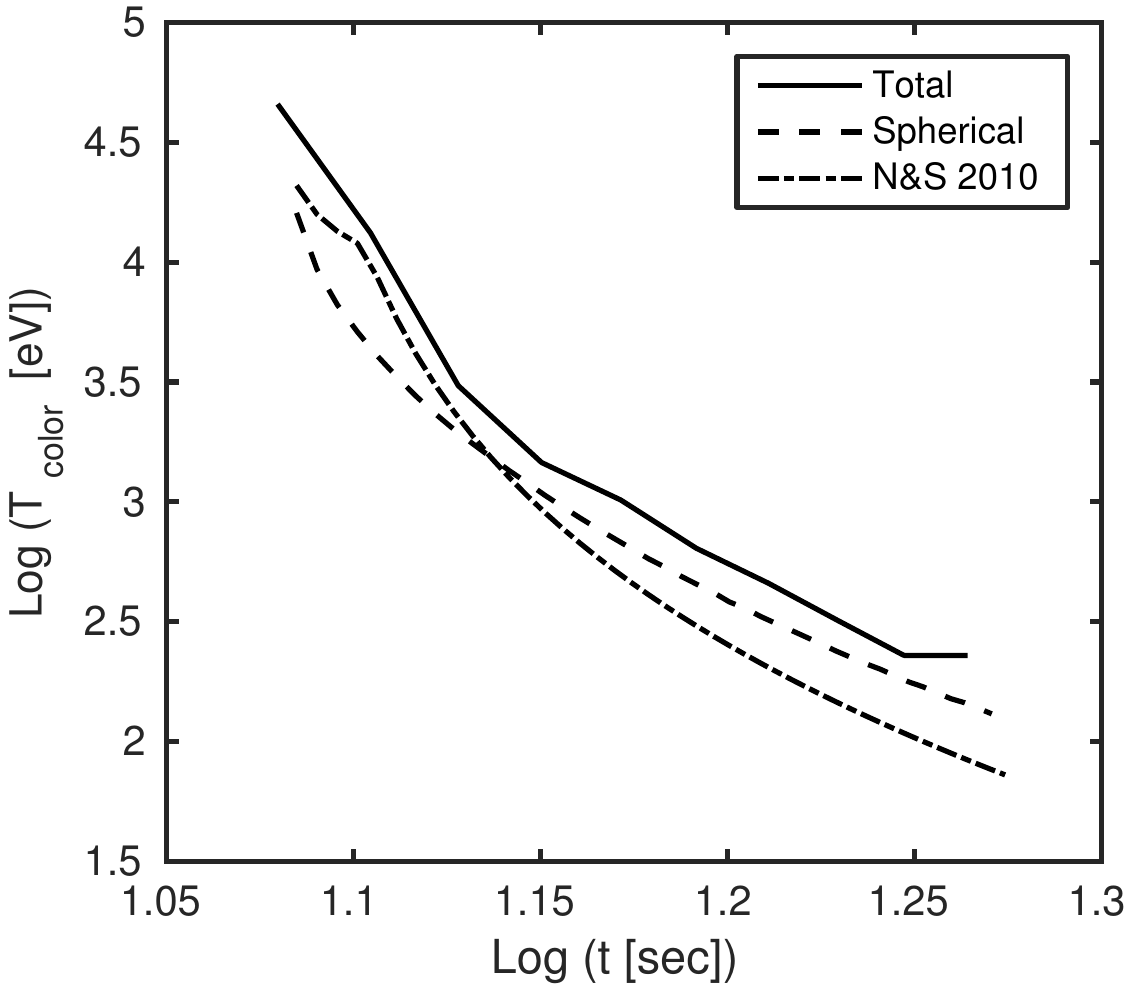}{0.32\textwidth}{(a) Type Ic model}
		  \fig{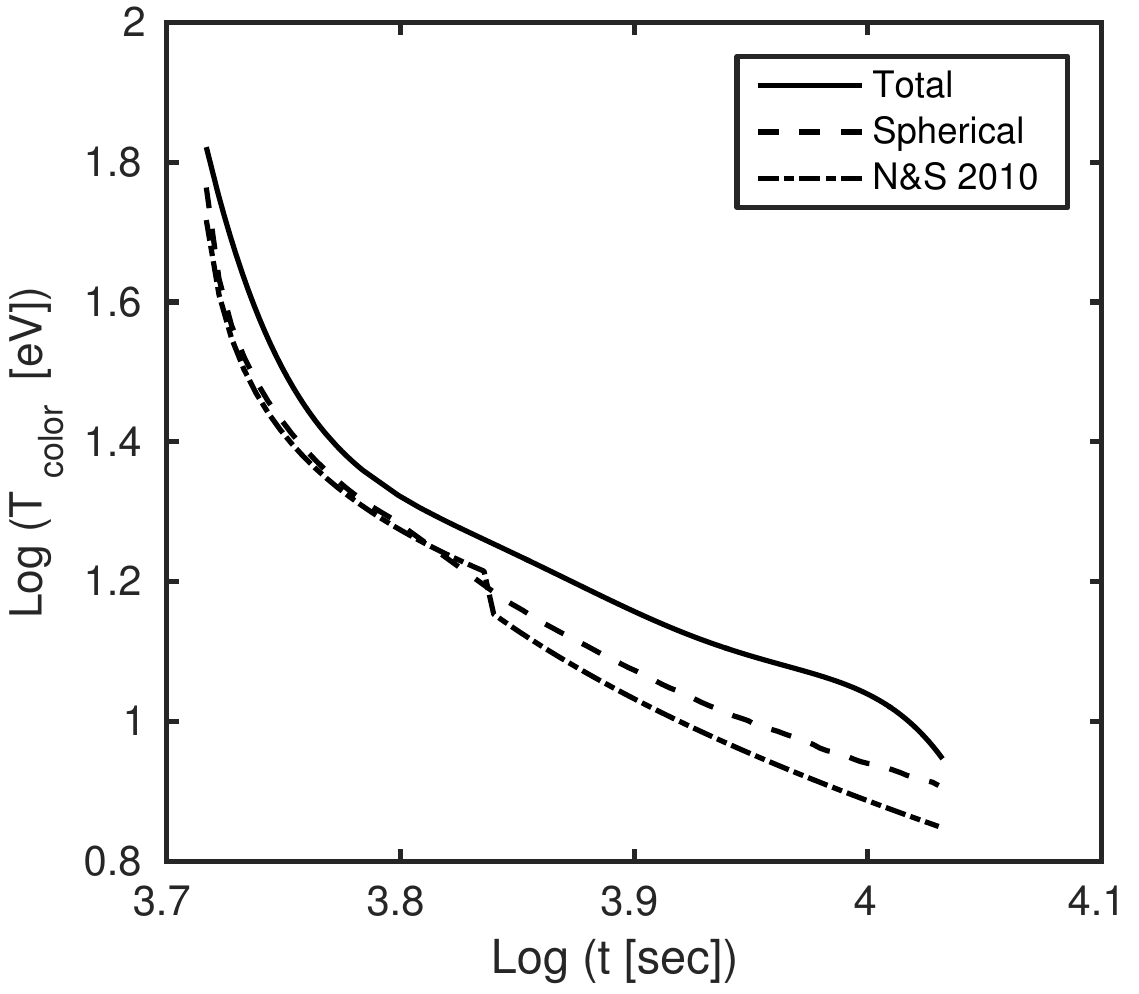}{0.32\textwidth}{(b) BSG model} 
		  \fig{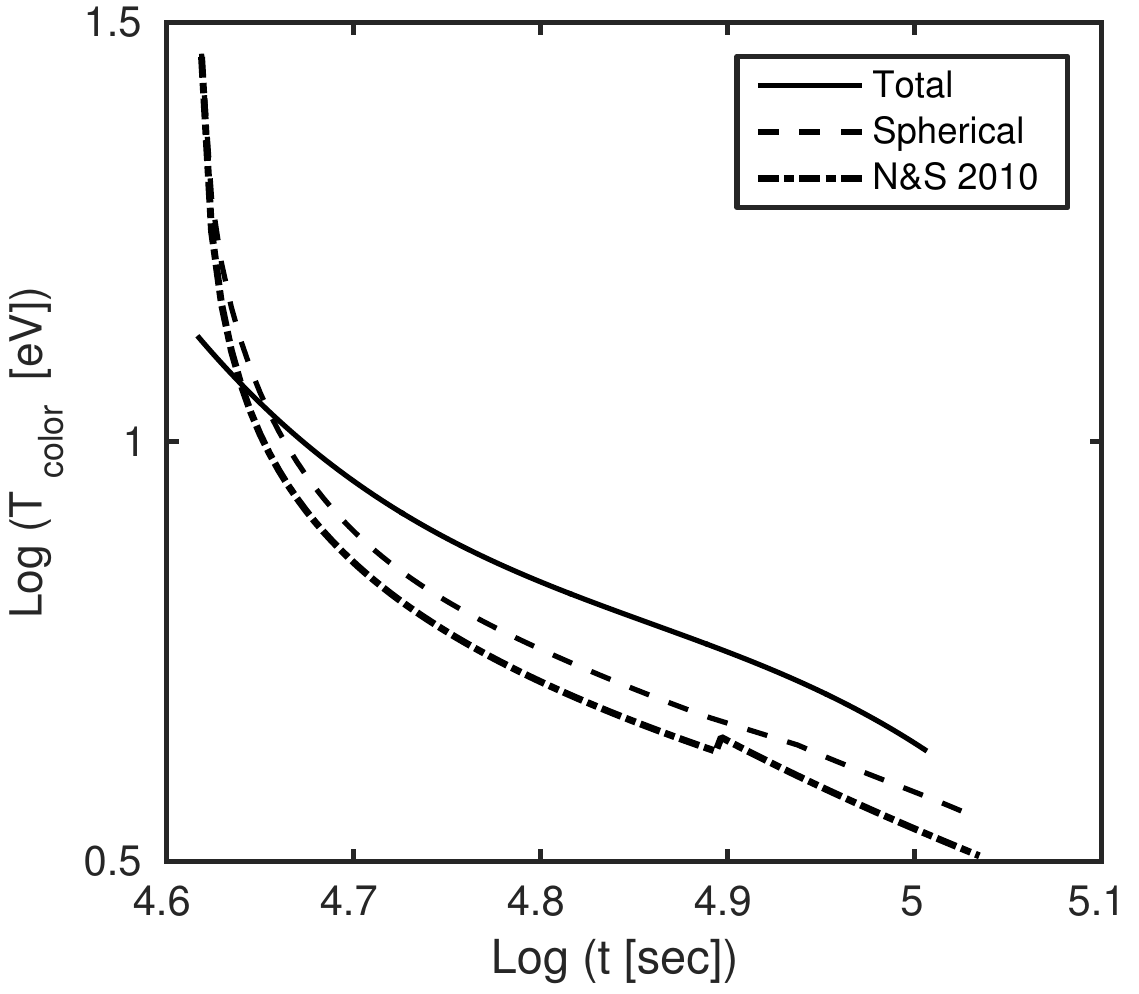}{0.32\textwidth}{(c) RSG model}}

	\caption{ \nil The color temperature evolution is depicted for the oblique simulation (solid line), the spherical \cite{Nakar2010} model (dash-dotted curve), and 1D spherical result (dashed curve). Color temperature is derived from Equation (\ref{eq:colortot}). Our type Ic model reveals non-thermal evolution for which the thermalization front is inside the diffusion front, while the RSG and BSG models maintain the thermalization state. 
	 }
	\label{fig:colorevol}
\end{figure*}

\begin{figure}
  \centering
    \includegraphics[scale=0.65]{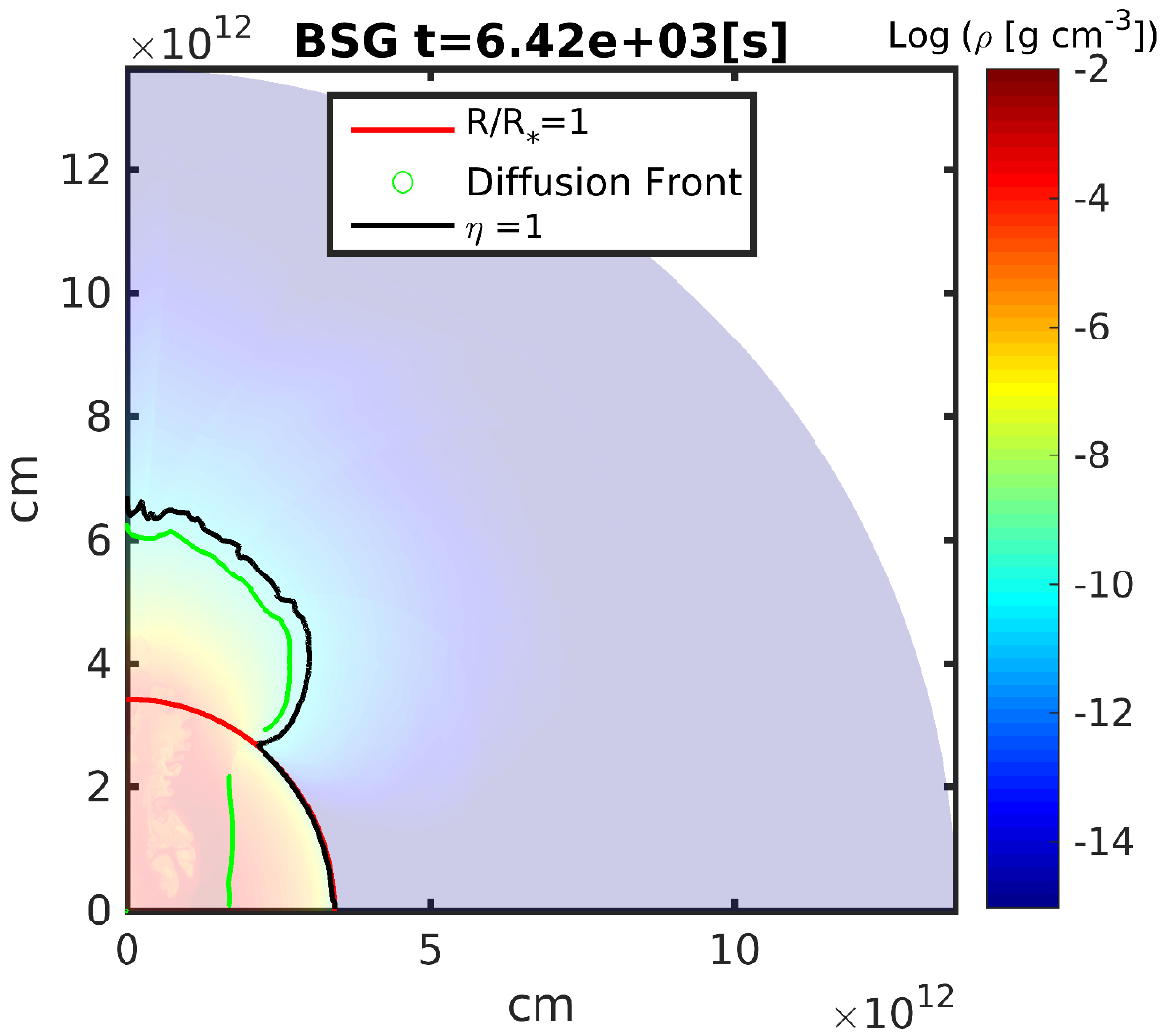}
    \caption{The position of diffusion front ($\rdiff$, green curve), thermalization front ($\eta=1$, black curve), and original progenitor (red curve) are shown for BSG at $t=6420$ s. The color bar represents the logarithm of density. As the diffusion front is well inside the thermalization front, the emission is thermal with color temperature being roughly equal to the blackbody temperature at the diffusion front.}
    \label{fig:rthBSG}
\end{figure}
\rev{
\subsection{Changing the asphericity}
 \label{sub:asphericityParam}
As we discussed in the Introduction, non-radial effects should be weakest in red supergiants, both because these appear to be least aspherical \citep{Leonard2001} -- presumably due to the dampening effect of the hydrogen envelope on blastwave perturbations -- and because more extended stars have stronger diffusion that limits the development of non-radial flows.  While our analysis accounts for the effects of diffusion within a single simulation, changes in asphericity require a comparison between simulations. 
%much stronger for compact progenitors than for red supergiant explosions, since it is much more difficult to keep large progenitors---like RSGs--- aspherical for the long shock propagation time-scale of such progenitors. Therefore, decreasing the degree of asphericity gives more relevant results for these progenitors. To evaluate the impact of asphericity on the early bolometric light curve,
\revsec{For this purpose we consider an intermediate case in which the bipolar momentum is introduced at $r_\text{v}=0.06 R_*$ (as oppose to $r_\text{v}=0.12 R_*$ in our fiducial). In this case, the ellipticity parameter is reduced to $\epsilon=0.09$.} Our diffusion analysis shows that non-radial flows develop only for the compact type-Ic progenitor, while BSG and RSG models do not exhibit non-radial flows according to the criteria in $\S$\ref{sec:oblique}. The absence of non-radial flows means that SBO is not hidden by an optically thick ejecta when the shock hits the equator and---unlike previous sections---the blending of SBO into early cooling emission is not expected to happen.  As shown in Figure \ref{fig:totlum}, this change leads to a strong second peak in luminosity for BSG and RSG models as the shock breaks from the equator. Our results for the mildly aspherical explosion is similar to the light curves of aspherical but \emph{non-oblique} simulations of \cite{Suzuki2016}. The minimum $\epsilon$ for producing oblique breakouts in certain progenitors is estimated in Table 1 of M13.  
} 

\subsection{Color Temperature}
\label{sub:coltemp}
The optical depth of the diffusion front is normally greater than {\cdmnow unity} and therefore the energy of released photons may change as they travel through the upper layers of ejecta. {\cdm If there is time for absorption and emission to act, the population of photons will equilibrate with the local energy density; the color temperature will therefore reflect the blackbody relation where this \revthird{has occurred.}}
%If there is enough time, the diffused photons will share their energy with photons generated by dominant free-free mechanism, which helps thermalizing the photons.
Otherwise, photons from the diffusion front are out of thermalization, with energies reaching as high as hundreds of keV.  \cite{Nakar2010} define the {\cdm thermalization parameter}
\begin{eqnarray}
\eta &=& \frac{\nBB}{\ndotem(\TBB) \min(\tdiff,\tdyn) }
\label{eq:eta}
\end{eqnarray}
where $\nBB= a\TBB^3/(2.7k_B)$ is the photon number density at equilibrium temperature  {\cdm $\TBB = (u_{\rm rad}/a)^{1/4}$,} and $\ndotem(T) = \Cff \rho^2 T^{-1/2}$ is {\cdmnow the effective} free-free photon production rate, where ${\cdmnow \Cff\simeq 8 \times 10^{37}\times \{1,0.5,1.5,2\},}$ {\cdmnow for \{H,He,C,O\} compositions,} is the appropriate coefficient.  {\cdm  We refer to $\eta$ as $\eta_{\rm dyn}$ when $\tdyn < \tdiff$  (which holds within the diffusion front) and as $\eta_{\rm diff}$ when $\tdyn > \tdiff$ (outside the diffusion front). 

\subsubsection{Case A: Weak thermalization at $\rdiff$}
Radiation is rapidly brought into thermal equilibrium if $\eta<1$, and thermal equilibrium is preserved by adiabatic expansion; this implies that $T=\TBB$ at the diffusion front if matter has ever experienced $\eta_{\rm dyn}<1$, even if $\eta(\rdiff)>1$ when photons are released. 
%  This sequence of events is entirely possible, as $\eta_{\rm dyn}$ is minimized during the transition to spherical flow (because for $\rho(m) \propto t^{-q}$, $\eta_{\rm dyn}\propto t^{5q/6 -1}$). 

However if this matter has never been brought into thermal equilibrium, then $T\simeq \min(\eta_{\rm dyn})^2 \TBB$, where the minimum refers to the peak time of photon production for the mass element in question.   In fact, $\eta_{\rm dyn}$ is minimized at the transition to spherical flow (because for $\rho(m,t) \propto t^{-q}$, $\eta_{\rm dyn}\propto t^{5q/6 -1}$; $q$ is typically $\sim 1$ at early times, steepening to 3).   This makes it sufficiently difficult to estimate the color temperature, especially in the non-spherical case, that we opt for a numerical evaluation. 

In Appendix A we show that the color temperature excess $x = T/\TBB$ evolves in each mass element (while it is within the diffusion radius) according to  
\begin{eqnarray}
\frac{d \ln x}{dt} &=& - \frac{\ndotem}{n_\text{ph}}\left(1-x^{-4}  \right). 
\label{eq:xeqn}
\end{eqnarray}
The time derivative is Lagrangian, i.e.\ evaluated along the fluid path. 
Using $n_\text{ph} = \nBB/ x$ and $\ndotem=\Cff \rho^2 T^{-1/2}= \Cff \rho^2 x^{-1/2}  T^{-1/2}_\text{BB}$, we find 
\begin{eqnarray}
\int_{x}^{x_0} \frac{d x}{x^{3/2}\big(1-x^{-4}  \big)} &=& \int_{t_s}^{t}  \frac{\Cff \rho^2 \TBB^{-1/2}}{\nBB}dt \label{eq:x}
\end{eqnarray} 
which we rewrite 
\begin{eqnarray} 
G(x_0)-G(x) &=&  \frac{\Cff \rho_*^2 T^{-1/2}_{*,\text{BB}}}{n_{*,\text{BB}}} \int_{\tsh/t_*}^{t/t_*} h(\tilde{t} )d\tilde{t} \equiv G_* \int dg,
\label{eq:Gx}
\end{eqnarray}
where $G(x)$ is the integral on the left hand side of Equation \ref{eq:x}, $\tilde{t}=t/t_* $ denotes normalized time, and $h(\tilde{t})\equiv dg/d\tilde{t}$ contains normalized simulation parameters.  Integration is along fluid trajectories.    The final color temperature is determined by finding $G(x)$ at $\rdiff$, then inverting to get $x$.  The function $G(x)$ is analytical but cumbersome, so we do not write it out, but we plot it in Figure \ref{fig:Gfun}. 

We integrate along fluid trajectories by making use of FLASH's `mass scalar' capability, meant for passive advection of quantities with the fluid motion.\footnote{We thank Paul Ricker for suggesting this approach.}  We define $g$ as a mass scalar, and update it according to the differential equation $dg=h\, dt/t_*$ between each hydro step.   (We set $dg=0$ for $t<\tsh$ to exclude the contribution to this integral before shock arrival, as this early contribution represents the initial thermal equilibrium of the star and is not involved in post-shock thermalization.  Furthermore Equation (\ref{eq:Gx}) assumes adiabatic flow, so it is invalid across the shock.)   % We plot $g({\mathbf r})$ in Figure ??.   [[when it's ready...]] 

 There are several points to note.  First, $G(x)$ has a logarithmic divergence at $x=1$, and this feature effects thermalization ($T\rightarrow \TBB$ when $G(x)\ll G(x_0)$).  Second, we must estimate the initial value $x_0$ corresponding to the post-shock state,  which we do in \S \ref{SS:Photon_starving} of the Appendix, although the end result is insensitive to this choice so long as $x_0\gg 1$.  Third, a single simulation suffices to determine the normalized thermalization field $g$ in terms of other normalized variables; the color temperature at the diffusion front of a particular explosion can then be determined by setting appropriate values for $G_*$ and $x_0$, and identifying $\rdiff$ as described in \S~\ref{sub:difffront}.  However this only represents the observed color temperature if thermalization is weak at $\rdiff$; we now consider the correction appropriate to thermalization outside the diffusion front. 

\subsubsection{Case B: Strong thermalization at $\rdiff$}

In the alternate case in which $\eta<1$ at the diffusion front, photons continue to be produced in the zone of diffusing radiation and the observed color temperature is set at the location where thermalization fails, i.e.\ where $\eta_{\rm diff}=1$.   We treat this in an approximate fashion, by examining the scaling of $T$ with $\eta_{\rm diff}$ in a diffusive region.   Between the diffusion radius and the photosphere, the diffusion equation $d P_{\rm rad} = -(F/c)\, d\tau$ implies, for a relatively constant flux, $P_{\rm rad} \propto \tau \sim \kappa \rho L_\rho$, while $\tdiff \sim \tau L_\rho/c$ where $L_\rho$ is the local density scale length.  Together with $P_{\rm rad} \propto T^4$ this implies $T\propto (\eta_{\rm diff}/L_\rho)^{-2/17}$; we  ignore the variation of $L_\rho$.  Since $\eta_{\rm diff}$ increases from its value at the diffusion front (which equals $\eta_{\rm dyn}(\rdiff)$) to unity, these approximations imply a color temperature $T = \eta(\rdiff)^{2/17} \TBB(\rdiff)$.    Note that, in this limit, $T(\rdiff) = \TBB(\rdiff)$.  This suggests that the single formula 
\begin{equation}\label{eq:colorTemperatureApproximation}
T_c \simeq {T(\rdiff) \over [1 + 1/\eta_{\rm dyn}(\rdiff)^{2}]^{1/17}   } 
\end{equation} 
captures the color temperature $T_c$ in both regimes.   Here $T(\rdiff)$ is meant to be evaluated according to the inversion of equation (\ref{eq:Gx}), which includes thermalization, and any additional reduction relative to this value takes place only if thermalization is strong at $\rdiff$.     We apply Equation (\ref{eq:colorTemperatureApproximation}) at every angle $\theta$. 

In our calculation the early SN light is determined in its bolometric luminosity by radiation diffusion of shock-deposited heat, and the mean photon energy is determined by the color temperature that results from photon production in the ejecta (or in the diffusion front).    Our post-processing of the adiabatic simulation is therefore similar in spirit to, but more accurate than, the analyses of \cite{Couch2009} (blackbody emission from the photosphere) and \cite{Couch2011} (blackbody emission from an estimated thermalization radius). 
}

\subsection{Average color temperature}

With {\cdm $T_c$} calculated as above, the luminosity-weighted angle-averaged color temperature is 
\begin{eqnarray}
 \label{eq:colortot}
 T_{c, \text{tot}}(t) & = & \frac{1}{L_\text{tot}(t)}\sum_\theta \Delta L_{\theta,t} T_c(\theta,t)
\end{eqnarray} 
where $\Delta L_{\theta,t}$ is the total photon luminosity for a small patch along $\theta$ at time $t$.   Here $\Delta L_{\theta,t}= 2 \pi \int_{\theta}^ {\theta+\Delta \theta}  {\rdiff^2}/(\hat{r}\cdot\hat{n}) |F_\text{rad}(\rdiff, \theta , t)|  \sin(\theta)d \theta$ as described in the integral of Equation \ref{eq:lumtot}.

\begin{figure*}[ht!]

\gridline{\fig{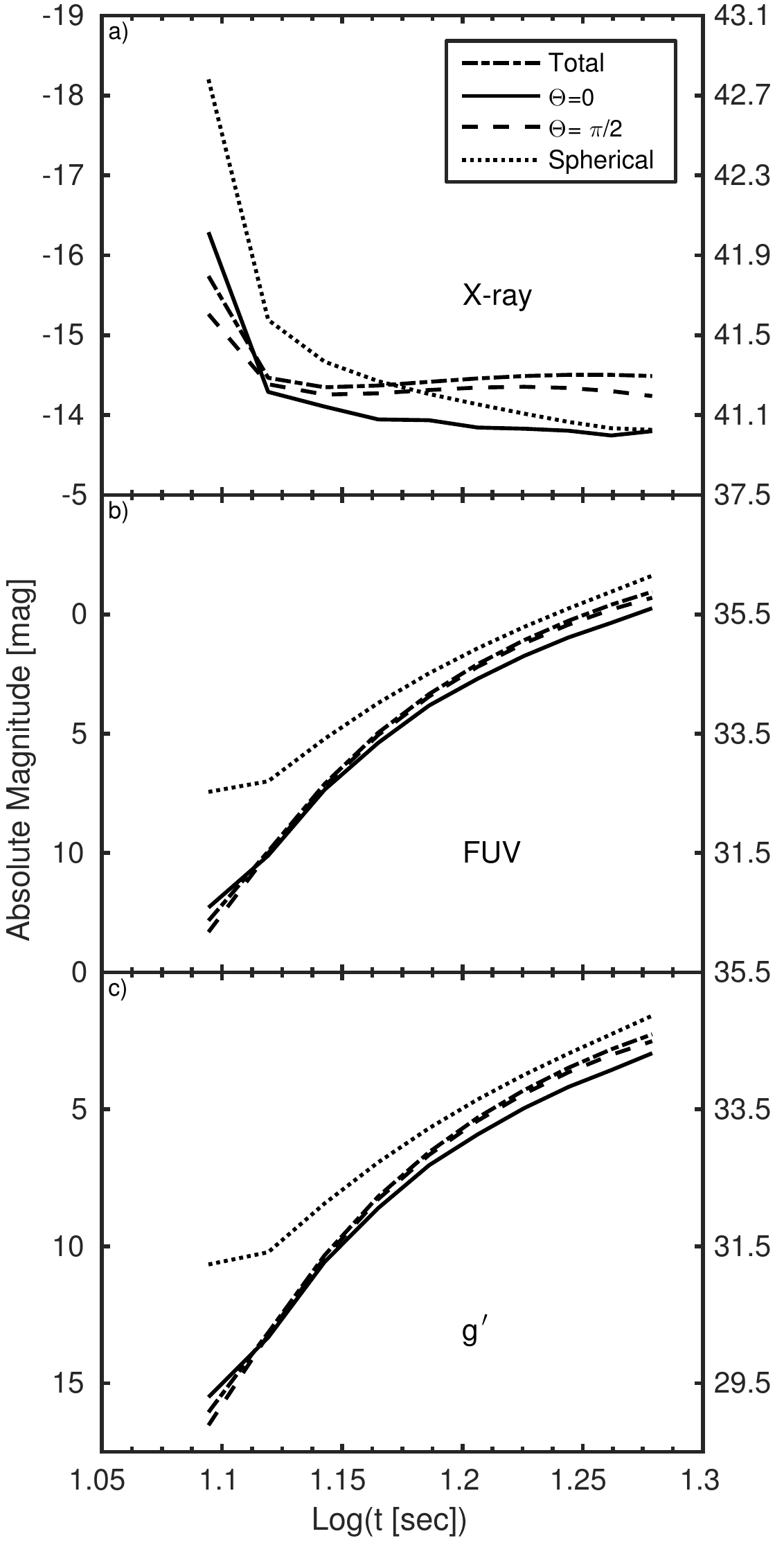}{0.32\textwidth}{ Type Ic model}
		  \fig{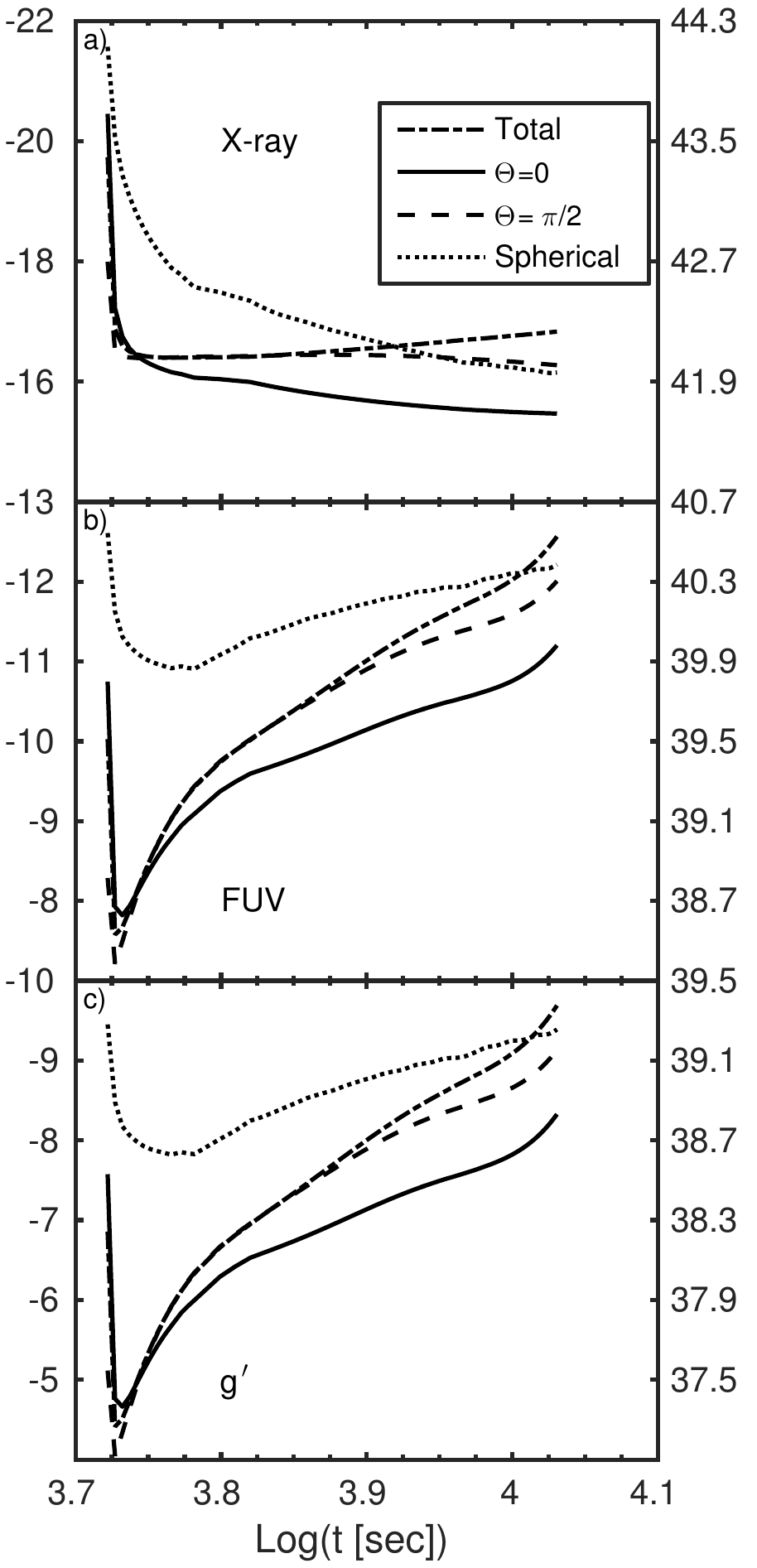}{0.305\textwidth}{ BSG model} 
		  \fig{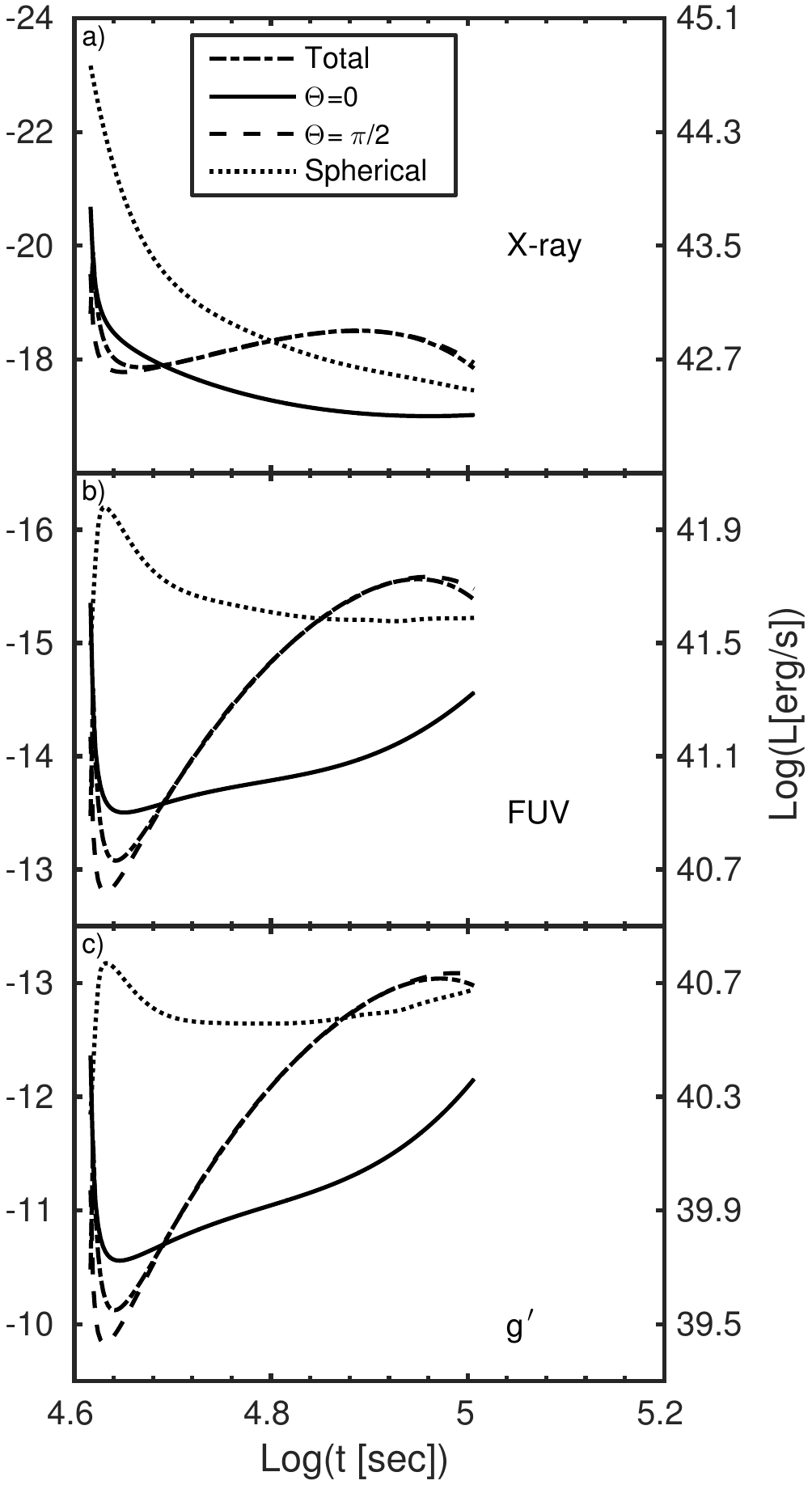}{0.345\textwidth}{ RSG model}}

	\caption{ \nil Multicolor light curves are shown for three progenitor models $\{$Ic, BSG, RSG$\}$ and in three filters: $g^\prime$, FUV, and X-ray (0.2keV-20keV). In each band, four different cases are depicted: (a) total aspherical luminosity and magnitude (dash-dotted curve), (b) an observer along the axis of symmetry, $\Theta=0$ (solid curve), (c) an observer along the equator, $\Theta=\pi/2$ (dashed curve), and (d) multicolor light curves of our 1D spherical explosion model (dotted curve).  All of the non-spherical light curves refer to the fiducial model with $r_\text{v}=0.12$ and $\epsilon=0.26$.  {\em Note:} Emission from the circumstellar collision zone, and some of the shock breakout emission, is not included; this will especially affect the FUV and X-ray bands. }
	\label{fig:bandlum}
\end{figure*}

{ \nil In Figure \ref{fig:colorevol}, the total color temperature evolution of the oblique simulation is shown for the progenitors Ic, BSG, and RSG (solid curve). Color temperature---like luminosity--- depends on the position of the observer, which can be derived by substituting the luminosities in Equation \ref{eq:colortot} by the observed luminosities; the difference between observer-dependent color temperature and $T_\text{c,tot}$, however, is found to be negligible. We also plot the color temperature predicted by   \cite{Nakar2010} for the spherical case (dash-dotted curve) along with the color temperature of our 1D simulation (dashed curve) for comparison; the methods agree well.  

For $T_\text{c,tot}(t)$, we first derive the location of $\rdiff$ and $r_{\eta=1}$ for the progenitors over time to check whether weak or strong thermalization regimes applies.  For the compact type Ic progenitor, we find $\rdiff>r_{\eta=1}$ giving rise to a non-thermal spectrum and thus the color temperature starts as high as 50\,keV (hard x-rays) and rapidly cools down to 100\,eV in a few seconds. We should, however, be cautious while comparing these results to observations; as pointed out by \cite{Couch2011}, the thermalization front of compact progenitors---like Wolf-Rayet stars---is located in the region between reverse shock and forward shock due to thick winds that these progenitors normally launch.  We do not address this possibility.  Our analysis of the BSG model also shows that \revthird{ $\rdiff<r_{\eta=1}$} during the evolution as can be seen in Figure \ref{fig:rthBSG} for $t=6440$s. For this model, the color temperature peaks at 65 eV--- slightly higher than its spherical counterpart --- during the breakout from the poles and continues to closely follow the spherical limit as an oblique flow develops. 
%Note that we plot the thermal BSG model of \cite{Nakar2010} in this panel. 
Similarly, for the RSG model, the color temperature is initially as high as $\sim$ 13 eV.  The influence of asphericity on $T_c$ is at most a factor of two.  As $T_c>1$\,eV, hydrogen recombination will not significantly affect our light curves. }

\subsection{Multicolor Light Curves}

Using color temperature and bolometric luminosities, derived in Subsections \ref{sub:coltemp} and \ref{sub:bolLC} respectively, we employ a blackbody photon distribution to find band-dependent light curves, ignoring complications such as comptonization, finite photon chemical potential, and light travel time effects. In Figure \ref{fig:bandlum}, the multicolor light curves are shown in four different cases: (a) total luminosity $L_\text{tot}$; (b) an observer along the axis of symmetry, $\Theta=0$; (c) an observer along the equator $\Theta=\pi/2$; and {\nil (d) multicolor light curves of our 1D spherical explosion}.

\begin{figure*}[ht!]

\gridline{\fig{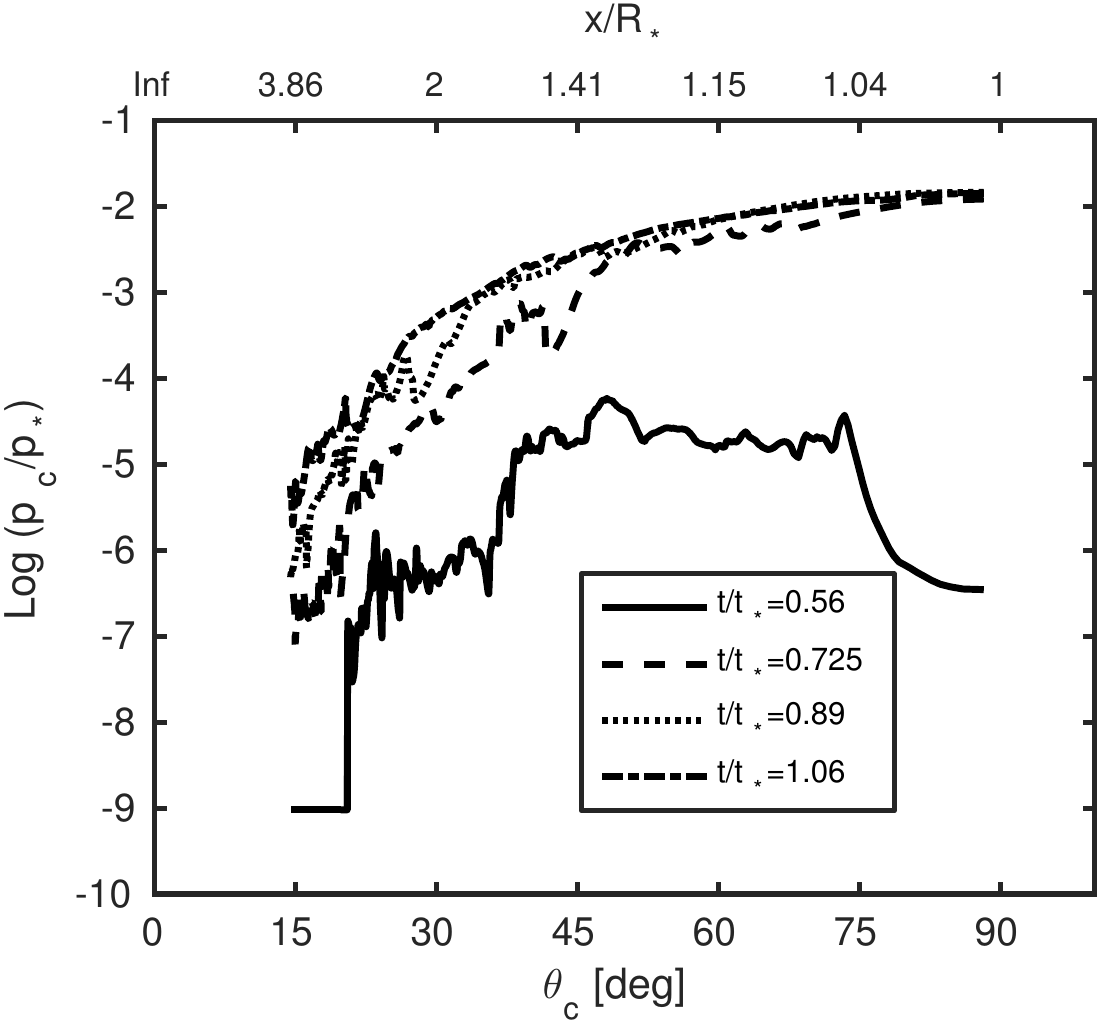}{0.33\textwidth}{(a) Pressure in collisional zone}
		  \fig{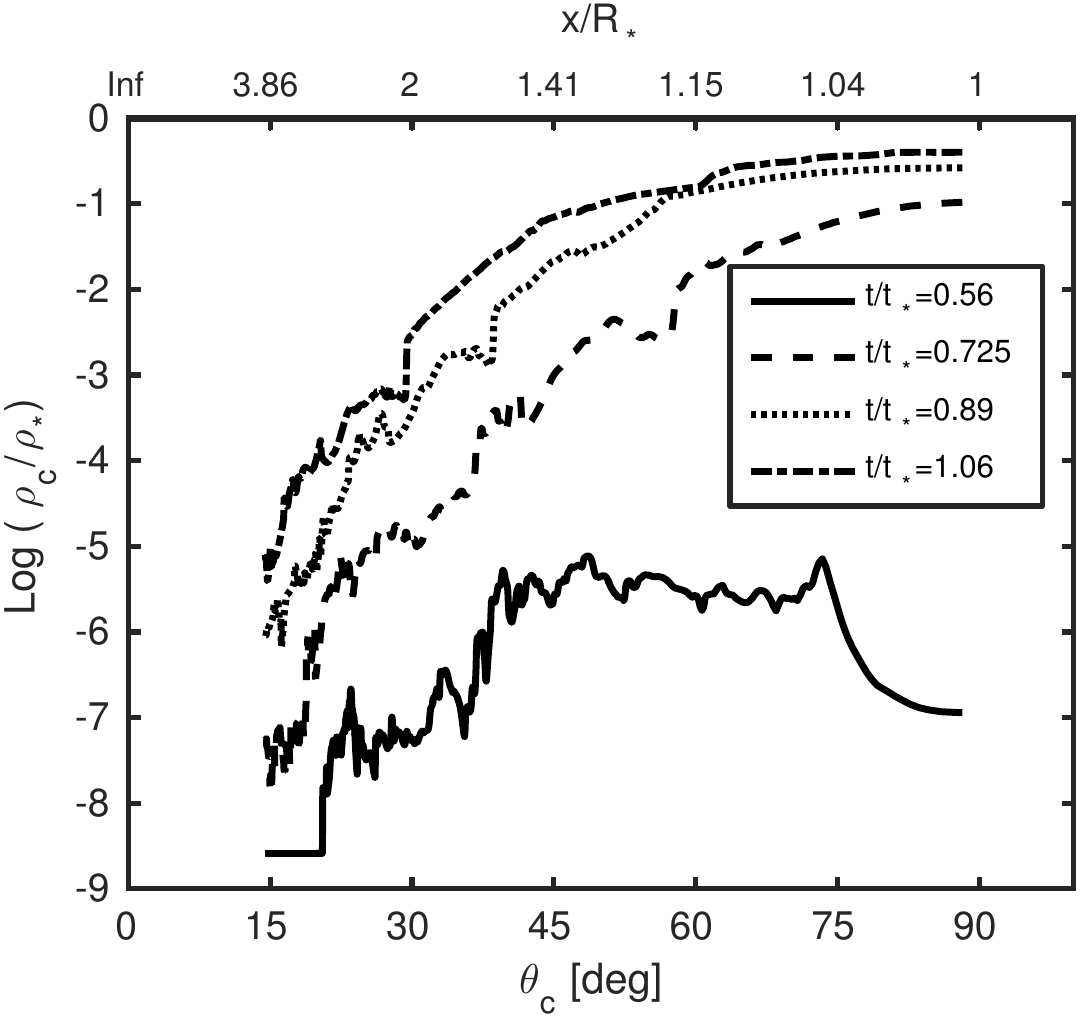}{0.32\textwidth}{(b) Density in collisional zone} 
 		  \fig{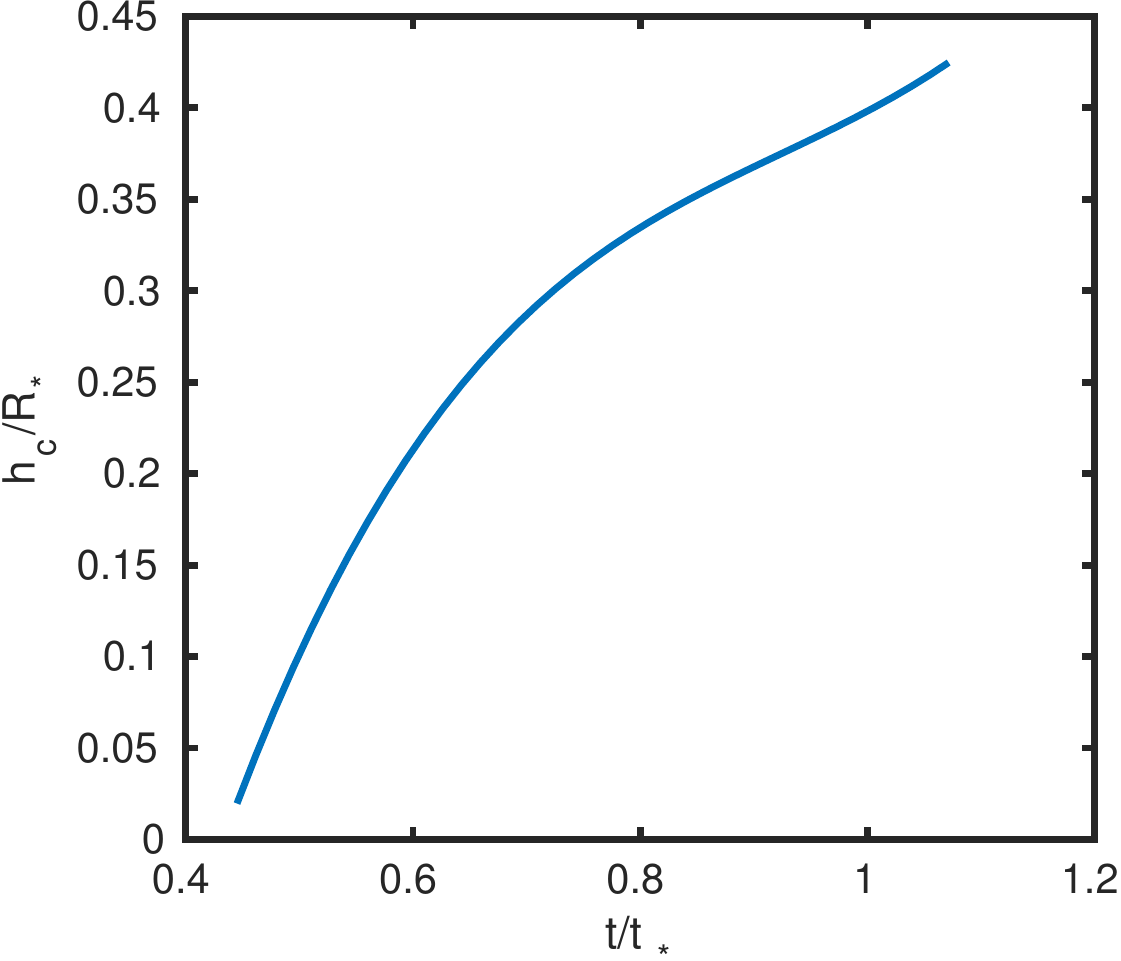}{0.34\textwidth}{ (c) Collisional zone height at $4R_*$} 
 }

	\caption{When non-radial flows from the two hemispheres collide with each other along the equator, they form an expanding wedge or disk of collided material. Left panel shows the normalized pressure in the collisional disk per $\theta_c$. Each curve represents ${\cdm \log} (p_c/p_*)$ at a specific time. The quantity $\theta_c= \sin^{-1}(R_*/x)$ is inversely related to the equatorial distance. In the middle panel, the normalized density in the collisional wedge is shown at different times. The right panel illustrates the height of the wedge at the border of simulation box, i.e., $R/R_*=4$, as a function of normalized time. }
	\label{fig:colquantity}
\end{figure*} 

\begin{figure*}[ht!]

\gridline{\fig{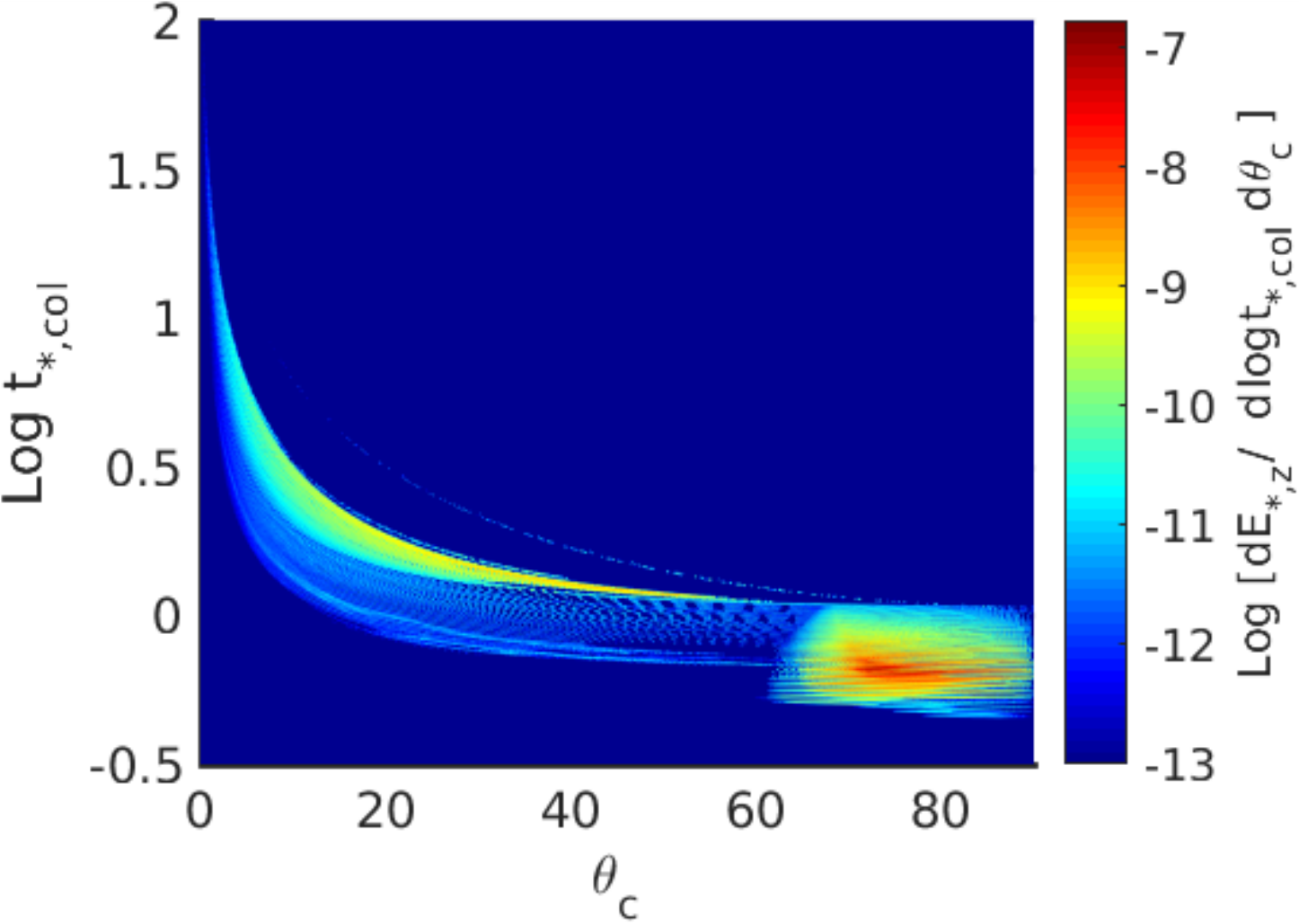}{0.49\textwidth}{(a) Map of energy per time per location \revthird{in the equatorial collision: $\theta_c= \sin^{-1}(R_*/x)$ is the angle to the tangent for a collision at radius $x$.}}
		  \fig{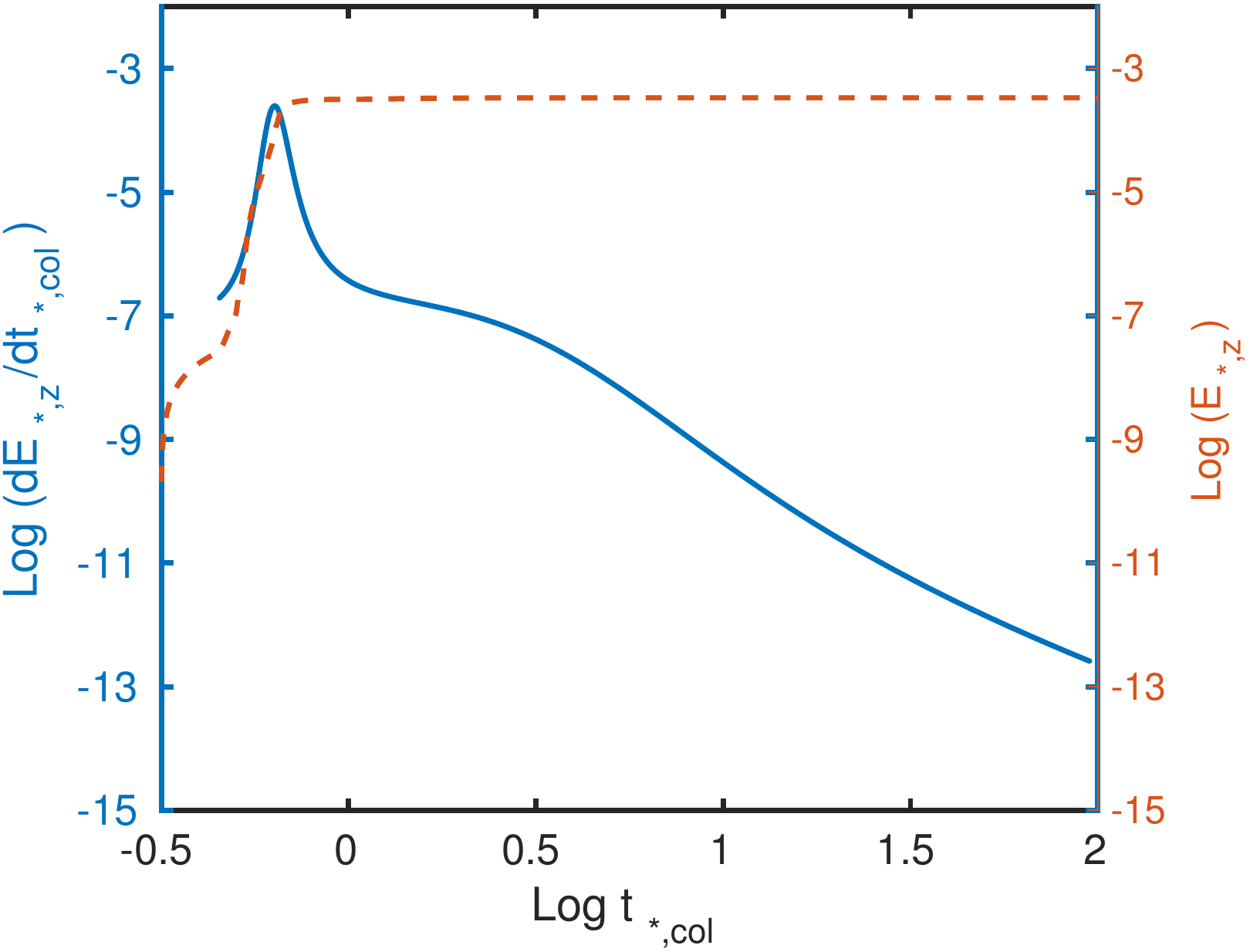}{0.45\textwidth}{(b) Collisional energy per time} 
}

	\caption{Left panel presents the rate at which the normalized kinetic energy along direction $-\hat{z}$ enters the collisional wedge at different times and locations along the equator. The right panel shows the cumulative  normalized kinetic $E_{*,z}$ energy that has entered the collisional wedge (orange dashed curve), as well as the logarithm of kinetic power entering the wedge (solid blue curve), over time. }
	\label{fig:colenerg}
\end{figure*} 

As shown in Figure \ref{fig:bandlum}, the early light curve of our type Ic model is \revthird{initially} dimmer than the spherical case by \revthird{several magnitudes} in the optical (i.e., $g^\prime$ filter) and far ultra violet (FUV) bands. There are two parameters determining the shape of these light curves: the bolometric light curve is less luminous for aspherical cases, and the color temperature is about 35\,keV higher in aspherical cases (shown in Figure \ref{fig:colorevol}), shifting the optical and FUV bands more towards the Rayleigh-Jeans tail of the spectrum, hence reducing the flux in these bands. In X-ray (0.2-20\,keV), the evolution approximately follows the bolomteric light curves of Figure \ref{fig:totlum}(a) with minor differences. The earliest part of the light curve takes time to rise in $g^\prime$ and FUV, so long that the observed band is in the Rayleigh-Jeans tail of the spectrum and the bolometric luminosity is decreasing. The rise stops when the band color matches the color temperature. 
%Therefore, the rise-time of the light curve is longer for aspherical cases for several seconds as the color temperature is higher.

{\nil The early  light curves of the aspherical BSG model have distinctive features in optical and FUV as shown in Figure \ref{fig:bandlum}.  During the early non-oblique breakout phase, $L_\text{tot}$ suddenly drops  $\gtrsim3$ magnitudes over $\sim$100 seconds. As a result, the early aspherical light curves exhibit a peak, despite the fact that $T_\text{band}\ll T_c$ and  $T_c$ is rapidly declining. The light curves then rise at a mostly decelerating rate. This evolution is somewhat different from the spherical case. First, the early part of the aspherical light curves ($t<10^{3.8}$s) is $\lesssim4$ mag dimmer  than the spherical one. Second, the early peak in aspherical cases is strong ($\gtrsim 3$ mag drop), compared to the spherical case  ($\lesssim2$ mag drop). The oblique phase ($t>10^{3.8}$s) of the polar light curve is similar to, but dimmer than, the expanding ejecta phase of a spherical explosion.  In X-ray, the light curves closely follow their bolometric counterparts.}

{\nil For the RSG model, the aspherical light curves are even more observer-dependent in optical and FUV. For example, the observer looking along the axis of symmetry sees a shallower early peak which follows by a slower rise. This is because the on-axis view of the isotropic equivalent light curve, $L_\text{obs}(\Theta=0)$, declines more slowly than the equatorial view,  $L_\text{obs}(\Theta=\pi/2)$ and is monotonically decreasing. The early peak in the  spherical model is associated with the shock breakout flash in optical and FUV. The spherical model's light curve then declines until $t=10^{4.7}$s and $t=10^{4.9}$s in FUV and $g^\prime$, respectively, and rises afterwards; this is when the planar phase ends and spherical evolution begins \citep{Nakar2010}. } The aspherical evolution does not exhibit these features and  the earliest light is much dimmer than in the spherical case. We must, however,  be cautious when interpreting the band-dependent light curves of the RSG model. For this model, {\cdm the simulation's imposition of adiabatic flow is incorrect in the outermost zones.  Because the diffusion front typically moves inward in mass coordinates, the adiabatic approximation tends to be valid up to the point that luminosity is generated; however any imprint of the outermost matter must be taken with a grain of salt.}

%\begin{eqnarray}
% \label{eq:eta}
%\eta \approx \frac{7 \times 10^5 \text{s}}{\text{min\{  }\tdiff,\tdyn\}} \Big( \frac{\rho}{10^{-10} \text{g cm}^{-3}} \Big)^{-2} \Big(\frac{kT_{BB}}{100 \text{eV}}\Big)^{3.5}
%\end{eqnarray} 

% if there is enough time for the free-free photons to be generated effectivl while diffusing out of the diffusion front \citep{Nakar2010}. The depth at which photons have enough time to be thermalized is called thermalization front 
 
%where $\tau_*=\sqrt{3 \tau_{\text{abs}} \tau_{\text{tot}}}$ is effective optical depth and $\tau_{\text{tot}}=  \tau_{\text{abs}} + \tau_{\text{es}}$ is the sum of absorption and scattering optical depths \citep{Rybicki}. If the thermalization front is exterior to the diffusion front $r_\text{th}>\rdiff$, the color temperature of the emitted photons $T_c$ is obtained at $r_\text{th}$ by $T_c=T(r_\text{th})$, otherwise the released photons from the diffusion front are not in thermal equilibrium and do not have enough time to be thermalized in the upper layers of the ejecta, in which case the color temperature can be significantly higher than the blackbody temperature at the diffusion front \citep{Katz2010}. 

\begin{figure}[t]
  \centering
    \includegraphics[scale=0.44]{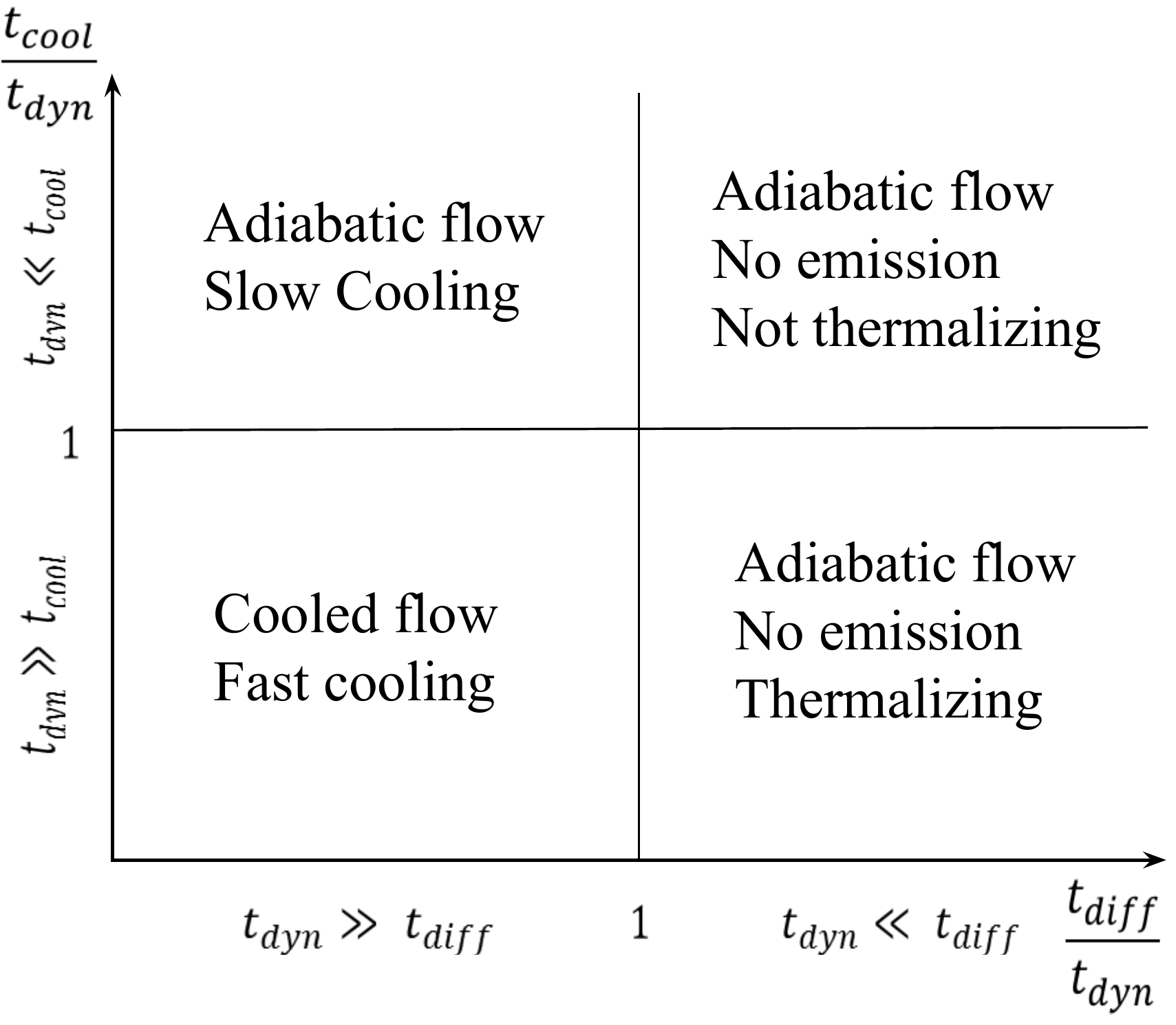}
    \caption{The emission observed from the collision wedge depends on diffusion time $t_\text{diff}$, cooling time $t_\text{cool}$ and dynamical time $t_\text{dyn}$. In the fast cooling regime the upper limit of luminosity is the rate at which the vertical kinetic energy enters the wedge. This leads to a collisional luminosity peak $L_{p,\text{col}}=$\{$10^{45.76},10^{43.14}$\}erg s$^{-1}$ at time $t_{p,\text{col}}=$\{$10^{1.45},10^{4.07}$\}s for type-Ic and BSG progenitor respectively.  }
    \label{fig:colregimes}
\end{figure}

\section{Circumstellar Collisions}
\label{sec:col}

Circumstellar collisions are a direct consequence of the formation of oblique flows. As shown in Figure \ref{fig:evolution}(d), the collisions happen along the equator as oblique flows run into each other and form an expanding wedge of material with excess pressure and density. According to Figure \ref{fig:colquantity}(a) and \ref{fig:colquantity}(b), the wedge pressure $P_c$ and wedge density $\rho_c$ are both increasing within our limited simulation time. The horizontal in these plots represents  $\theta_c\equiv \sin^{-1}(x/R_*)$ which {\cdm corresponds to the latitude of a point on the stellar surface tangent to a line crossing the equator at $x$. This labels equatorial radii in a compact fashion.} The tip of this axisymmetric wedge is roughly located at $r/R_*=1$, equal to the original radius of the progenitor. The height of this wedge $h_c$ increases with time as shown in Figure \ref{fig:colquantity}(c) due to increasing internal energy in the region. 

Figure \ref{fig:colenerg}(a) shows the energetics of the collision zone at different times and locations along the equator. In order to make this plot, the region where the fluid elements either enter the wedge or leave the simulation box between two consecutive temporal checkpoints is first identified. In this step, we assumed that the fluid elements move along straight lines at constant speed. Next, the kinetic energy toward the equator (along $\mp \hat{z}$ for the upper/lower hemisphere) is computed for each cell in that region, and subsequently we obtained the location and time at which each fluid element hits the equator according to its velocity vector. This way, we ensure that the energy of a fluid element is not summed multiple times between temporal checkpoints and that all energy in the colliding region is taken into account, even the energy of  material that enters the collisional wedge after the simulation end time. As shown in  Figure \ref{fig:colenerg}(a) the input power peaks in the interval $\theta_c=(65,90^\circ)$, {\cdm i.e. $r<1.1 R_*$} over the period $0.4<t/t_*<1$.  After this peak,  the distance of the equatorial collision increases with collision time {\cdm because the horizontal velocity is at most $2 \vphi$}. Figure \ref{fig:colenerg}(b) depicts the radially-integrated kinetic power into the equatorial wedge, as well as its cumulative energy.  Its final value is $E_z = 10^{-3.17} E_*\simeq10^{47.8}$\,erg considering both hemispheres. Note that the characteristic energy for all three progenitors is $10^{51}$erg but the energy that ends up in non-radial flows depends on the progenitor model as discussed in $\S$\ref{sec:oblique}. The physical value of $E_z$ is, thus, mostly accurate for type Ic and BSG models. For RSG progenitors, $E_z$ is suppressed as non-radial flows are inhibited by the early onset of radiation diffusion.  \revthird{(Furthermore, RSGs are likely to be less aspherical than our fiducial case;  S14 found $E_z\propto \epsilon^{4.5}$ within adiabatic simulations.) }

Deriving the emission from the collision zone is not trivial and is a subject for future work. Here, we only describe possible solution regimes. As shown in Figure \ref{fig:colregimes}, the dividing lines for these regimes are set by parameters {\cdm ${t_\text{diff}}/{t_\text{dyn}}$ and ${t_\text{cool}}/{t_\text{dyn}}$}, where $t_\text{cool}$ is the local cooling timescale. When $t_\text{diff}>t_\text{dyn}$, no emission is expected to be released as the photons are trapped in the collision wedge and the adiabatic condition holds; this situation resembles SN before SBO, {\cdm or ejecta before the arrival of the diffusion front}. Adiabatic flow is also maintained when $t_\text{cool}\gg t_\text{dyn}$, so there is not enough time to cool. In the fast cooling regime, i.e., $t_\text{cool}<t_\text{dyn}$ and $t_\text{diff}<t_\text{dyn}$, the luminosity is constrained by the rate at which the vertical kinetic energy $E_z$ enters the collision wedge. This rate is shown in Figure \ref{fig:colenerg}(b) in normalized units by a solid curve. {\cdm For the other regimes, \revthird{the kinetic luminosity} is only an \revthird{upper} limit \revthird{to the radiative luminosity.}} Scaling the normalized luminosity in Figure \ref{fig:colenerg}(b), we find the peak of the collisional \revthird{luminosity} $L_{p,\text{col}}=$\{$10^{45.76},10^{43.14}$\}erg s$^{-1}$ occurs at time $t_{p,\text{col}}=$\{$10^{1.45},10^{4.07}$\}s for type-Ic and BSG progenitor respectively. This {\cdm potential luminosity source is significant relative to the diffusive light curve shown in Figure \ref{fig:totlum}.   However, we postpone any prediction of the actual emission to a future paper.}

\section{Summary and Discussion}
\label{sec:summary}

{\cdmnew  We summarize this work and our major findings as follows.   

\noindent {\em - Goals and Strategy}: Our goals are to highlight the effects of aspherical geometry on the dynamics of shock breakout and on supernova early light; to test theoretical predictions regarding oblique shock breakout in a global simulation; and to advance the art of using adiabatic simulations to predict the outcome of radiation hydrodynamics and the band-dependent SN display.   We do not attempt to study the mechanism that breaks spherical symmetry,  nor do we survey realistic progenitors or include complicated physics (gravity, initial thermal energy, relativity, nuclear reactions, etc.) that are important for the central engine and that remain important in certain classes of explosions.  Instead we consider strong, bipolar \revthird{explosions} within a single, simple, polytropic progenitor.  We carefully study the effect of resolution on the explosion and on the breakout dynamics.  We use the scale-free nature of adiabatic non-self-gravitating hydrodynamics to scale our results to several SN types.  Then, we inspect our results to identify features that are invalid (for the chosen progenitor) due to the influence of photon diffusion; this is especially important in limiting the production of non-radial ejecta in diffuse red supergiant explosions.  Next we trace the progress of a photon diffusion front through the ejecta, acquiring the local energy flux and the bolometric light curve.  Finally we track photon production within the ejecta, in order to identify a color temperature for each patch of the diffusion front and  hence to derive band-dependent light curves.  In each step we rely on the fact that the rates of photon diffusion and production are  power laws of the local fluid quantities, and therefore can be scaled from the simulations through our variables $\calD$ (P\'eclet number for non-radial flows; eq.~[\ref{eq:calD_Peclet}]), $g_\varepsilon$ (dimensionless, Galilean-invariant diffusion parameter, eq.~[\ref{eq:diffvel1}]), and $G$ (photon thermalization parameter, eq.~[\ref{eq:Gx}]).

\noindent {- \em Dynamical evolution: }
We break spherical symmetry by adding a bipolar momentum to the hydrodynamic solution of an early spherical explosion. The shock first breaks out from the poles  but then develops laterally, giving rise to a spray of ejecta in different directions.  The asphericity in the simulation is sufficient to form highly non-radial flows during the shock breakout, limiting the ejection speed and strongly affecting the shock breakout emission and early light. It also engenders ejecta-ejecta collisions in the equatorial disk.  All of these features conform to the theoretical expectations of \citet{Matzner2013} and \citet{Salbi2014}, and are consistent with the outcome of earlier numerical works such as those by \citet{Couch2009}. \rev{We note that reducing the aphericity can lead to mildy aspherical explosions which are more realistic for larger progenitors such as RSGs. 
Non-radial motions are inhibited by radiation diffusion in such explosions. 
%Such explosions inihibit non-radial flows and thus have different light curve characteristics than the main simulation of this work as discussed in $\S$\ref{sub:asphericityParam}. 
}

\noindent{- \em Resolution dependence: } 
We investigate the effects of resolution on the results and find that the speed and structure of fastest ejecta is relatively robust for sufficiently high resolution runs, but Kelvin-Helmholtz instabilities affect the inner ejecta in an inevitably resolution-dependent fashion.   We find that the creation of non-radial ejecta requires that the characteristic turning depth $\lphi$ is resolved by about three zones.   While this is far below the 683-zones-per-$\lphi$ resolution of \citeauthor{Salbi2014}'s fiducial run, it demands significant resolution of $R_*$ in a global simulation.  

\noindent{- \em Validity of adiabatic simulations: }  We use adiabatic simulations for a problem involving photon diffusion and radiative processes, so it is important to establish a regime of validity.    This is especially important for the non-radial flow from oblique shock breakout: we find that diffusion significantly inhibits this flow ($\calD \ll 1$) for red supergiant progenitors ($400 R_\odot$); marginally affects  blue supergiant explosions \revthird{($49 R_\odot$)}; and is negligible for more compact progenitors.   Again, this accords with the expectations set by M13 and S14.   

\noindent{- \em Shock breakout emission: } Whereas several prior works have assumed SBO emission is \revthird{extended but otherwise} unaffected by the manner in which the shock reaches the stellar surface, M13 and S14 point out that strongly non-radial flow traps and hides this emission.     We derive a simple estimate (eq.~\ref{eq:L_SBO_estimate}) for the SBO luminosity in the case that photons are partially trapped.  Those photons  are  trapped in optically thick ejecta until they reach radii at which diffusion is important; for this reason the SBO emission blends into the early envelope cooling luminosity. 

\noindent{- \em  Bolometric light curve: }  We  trace the progress of a photon diffusion front through the ejecta,  acquiring the local energy flux at the diffusion radius and observer-dependent and spherically-averaged bolometric light curves.   We note that this method is robust against a breakdown of the adiabatic approximation, as the diffusion front moves inward relative to the matter -- so the flow is essentially adiabatic until radiation is released.    For our fiducial example (asphericity factor $\epsilon=0.26$) we find that (1) the early peak in the luminosity is comparable to spherical luminosity peak but briefer in time; (2) the peak is followed by a plateau phase for which the shock has become oblique -- this phase is comparable to envelope cooling phase of spherical explosion but somewhat dimmer; (3)  geometrical effects make the light curves observer dependent. In particular, the plateau is brighter for equatorial observers than polar observers. 

\noindent{- \em Thermalization, color temperature, and band-dependent light curves:}   Deviations from thermal equilibrium between radiation and matter have a controlling influence on early SN brightness in observed bands. Photon \revthird{production} can either precede the arrival of the diffusion front, or occur as radiation diffuses through the ejecta \citep{Nakar2010}.  We develop a technique to trace the thermalization of the photon population in each fluid element of the post-shock flow, up to the arrival of the diffusion front, and employ the results (along with an approximate treatment of the strongly thermalized case) to predict the color temperature of each sector of the ejecta.  From this we build band-dependent light curves.  Our analysis indicates that the radiation field in type Ic explosions is poorly thermalized, whereas thermalization is strong in RSG explosions.  Again, BSGs represent an intermediate case.  Because asphericity increases the color temperature, we see distinctive features in the optical and FUV.  For example, the rise time is extended, and the first peak in magnitude is dimmer but deeper than for the spherical case.  

\rev{ Our analysis is somewhat different from that of \citet{2017ApJ...845..168W} and \citet{2017arXiv170802630B}, who also conduct non-spherically-symmetric hydrodynamical simulations and then predict band-dependent light curves based on the result.   Whereas these authors assume a state of homologous free expansion and conduct Monte Carlo radiative transfer within the ejecta, accounting for radioactive heating, our approach addresses the emission of shock-deposited heat before homologous expansion is established.   We anticipate that the two approaches will ultimately be combined.  }

\noindent{- \em Circumstellar ejecta collisions: } A striking consequence of the non-radial nature of oblique breakout is the collision between ejecta outside the star. In the bipolar explosion considered here, ejecta-ejecta collisions occur in an expanding wedge around the equator of the explosion. We postpone a detailed analysis of radiative processes in this region to a subsequent paper,  but we derive the energetics and constrain luminosity in the fast cooling regime to show the observational importance of these collisions. We find that  a factor of $\sim 10^{-3.17}$ of the total energy ends up in the collisions and if this kinetic energy gets converted to radiation efficiently, the upper limit to the luminosity peak is $L_{p,\text{col}}=$\{$10^{45.76},10^{43.14}$\}erg s$^{-1}$ for type-Ic and BSG progenitor respectively. We note that this estimate does not apply to RSG model as radiation diffusion breaks the adiabatic assumption made in the simulation.

}

\section{Conclusion} \label{sec:conc}

Rich as it is, the parameter space of spherical models does not cover all of the dynamics and radiation processes that apply to supernova explosions, and this holds for SN breakout emission, early light, and circumstellar interactions just as it does for the central engine.   Non-spherical effects on the early light range from mild (extending the shock breakout emission but not altering it; \citealt{Suzuki2010}) to extreme (essential changes to the breakout emission and early diffusive light, and a new source of dense circumstellar interactions), depending on the departure from spherical symmetry and the importance of radiative diffusion in the explosion.   In general one should suspect non-spherical effects, like those we consider here, if there is independent evidence of aspherical flow.  \revthird{Examples} would include a declining early linear polarization (as seen in SN 2008ax by \citealt{Chornock2011} and in SN 2011dh by \citealt{Mauerhan2015}) or high-velocity nickel/iron or other `inner' ejecta (as seen up to 3500\,km\,s$^{-1}$ in SN 1987A by \citealt{1990ApJ...360..257H}).    In other cases there have been discrepancies between modelling of the early light and constraints on the progenitor radius (SN 2011dh; \citealt{Soderburg2012}); extra $u$-band emission and transient narrow absorption features (SN 2013ge; \citealt{Drout2016}); and differences between observed early light curves and what is expected from spherical theory \citep[e.g.,][]{Taddia2015,Garnavich2016}.   While some of these features might be attributable to factors like extended stellar envelopes \citep{Piro2015} or intense pre-SN mass loss  \citep[e.g.,][]{Smith2011,Margutti2017}, one cannot appeal to early radioactive heating from very high-velocity $^{56}$Ni \citep{Taddia2015} without  accounting for non-spherical effects as well.   

Moreover, the central engine is itself likely to be non-isotropic \citep{Blondin2003,Akiyama2003} and binary interactions are likely to spin up or tidally distort a fraction of SN progenitors \citep{Smith2011,Sana2012}.  Deviations from spherical symmetry will be weaker  for more extended progenitors, both because their explosions are known to be more spherical, and because rapid photon diffusion tends to prevent the development of non-radial motions.  However, in the most compact progenitors the early phase of emission (a few $t_*$ after shock breakout) is optically dim and ends rapidly.  We therefore posit that this particular signature of aspherical explosions will be most detectable in progenitors of intermediate radius, i.e. blue or yellow supergiant stars.   Circumstellar collisions might then be the most notable hallmark of asphericity in compact Type Ibc events, but any firm conclusion awaits future study.  

While this work demonstrates what types of changes one might attribute to an aspherical event, it is very limited in considering only a single structure for the progenitor and \revthird{a restricted explosion geometry}, and in separating the dynamical problem from the radiation transfer problem (even when analyzing shock breakout, where the two are clearly linked).   Radiation-hydrodynamics codes, especially with multi-group radiation transfer, will offer far more sophisticated solutions.   In this regard we note that \citet{Suzuki2016} recently studied an aspherical blue supergiant explosion within a radiation hydrodynamics code (M1 closure scheme) and did not see the development of strong non-radial flows.   This is consistent with our theory given the smaller departure from spherical symmetry in the  \citeauthor{Suzuki2016} model explosion. 

\software{FLASH4.2}

{\nil \acknowledgements}  {\cdmnew We thank Paul Ricker for advice regarding the capabilities of the Flash code, and Stephen Ro, Yuri Levin, and Maria Drout for suggestions and comments. We also thank the anonymous referee for the helpful comments on the original draft of this work. This work was supported by an NSERC Discovery Grant (CDM) and a QEII-GSST Fellowship (NA).  As the Appendix and parts of \S \ref{sub:bolLC} were written at KITP Santa Barbara, we derive partial support  from the U.S.\ National Science Foundation under Grant No.\ NSF PHY-1125915.  Our simulations were carried out on Compute Canada resources.  CDM thanks the Monash Centre for Astrophysics and the organizers of  the KITP program {\em The Mysteries and Inner Workings of Massive Stars} for hospitality and support.   }

\appendix 
\section{Photon production and temperature below the diffusion front} \label{Appendix}
We find the color temperature evolution in a region of trapped matter and radiation by writing the conservation laws for photon number, internal energy, and mass in Lagrangian form:
\begin{eqnarray}
\frac{dn_\text{ph}}{dt} &=& -n_\text{ph} \nabla \cdot \vvec + \ndotem - \dot{n}_\text{abs} \label{eq:nph}  \\
\frac{du}{dt} &=& -(u+p) \nabla \cdot \vvec = - \gamma u \nabla \cdot \vvec \label{eq:energycont}\\ 
\frac{d \rho}{dt} &=& -\rho \nabla \cdot \vvec
\label{eq:masscont}
\end{eqnarray}
where $n_\text{ph}$ is photon number density,  $\ndotem$ and $\dot{n}_\text{abs}$ are photon emission and absorption rates respectively, $u$ denotes the internal energy density, and  $p=(\gamma-1)u$ according to gamma-law equation of state.  The operator $d/dt = \partial/\partial t + {\mathbf v} \cdot \nabla$ is the Lagrangian time derivative.  For photon-dominated gas $\gamma=4/3$ and $u=2.7n_\text{ph}kT$, we find
\begin{eqnarray}
\frac{d \ln T}{dt} &=& -\frac{d \ln n_\text{ph}}{dt} +\gamma \frac{d \ln \rho}{dt} \nonumber \\
&=& (\gamma-1)  \frac{d \ln \rho}{dt} - \frac{\ndotem - \dot{n}_\text{abs}}{n_\text{ph}}.
\label{eq:dlnT}
\end{eqnarray}
Equation \ref{eq:dlnT} is obtained by plugging $\nabla \cdot {\mathbf v}$ from Equation \ref{eq:masscont} into Equation \ref{eq:nph}. Taking $T_\text{BB} \propto u^{1/4} \propto \rho^{\gamma/4}$ and normalizing the left-hand side of Equation \ref{eq:dlnT} by $T_\text{BB}$ give
\begin{eqnarray}
\frac{d \ln (T/T_\text{BB})}{dt} &=& \left(\frac{3}{4}\gamma-1\right)  \frac{d \ln \rho}{dt} - \frac{\ndotem - \dot{n}_\text{abs}}{n_\text{ph}} \nonumber \\
&=& - \frac{\ndotem - \dot{n}_\text{abs}}{n_\text{ph}}.
\label{eq:dlnTTBB}
\end{eqnarray}

Using $\ndotem=\int {4 \pi j_\nu}/(h\nu) d\nu$ and $j_\nu=\alpha_\nu B_\nu(T)$ from  Kirchhoff's law, we find $\ndotem=\bar{\alpha}c \nBB(T)$. Similarly, $\dot{n}_\text{abs}=\int \alpha_\nu c \frac{dn_\nu}{d\nu}d\nu=\bar{\alpha}c n_\text{ph}$. Plugging $\ndotem $ and $ \dot{n}_\text{abs}$ into the right hand side of Equation \ref{eq:dlnTTBB}, we get
\begin{eqnarray}
\frac{\ndotem - \dot{n}_\text{abs}}{n_\text{ph}} &=& \frac{\ndotem}{n_\text{ph}}\left[1-\frac{n_\text{ph}}{\nBB(T)}  \right] \nonumber \\
&=& \frac{\ndotem}{n_\text{ph}}\left[1-\left(\frac{T}{T_\text{BB}(u)}\right)^{-4}  \right],
\label{eq:nratio}
\end{eqnarray}
where the last equation is found using $u=2.7n_\text{ph}kT=aT^4_\text{BB}(u)$ and $\nBB(T)=aT^3/2.7k$. Combining Equation \ref{eq:dlnTTBB} and \ref{eq:nratio} and defining $x=T/T_\text{BB}(u)$ we arrive at Equation (\ref{eq:xeqn}) for the evolution of the temperature.

\subsection{Initial photon starving factor}\label{SS:Photon_starving}
 
\rev{ Our analysis requires an initial value for the photon starving parameter $x$.  For this we appeal to the theory of non-relativistic, radiation-dominated shocks.  For conditions relevant to core-collapse supernovae, the shock jump is entirely controlled by radiation diffusion, with no hydrodynamic discontinuity \citep{2015ApJ...811...47T}.   In steady state this diffusive shock has the  form described by \citet{1976ApJS...32..233W}, in which changes occur on a length scale $c/(\vsh\kappa \rho_0)$ and time scale $\tshock = c/(3\vsh^2\kappa\rho_0)$, where $\rho_0$ is the upstream density and $\vsh$ the shock speed.}

\rev{Free-free photon production in such shocks has a diffusive shock transition in which the temperature reaches its maximum, and a relaxation region in which diffusion is negligible. The relaxation zone is covered by equation (\ref{eq:xeqn}), so we use an estimate of the peak temperature to give $x_0$.   The time to build up a photon density $n_{\rm ph}$ is  $\tff = n_{\rm ph}/\ndotem$.  Therefore a shock that builds up photons in time $\tshock$, assuming absorption and the initial photon population are both negligible, reaches a peak temperature $T_p$ set by $\tff\simeq\tshock$, given conditions at the shock: $n_\text{ph} k T_p  \simeq \rho_0 \vsh^2$ and $n_\text{ph} T_p^{1/2}/(\Cff \rho_0^2) \simeq c/(\kappa \rho_0 \vsh^2)$, implying $k T_p  \sim k\kappa^2 \vsh^8/(c^2 \Cff^2)$, or using the prefactor estimated by  \citet{Katz2010} (which agrees with} \citealt{1976ApJS...32..233W}), 
\begin{equation}\label{eq:peak-Shock-Temp} 
k T_p \simeq 10\,{\rm keV} {A^2\over Z^4} \left(\vsh\over 0.2\,c\right)^8. 
\end{equation} 
\rev{We have introduced the composition dependence, given by $A^2/Z^4 \simeq \{1,1,1/9,1/16\}$ for \{H\,,He\,,C\,,O\}.   To compute $x_0$ we must compare $T_p$ to the equilibrium temperature at the post-shock pressure: }
\begin{equation} \label{eq:equilibrium-Shock-Temp} 
k T_{\rm eq} = 15.4\,{\rm keV}  \left( E_{\rm in} \over 10^{51}\,{\rm erg}  \right)^{1/4} \left(R_\odot \over R_*\right)^{3/4}  \left(\rho_0 \vsh^2\over \rho_* v_*^2\right)^{1/4}. 
\end{equation} 
\rev{In the case that $T_p<T_{\rm eq}$, thermal equilibrium is achieved within the shock itself and so $x_0=1$.  Furthermore, an upper limit ($x_0<x_{0,i}$) is set by the pre-shock photon population.  Therefore} %$ x_0 \simeq \max({T_p/ T_{\rm eq}}, 1)
\begin{equation}\label{eq:x_0}  
x_0 \simeq \max\left[\min\left( x_{0i}, ~ 
 {T_p\over T_{\rm eq}}\right), 1
\right]. 
\end{equation} 

\rev{As a practical matter, we do not track the initial density and shock velocity of the ejecta in the simulation; these must be reconstructed from conditions at the diffusion front to evaluate $T_p$ and $T_{\rm eq}$.   To estimate $\vsh$, we note that both planar self-similar breakouts, and strongly oblique breakouts, cast away matter  at twice the shock velocity (2.03$\vsh$ and $2 \vsh$ respcectively, to be precise: see \citealt{Matzner1999} and \citealt{Matzner2013}).  So, we estimate $\vsh = f_v v/2$ with $f_v\simeq 1$, and adjust $f_v$ to best fit the simulation results.  Then $\rho_0$ can be obtained from entropy conservation: $s = p/\rho^{4/3} = (6/7^{7/3}) \vsh^2\rho_0^{-1/3}$.   Eliminating $\rho_0$, we find }
\begin{equation}
{T_p\over T_{\rm eq}} = 0.25 {A^2\over Z^4}  \left( E_{\rm in} \over 10^{51}\,{\rm erg}  \right)^{15/4} \left( R_*\over R_\odot \right)^{3/4} \left( 10\,M_\odot \over M_{\rm ej} \right)^{4} \left(s\over s_*\right)^{3/4}  \left(f_v v\over 10 v_*\right)^6. 
\end{equation} 
\rev{We see that the deviation from thermal equilibrium is a strong function of the shock velocity,. Therefore $x_0$ is quite uncertain.  However, the band-dependent light curve is much less uncertain because of the insensitivity of the color temperature to $x_0$.}  

\rev{For the upper limit, we assume that the pre-shock photons have time to be Compton scattered up to energy $kT$ long before being released, even if this did not occur during the shock transition.  (This is valid so long as $\ln[m_e c^2/(7.6kT_i)]< \int \kappa \rho c\,dt\sim \kappa\rho(\delta r)c/v$, i.e., for layers not involved in a breakout flash.) Then  }
\begin{equation}\label{eq:x_0_photonstarving} 
x_{0i} = {(n_\gamma/n_b)_{\rm eq.,2} \over (n_\gamma/n_b)_i }= {(P_{\rm rad}/P_{\rm gas})_{\rm eq., 2}\over(P_{\rm rad}/P_{\rm gas})_i }. 
\end{equation}
\rev{Here the subscripts `$i$' and `eq., 2' mean the initial hydrostatic state and an ideal post-shock state of thermal equilibrium, respectively, and `$\gamma$' and `b' mean photons and baryons.  Equation (\ref{eq:x_0_photonstarving}) defines $x_{0i}$ as the ratio between the equilibrium population of photons (expressed as photons per baryon) and the initial population; the latter expression involving pressures is valid if the mean molecular weight $\mu$ does not change across the shock front. }

\rev{The denominator of this expression can be obtained by reference to the progenitor stellar model.  It is particularly simple to evaluate within our $n=3$ polytrope progenitors, as $(P_{\rm rad}/P_{\rm gas})_i$ takes the uniform value $(1-\beta_i)/\beta_i$, where $\beta_i$ is the solution to \citeauthor{1926ics..book.....E}'s (\citeyear{1926ics..book.....E}) quartic equation }
\begin{equation} \label{eq:PradOnPgas_initial}
{(1-\beta_i)^{1/4}\over\beta_i} = \left[M\over 0.618 {(ch/G)^{3/2}/ \mu^2}\right]^{1/2} = \left[M\over 50 (0.6\,m_p/\mu)^2 M_\odot \right]^{1/2}. 
\end{equation} 
\rev{With  $\mu = (0.61, 0.61, 1.71)m_p$ for the (RSG, BSG, WR) progenitors, respectively, we calculate $(P_{\rm rad}/P_{\rm gas})_i = (0.08, 0.09,0.40)$.} 

\rev{In the numerator, the ratio $(P_{\rm rad}/P_{\rm gas})_{\rm eq., 2}$ can be easily inferred from the entropy profile within our simulations, as this ratio is conserved in adiabatic flow.  Specifically, it is $(1-\beta_f)/\beta_f$ where $\beta_f$ is the solution to }
\begin{equation}\label{eq:PradOnPgas_final}
{(1-\beta_f)^{1/4} \over \beta_f} = {\mu \over k_B} \left(a\over 3\right)^{1/4} {p_*^{3/4}\over \rho_*} \left(s\over s_*\right)^{3/4}
 = 2.0 {\mu \over 0.6\,m_p} \left({E_{\rm in}\over 10^{51}\,{\rm erg}} {R\over R_\odot}\right)^{3/4} {10\,M_\odot\over M_{\rm ej}} \left(s\over s_*\right)^{3/4}.  
\end{equation} 
\rev{In fact, as $\beta_f\ll 1$, the right-hand side of this expression is a good estimate for $(P_{\rm rad}/P_{\rm gas})_{\rm eq., 2}$. }

\rev{All together, then,  $x_{0i} \simeq (10^{3.3}, 10^{2.4}, 8.4)\times (s/s_*)^{3/4}$ for the (RSG, BSG, Ic) progenitors, where $s$ must be sampled at the radiation diffusion front. }

\bibliography{paperDraft}

\end{document}